\documentclass[%
 reprint,
superscriptaddress,
 amsmath,amssymb,
 aps,
prb,
citeautoscript,
]{revtex4-2}

\usepackage{graphicx}
\usepackage{dcolumn}
\usepackage{bm}
\usepackage{siunitx}
\usepackage{braket}
\usepackage{tikz}
\usetikzlibrary{quantikz2}
\usepackage{floatrow} 
\usepackage{multirow} 
\usepackage{csquotes}
\usepackage{booktabs}
\usepackage{algorithm}
\usepackage{algorithmic}
\usepackage{orcidlink}  
\usepackage[caption=false]{subfig}   
\usepackage{hyperref} 

\hypersetup{
  plainpages=true,
  breaklinks=true,
  hypertexnames=false,
  pageanchor=true,
  colorlinks=true,
  linkcolor=blue,
  citecolor=magenta,
  urlcolor=blue,
  pagecolor=black,
  anchorcolor=black
}

\usepackage{microtype}  
\usepackage{mleftright} 

\newcommand{\figref}[1]{\mbox{Fig.~\ref{#1}}}

\newcommand{\secref}[1]{\mbox{Sec.~\ref{#1}}}

\newcommand{\appref}[1]{\mbox{Appendix~\ref{#1}}}
\renewcommand{\eqref}[1]{\mbox{Eq.~(\ref{#1})}}
\newcommand{\figpanel}[2]{Fig.~\hyperref[#1]{\ref*{#1}(#2)}} 

\newcommand{\figpanels}[3]{Fig.~\hyperref[#1]{\ref*{#1}(#2)--(#3)}} 
\newcommand{\figpanelNoPrefix}[2]{\hyperref[#1]{\ref*{#1}(#2)}} 
\newcommand{\figpanelsNoPrefix}[3]{\hyperref[#1]{\ref*{#1}(#2)--(#3)}} 

\newcommand{\figurepanel}[2]{Figure~\hyperref[#1]{\ref*{#1}(#2)}} 

\makeatletter

\begin{document}

\preprint{APS/123-QED}
 
\title{Simulation-Based Quantum System Inference with Neural Posterior Estimation}   

\author{Hang Zou}
\affiliation{Department of Computer Science and Engineering, Chalmers University of Technology and University of Gothenburg, 41296 Gothenburg, Sweden} 

\author{Anton Frisk Kockum}
\affiliation{Department of Microtechnology and Nanoscience, Chalmers University of Technology, 41296 Gothenburg, Sweden} 

\author{Martin Rahm}
\affiliation{Department of Chemistry and Chemical Engineering, Chalmers University of Technology, 41296 Gothenburg, Sweden} 

\author{Simon Olsson}
\email{simonols@chalmers.se}
\affiliation{Department of Computer Science and Engineering, Chalmers University of Technology and University of Gothenburg, 41296 Gothenburg, Sweden} 

\begin{abstract}
Models of quantum systems faithfully map system parameters to observations, but the inverse problem of parameter inference from measurement data presents a fundamental challenge: computationally intractable likelihoods due to an exponentially large Hilbert space. Here, we introduce simulation-based quantum system inference, a unified, likelihood-free framework that learns parameter posteriors directly from classical simulation data. The central idea is to pair polynomial-cost classical simulators, such as Pauli propagation and tensor networks, with normalizing flows or other neural density estimators for accurate, reusable inference. A single model, trained once, maps any new measurement record to its posterior in one forward pass---turning per-experiment inference into a fixed, up-front cost. We numerically demonstrate the framework’s versatility across Pauli noise learning, quantum error mitigation, quantum state tomography, and Hamiltonian learning, with examples involving 81-qubit shallow circuits and 735-parameter inference. In each case, the approach yields accurate estimates of identifiable parameters, while posterior uncertainty provides additional diagnostics of non-identifiability and indicates where further characterization is needed. Our framework reduces data-acquisition requirements in quantum experiments and accelerates parameter inference, providing a practical route to characterizing and improving large-scale quantum systems.\looseness=-1
\end{abstract}
 
\maketitle

\section{Introduction} 
Quantum theory provides a precise forward map, predicting how the expectation values of chosen observables evolve from an initial state under a Hamiltonian. Quantum characterization demands the inverse---inferring Hamiltonians, states, or noise processes from limited measurement data. Such \textit{inverse problems} pervade modern quantum science, from the reconstruction of entangled states~\cite{cramer2010efficient,torlai2018neural, PhysRevLett.127.140502,PRXQuantum.6.030202,gebhart2023learning} and the benchmarking of quantum processors~\cite{PRXQuantum.6.030202,proctor2025benchmarking, Nielsen2021gatesettomography,gebhart2023learning} to the identification~\cite{PhysRevLett.112.190501,wang2017experimental,gentile2021learning,f58h-zxs3} and design~\cite{PhysRevX.8.031029,PhysRevResearch.6.033080, kokail2026inversequan} of many-body Hamiltonians. Yet this inversion is obstructed by an intractable likelihood function $p(x|\theta)$, which quantifies how probable the observed data $x$ are under candidate parameters $\theta$. Evaluating it generally requires simulating the full quantum dynamics~\cite{PhysRevLett.112.130402}, the cost of which grows exponentially with system size. Conventional likelihood-based inference is therefore impractical beyond small quantum systems~\cite{PhysRevLett.112.130402,catana2014maximum,PhysRevApplied.23.044040}.

Simulation-based inference (SBI) side-steps intractable likelihood evaluations by treating the forward simulator as the statistical model~\cite{cranmer2020frontier,deistler2025simul}. The sampling distribution induced by the simulator, $x\sim p(x|\theta)$, implicitly specifies the likelihood, thereby enabling inference of the posterior $p(\theta| x)$---the probability distribution over the unknown parameters given observed data. This paradigm has found success across diverse scientific disciplines, from astrophysics to molecular biophysics~\cite{dax2025real,DINGELDEIN2025102988, atlas2025implementation,zhang2026discovery,doi:10.1073/pnas.2420158122}.

To date, traditional SBI techniques based on approximate Bayesian computation have been applied to quantum parameter inference~\cite{PhysRevApplied.23.044040,PhysRevLett.112.130402,catana2014maximum}, but major computational barriers limit their practical deployment. The curse of dimensionality manifests in two ways. In forward simulation, even a single evaluation of a quantum model becomes prohibitive for large systems. In statistical inference, learning high-dimensional quantum systems generally requires many samples, further increasing the computational burden. Beyond these scaling barriers, the computational effort invested in one inference does not accelerate the next: the procedure must be rerun from scratch for every new observation.

Against this backdrop, advances on two complementary fronts now provide the key ingredients for efficient high-dimensional quantum-system inference. On the statistical side, deep generative modeling has transformed high-dimensional density estimation~\cite{9555209,9089305}. Building on this progress, neural posterior estimation (NPE) trains a conditional generative model to learn complex mappings between parameters and observations~\cite{10.5555/3294771.3294894, 10.5555/3157096.3157212,greenberg2019automatic,10.5555/3666122.3666859,zammit2025neural}. NPE provides amortized inference by reusing this model for rapid, high-throughput inference on new data, thereby offsetting its initial training cost over time~\cite{zammit2025neural}. This amortization is especially valuable in the quantum domain, where hardware access is costly, but characterization tasks such as real-time noise tracking~\cite{proctor2020detecting,kim2025error} and cross-device benchmarking~\cite{zhu2022cross} must be carried out repeatedly. Yet harnessing this fit at scale requires more than importing NPE: it demands a forward simulator cheap enough to generate training data for large quantum systems.

On the simulation side, the frontier of classical simulability for quantum systems has expanded. Realistic noise intrinsically restricts state complexity, giving rise to a range of polynomial-cost classical simulation algorithms for noisy quantum circuits~\cite{PhysRevX.10.041038,PRXQuantum.5.010308, doi:10.1126/sciadv.adk4321,PhysRevLett.133.120603,aharonov2023polynomial,schuster2025polynomial,rudolph2025pauli}.  In addition, exploitable \textit{structures} can further reduce simulation complexity, including symmetry reductions~\cite{anschuetz2023efficient,nnzz-481j,chang2026permutationequivariant}, stabilizer circuits~\cite{PhysRevA.70.052328}, free fermions~\cite{jozsa2008matchgates}, dynamical Lie algebras~\cite{3y65-f5w6}, and limited state or operator entanglement~\cite{PhysRevLett.91.147902, dowling2026operatorentanglement,PhysRevLett.131.160402}. 
Numerical techniques such as Pauli propagation~\cite{PhysRevLett.133.120603,doi:10.1126/sciadv.adk4321,schuster2025polynomial,aharonov2023polynomial,rudolph2025pauli} and tensor networks~\cite{PhysRevX.10.041038,doi:10.1137/050644756,PRXQuantum.5.010308,doi:10.1126/sciadv.adk4321,hartnett2026fastaccurate,deger2026efficiently} can leverage these features to efficiently approximate systems exceeding 100 qubits. This scale is comparable to those of leading quantum hardware accessible to external researchers, such as Google’s Willow~\cite{acharya2025willow}, IBM’s Heron~\cite{mckay2023benchmarking}, and QuEra’s Aquila~\cite{wurtz2023aquilaqueras256}. For suitably noisy or structured problems, state-of-the-art classical simulators can surpass contemporary error-mitigated quantum computers in both runtime and precision~\cite{doi:10.1126/sciadv.adk4321,PhysRevLett.133.120603,PRXQuantum.5.010308,hartnett2026fastaccurate}. Accelerated by modern hardware backends~\cite{bayraktar2023cuquantum,10.1145/3762672,PRXQuantum.3.020331}, these simulators make it feasible to generate the large datasets required for NPE training, opening a path to scalable, likelihood-free quantum inference. \looseness=-1

Here, we bring these two developments together and develop \textit{simulation-based quantum system inference} into a general framework (\figref{fig:sbi}): a single template that pairs polynomial-cost classical simulators with neural posterior estimation to unify otherwise tailor-made characterization tasks. Trained entirely on classical simulations, the estimator approximates posterior distributions for new experimental observations without retraining, amortizing inference across experiments. Posterior samples provide parameter estimates together with uncertainty information, subject to the assumed model and the accuracy of the learned posterior. We demonstrate the framework on four representative inverse problems (\figref{fig:sbi}, bottom).

First, we show that our amortized models efficiently learn high-dimensional Pauli noise, scaling to 50-qubit shallow circuits while accurately tracking parameter drift across instances. The resulting noise posteriors directly diagnose which Pauli-noise parameters are identifiable under the chosen characterization protocol~\cite{chen2023learnability,69wc-gzl6}. We further find that the training-data budget required for high-accuracy inference grows only approximately linearly with system size, indicating a route toward characterization beyond 100 qubits for the sparse Pauli noise model. 

Next, we apply the inferred noise posterior to quantum error mitigation (QEM) protocols that require an explicit noise model, including probabilistic error cancellation (PEC)~\cite{van2023probabilistic,temme2017error} and zero-noise extrapolation (ZNE)~\cite{temme2017error,li2017efficient,kim2023evidence}. Both protocols achieve mitigation accuracy comparable to that obtained using the ground-truth noise parameters predefined in the simulations. We further introduce a digital-twin strategy, in which posterior-driven simulators generate training data for machine-learning QEM (ML-QEM)~\cite{liao2024machine,czarnik2021error}, effectively overcoming the bottleneck of quantum data scarcity. Applied to Trotterized Ising dynamics, it outperforms the ZNE baselines by over twofold in error reduction, and extrapolates beyond the trained circuit depths.

Beyond noise analysis, we apply the framework to quantum state tomography (QST), constructing a reusable estimator for 12-qubit states through structured parameterized quantum circuits~\cite{PhysRevA.101.052316}. The model reconstructs a range of randomly sampled states with fidelity typically exceeding 99\%, without per-state optimization. 

Finally, we demonstrate Hamiltonian learning in simulated $9\times 9$ Rydberg atom arrays~\cite{f58h-zxs3} by inferring atomic displacements from local observables evaluated after short-time Hamiltonian evolution. The inferred displacements agree closely with the prescribed values (coefficient of determination $R^2=0.983$), and the site-resolved posterior uncertainties highlight locations that may warrant further calibration. \looseness=-1

Collectively, our results span systems of up to 81 qubits and inference over as many as 735 latent parameters. This parameter dimensionality lies at the frontier of both Bayesian quantum parameter estimation~\cite{belliardo2025multi} and SBI in other domains~\cite{moss2026fnope,deistler2025simul}. Our framework provides a principled and reusable tool for high-dimensional quantum characterization. We anticipate that this paradigm will extend to broader applications in quantum technologies, such as non-Markovian noise learning~\cite{WhiteNonMarkovian,PhysRevA.97.012127,1ncg-11hz}, quantum sensor calibration~\cite{PhysRevLett.123.230502,PhysRevApplied.15.044003} and feedback control~\cite{vepsalainen2022improving,shulman2014suppressing}.

The remainder of this paper is organized as follows. In \secref{sec:sbi}, we introduce our inference framework and lay out the foundations of normalizing-flow density estimation. We then apply the framework to sparse Pauli-noise learning in \secref{sec:pauli_noise} and use the learned noise models for QEM in \secref{sec:qem}, where we also introduce the digital-twin strategy. Next, we develop a neural estimator for QST in \secref{sec:qst}. In \secref{sec:rydberg}, we present Hamiltonian learning of Rydberg atom arrays. Finally, we conclude in \secref{sec:discussion} and highlight promising avenues for future research.

\section{Methods \label{sec:sbi}}
In this section, we first present our framework for simulation-based quantum-system inference from a Bayesian perspective. We then introduce conditional normalizing flows as the posterior density estimators used in this work.

\subsection{Simulation-based quantum system inference}
\begin{figure*}[t]
    \centering
    \includegraphics[width=\linewidth]{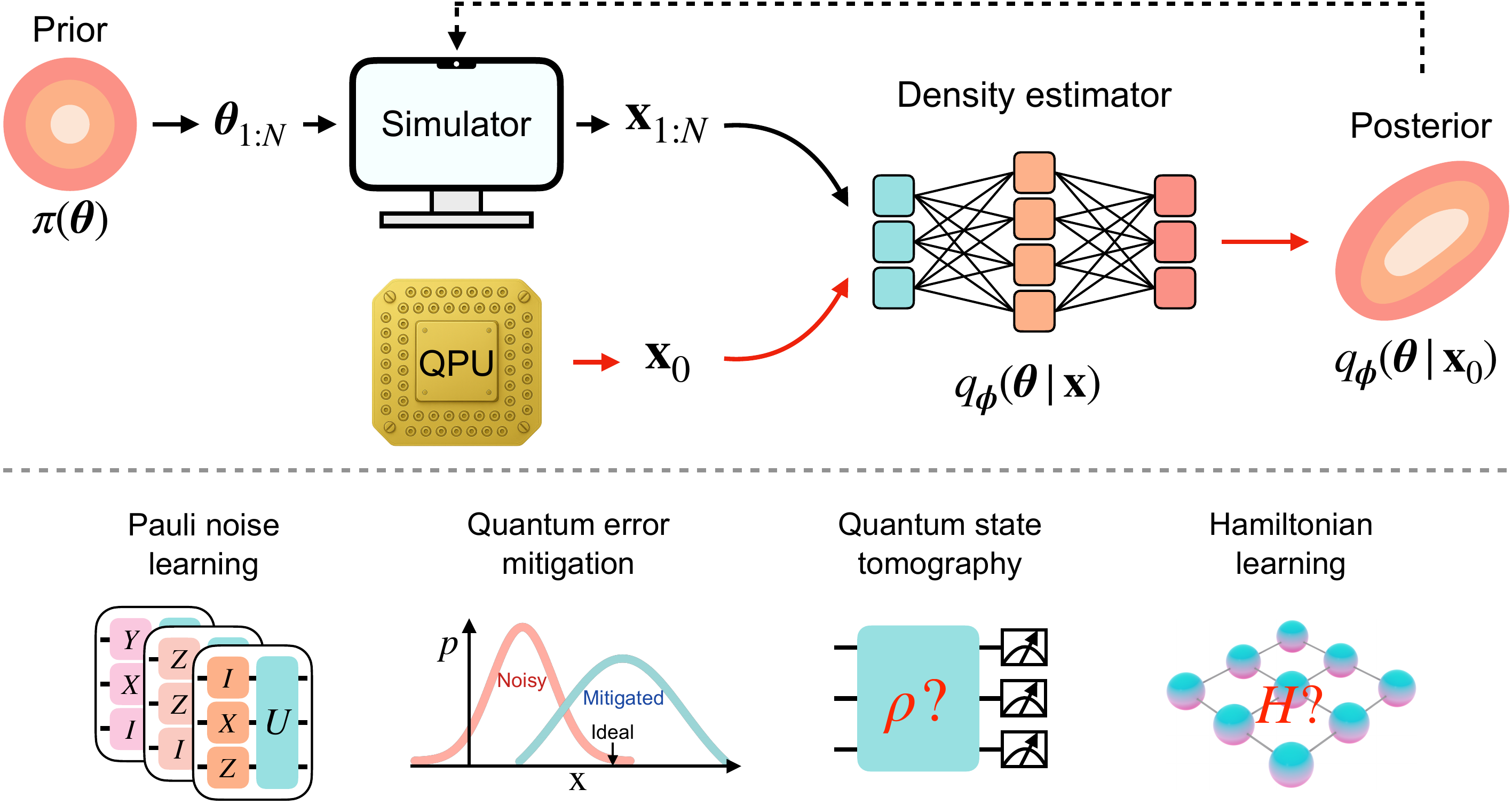}
    \caption{Simulation-based quantum system inference.
        Top, schematic of the framework. 
        Training phase (solid black arrows): parameters $\boldsymbol{\theta}_{1:N}$ sampled from a prior $\pi(\boldsymbol\theta)$ are passed through a forward simulator to produce synthetic observations $\mathbf x_{1:N}$, on which a neural density estimator $q_{\boldsymbol{\phi}}(\boldsymbol{\theta}| \mathbf x )$ learns to approximate the posterior. 
        Inference phase (red arrows): given a target observation $\mathbf x_0$ from a quantum processing unit (QPU), the trained estimator efficiently evaluates the posterior $q_{\boldsymbol\phi}(\boldsymbol\theta| \mathbf x_0)$ without further simulation.
        Sequential training (dashed black arrow): an optional sequential refinement round that reuses the posterior as a new proposal. 
        Bottom, the representative applications demonstrated in this work, including Pauli noise learning, quantum error mitigation, quantum state tomography, and Hamiltonian learning.
        }
    \label{fig:sbi}
\end{figure*}

We cast a broad class of quantum-system learning tasks as a unified Bayesian inverse problem. Given an observation $\mathbf{x}$, Bayes' formula yields the posterior over the latent parameters $\boldsymbol{\theta}$, $p(\boldsymbol{\theta}|\mathbf{x})\propto p(\mathbf{x}|\boldsymbol{\theta})\pi(\boldsymbol{\theta})$. Here, $\pi(\boldsymbol{\theta})$ encodes the prior over physically admissible parameter values. The quantum process defines the target likelihood \(p(\mathbf{x}|\boldsymbol{\theta})\) through its exact forward map from the latent parameters to the chosen observable values. Although this likelihood is generally intractable, SBI performs inference using samples generated by scalable classical simulators $\mathcal{S}$ that approximate the quantum forward map. Accordingly, each SBI task is specified by $(\boldsymbol{\theta},\pi,\mathbf{x},\mathcal{S})$.

As illustrated in the upper part of Fig.~\ref{fig:sbi}, parameters $\boldsymbol{\theta}_{1:N}$ are drawn from the prior $\pi(\boldsymbol{\theta})$ and propagated through $\mathcal{S}$ to generate simulated data $\mathbf{x}_{1:N}$, thereby encoding the likelihood implicitly without closed-form evaluation. NPE trains a conditional normalizing flow $q_{\boldsymbol\phi}(\boldsymbol{\theta}|\mathbf{x})$ (see \secref{nf}) on the resulting pairs $\{(\boldsymbol{\theta}_i,\mathbf{x}_i)\}_{i=1}^N$ via maximum likelihood,  
\begin{equation}
\boldsymbol\phi^\star=\arg\min_{\boldsymbol\phi}\,\mathbb{E}_{(\boldsymbol{\theta},\mathbf{x})}\mleft[-\log q_{\boldsymbol\phi}(\boldsymbol{\theta}|\mathbf{x})\mright].
\end{equation}
Once trained, the NPE model provides a tractable surrogate for the posterior $p(\boldsymbol{\theta}|\mathbf{x}_0)\approx  q_{\boldsymbol\phi^\star}(\boldsymbol{\theta}|\mathbf{x}_0)$, enabling direct generation of posterior samples for any $\mathbf{x}_0$ without further simulation or iterative sampling. 

When only a single $\mathbf{x}_0$ is of interest, sequential NPE (SNPE) reduces the total simulation cost at the expense of a model that cannot be reliably reused across observations. At each round $r$, the previous posterior serves as the proposal, $\tilde{\pi}_r(\boldsymbol{\theta}) = q_{\boldsymbol\phi_{r-1}^{\star}}(\boldsymbol{\theta}|\mathbf{x}_0)$; samples from $\tilde{\pi}_r(\boldsymbol{\theta})p(\mathbf{x}|\boldsymbol{\theta})$ are appended to the cumulative training set, and training is continued under the same objective. 
 

\subsection{Normalizing flows \label{nf}}
To model the posterior distribution $q_{\boldsymbol{\phi}}(\boldsymbol{\theta} | \mathbf{x})$ over the parameter space $\boldsymbol{\theta}\in\mathbb{R}^n$, we employ a conditional normalizing flow~\cite{dinh2017density,10.5555/3045118.3045281,10.5555/3294771.3294994,10.5555/3454287.3454962}. This flow constructs a smooth and invertible map $f_{\boldsymbol{\phi}}(\cdot; \mathbf{x}): \mathbb{R}^n \to \mathbb{R}^n$ that transforms a latent variable $\mathbf{z}$ drawn from a simple base density $q_{0}(\mathbf{z})$ into the target parameter $\boldsymbol{\theta}$, conditioned on the observation $\mathbf{x}$. Through the change-of-variables formula, the posterior density evaluates exactly to
\begin{equation}
    q_{\boldsymbol{\phi}}(\boldsymbol{\theta} | \mathbf{x}) = q_0\mleft(f_{\boldsymbol{\phi}}^{-1}(\boldsymbol{\theta};\mathbf{x})\mright) \mleft|\det \frac{\partial f_{\boldsymbol{\phi}}^{-1}(\boldsymbol{\theta};\mathbf{x})}{\partial \boldsymbol{\theta}}\mright|.
\end{equation}
The invertibility and exact likelihood computation have established normalizing flows as powerful tools across classical and quantum many-body problems~\cite{noe2019boltzmann,asghar2024efficient,PhysRevD.100.034515,zou2025generative}.

In this work, we use neural spline flows~\cite{10.5555/3454287.3454962}, which combine rational-quadratic splines for strictly monotonic, invertible mappings with autoregressive transformations for tractable Jacobian determinants for fast training and inference. Details of the flow architecture and computational hyperparameters are provided in \appref{spline_flows} and \appref{hyperparameters}, respectively. 

\section{Learning sparse Pauli noise\label{sec:pauli_noise}} 
Learning noise models is essential for benchmarking quantum devices, guiding calibration, and enabling quantum error mitigation~\cite{PRXQuantum.6.030202,proctor2025benchmarking}. Noise in two-qubit gates is a major contributor to errors in quantum circuits on current hardware~\cite{kim2023evidence,van2023probabilistic}. 
We therefore first present an application of our framework to inferring gate-by-gate Pauli error probabilities for two-qubit gates from a single set of circuit measurements. The inferred noise models can be used for several purposes, e.g., in quantum error mitigation protocols, as we will demonstrate in \secref{sec:qem}.

\begin{figure}[t]
    \centering
    \includegraphics[width=0.8\linewidth]{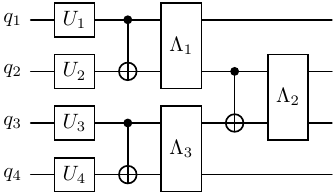}
    \caption{Brickwall circuit template used in this study, shown for $n=4$. Each qubit $q$ is initialized by a single-qubit rotation $U_q$, and each gate $\text{CNOT}_{k,k+1}$ is followed by a 2-local Pauli noise channel $\Lambda_k$.}
    \label{fig:circuit_layout}
\end{figure}

We consider an $n$-qubit register subjected to a circuit template over a gate set 
\begin{equation}
    G = \{P_0, U_\mathrm{prep}, U_{\text{BW}}, \mathcal{M}\}.
    \label{eq:gateset}
\end{equation} 
The system is initialized in $\rho_0=\ket{0}\bra{0}^{\otimes n}$ via $P_0$, followed by a non-commuting local preparation layer $U_\mathrm{prep} =\bigotimes_{q=1}^{n}U_q$ with $U_q=R_z^{(q)}(\pi/4) R_y^{(q)}(\pi/4)$, and a brickwall entangling layer $ U_{\text{BW}}$ composed of $n-1$ nearest-neighbor CNOT gates split into an odd-control and an even-control sublayer (see \figref{fig:circuit_layout}). The measurement ensemble $\mathcal{M}$ consists of all weight-1 and nearest-neighbor weight-2 Pauli observables, regardless of the circuit topology. 

Each gate $\text{CNOT}_{k,k+1}$ ($k = 1, \ldots, n-1$) is accompanied by a 2-local Pauli channel,
\begin{equation}
    \Lambda_k (\rho) = \mleft(1 - \sum_{P \in \mathcal{P}} p_{k,P} \mright)\rho + \sum_{P \in \mathcal{P}} p_{k,P} P \rho P,
\end{equation}
where $\mathcal{P}=\{I,X,Y,Z\}^{\otimes 2}\setminus \{II\}$ spans the 15-dimensional basis of two-qubit Pauli errors and $p_{k,P}$ are the corresponding error probabilities. The focus on Pauli noise is broadly applicable in practice, as randomized compiling can reshape arbitrary noise into Pauli channels~\cite{PhysRevA.94.052325}. The 2-local restriction further balances model sparsity against the ability to capture short-range crosstalk~\cite{van2023probabilistic}.
We treat state preparation and measurement (SPAM) and single-qubit gates as ideal, restricting characterization to the CNOT layers. For deeper circuits with repeated blocks, we assume stationary Markovian noise, keeping the noise parameters fixed across repetitions, as implemented in \secref{sec:dt_mlqem}.

Within the SBI framework, the latent vector $\boldsymbol{\theta}=\{p_{k,P}\} \in \mathbb{R}^{15(n-1)}$ covers the $n-1$ CNOT gates in the brickwall circuit; unused connections remain uncharacterized when this noise model is applied to real hardware. We adopt a uniform prior $p_{k,P}\sim\mathcal{U}[0,\theta_{\text{max}}]$ ($\theta_{\text{max}}=0.001$), yielding a maximum two-qubit gate error of 1.5\%, consistent with current superconducting processors~\cite{kim2025error,van2023probabilistic}. The observation $\mathbf{x}\in\mathbb{R}^{12n-9}$ collects all expectation values under $\mathcal{M}$. The simulator $\mathcal{S}$ evaluates these expectation values exactly via Pauli propagation at $\mathcal{O}(n^2)$ cost (see Appendices~\ref{pauli_propagation} and \ref{pauli_noise_pp}), requiring approximately \SI{0.1}{\second} at $n=50$ on a single CPU thread [see \figpanel{fig:pauli_noise_pp}{d} in \appref{pauli_noise_pp}]. All observations for both training data and ground-truth benchmarks are generated by this simulator. \looseness=-1

\begin{figure*}[t]
    \centering
    \includegraphics[width=1\linewidth]{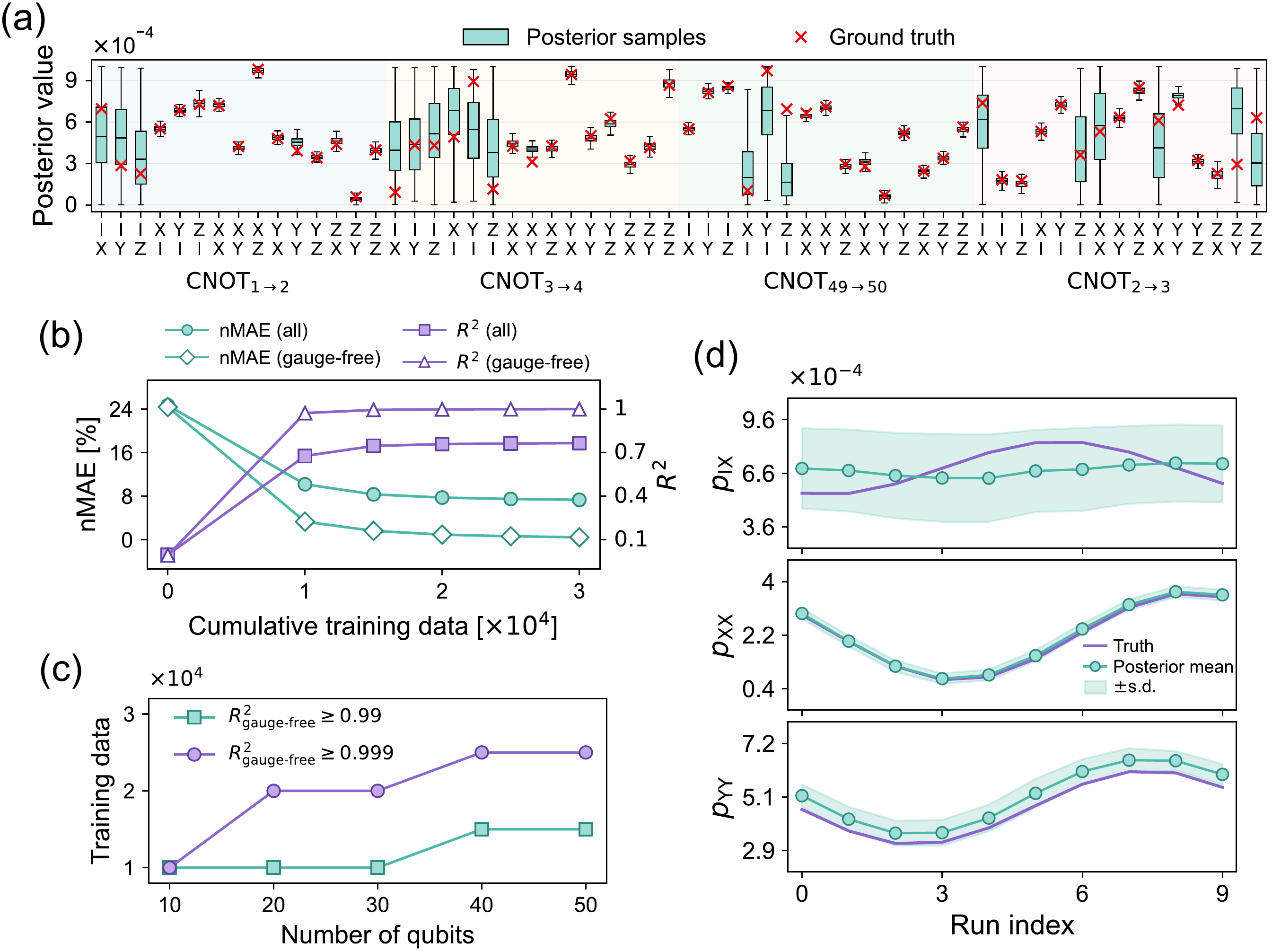}
    \caption{{Scalable Pauli noise learning with NPE.} 
    (a) Posterior distributions of Pauli error coefficients for four representative CNOT gates within a 50-qubit brickwall circuit, shown as box plots over 2,000 posterior samples. Gates are ordered by brickwall layer, with odd-control gates first, rather than by qubit index. The inference is performed using an amortized NPE model trained on 30,000 simulations. Red crosses denote ground-truth values. Broad posteriors on specific parameters reflect gauge ambiguity (see main text and \appref{gauge}).
    (b) Convergence of nMAE and $R^2$ as a function of cumulative training data across SNPE rounds for the 50-qubit model, starting from 10,000 simulations and incrementing by 5,000 per round. Metrics are shown for both the full parameter space and the gauge-free subset.
    (c) Training-data budget required to reach high-accuracy thresholds ($R^2_{\text{gauge-free}} \geq 0.99$ and $\geq 0.999$) as a function of system size, extracted from SNPE training analogous to (b) for each system size.
    (d) Tracking selected drifting Pauli noise coefficients for the CNOT$_{1\rightarrow 2}$ gate shown in panel (a). Prescribed sinusoidal ground-truth profiles (purple curves) are compared with posterior means (teal circles), with shaded bands indicating $\pm 1$ posterior standard deviation.}
    \label{fig:noise}
\end{figure*}

For $n=50$ qubits (a $735$-parameter noise model), \figpanel{fig:noise}{a} displays the posterior distributions of Pauli error coefficients for four representative CNOT gates, using 2,000 posterior samples. Most parameters concentrate tightly around the ground-truth values (red crosses), while a subset exhibits broad posteriors. The high variance is a diagnostic: it isolates exactly those parameters that are unidentifiable from composite circuit measurements. Pauli errors propagate through subsequent CNOT gates, coupling error coefficients from adjacent gates into gauge-equivalent combinations whose individual contributions cannot be resolved~\cite{Nielsen2021gatesettomography}. This ambiguity persists even with tomographically complete inputs and measurements, since noise contributions from adjacent gates cannot be separated, as illustrated in \appref{gauge}. In the $n$-qubit brickwall circuit considered here, $6(n-2)$ error coefficients are affected by this gauge freedom, consistent with the complete posterior distributions in \appref{complete_posterior}. This numerical approach complements the analytic analysis~\cite{chen2023learnability} and extends naturally to complex circuits where the algebra becomes cumbersome.  

\figurepanel{fig:noise}{b} tracks the convergence of two accuracy metrics as functions of cumulative training data across successive SNPE rounds: the normalized mean absolute error (nMAE), defined as the MAE between posterior means ($\bar{\theta}_i$) and ground-truth values ($\theta_i^{\text{true}}$) normalized by the prior range,
$\text{nMAE} = \frac{1}{d} \sum_{i=1}^{d} {|\bar{\theta}_i - \theta_i^{\text{true}}|}/{\theta_{\text{max}}} \times 100\%$, with $d$ the parameter dimension, and the coefficient of determination $R^2$. Training begins with 10,000 simulations and increments by 5,000 per round. For the full $d$-dimensional parameter space, both metrics plateau---nMAE at a nonzero floor and $R^2$ below unity---as a consequence of gauge freedom. Within the gauge-free subspace, both metrics converge with high accuracy, confirming that the learnable noise structure is faithfully captured. \figurepanel{fig:noise}{c} examines how the training-data budget required to reach high-accuracy thresholds scales with system size. The data requirement grows approximately linearly from $n=10$ to $n=50$, as estimated from the coarse training increments. Since both parameter and observation dimensions grow only linearly with system size---and the training-data budget scales likewise---this favorable scaling positions the approach for application on systems well beyond 50 qubits.  

Beyond static noise, we evaluate the amortized estimator on drifting noise parameters. \figurepanel{fig:noise}{d} tracks three representative coefficients across ten runs under prescribed sinusoidal drifts. Each run index represents either temporal drift or cross-device variation. Consistent with \figpanel{fig:noise}{a}, these examples show accurate tracking of $p_{XX}$ with low posterior uncertainty, a systematic overestimate of $p_{YY}$ despite capturing its phase, and a broad, out-of-phase posterior estimate for $p_{IX}$. The estimator provides consistent inference behavior across varying instances without per-instance calibration. Moreover, rapid tracking of parameter drifts could turn noise characterization into actionable feedback for stabilizing device performance~\cite{proctor2020detecting,vepsalainen2022improving}. 


\begin{figure*}[t]
    \centering
    \includegraphics[width=1\linewidth]{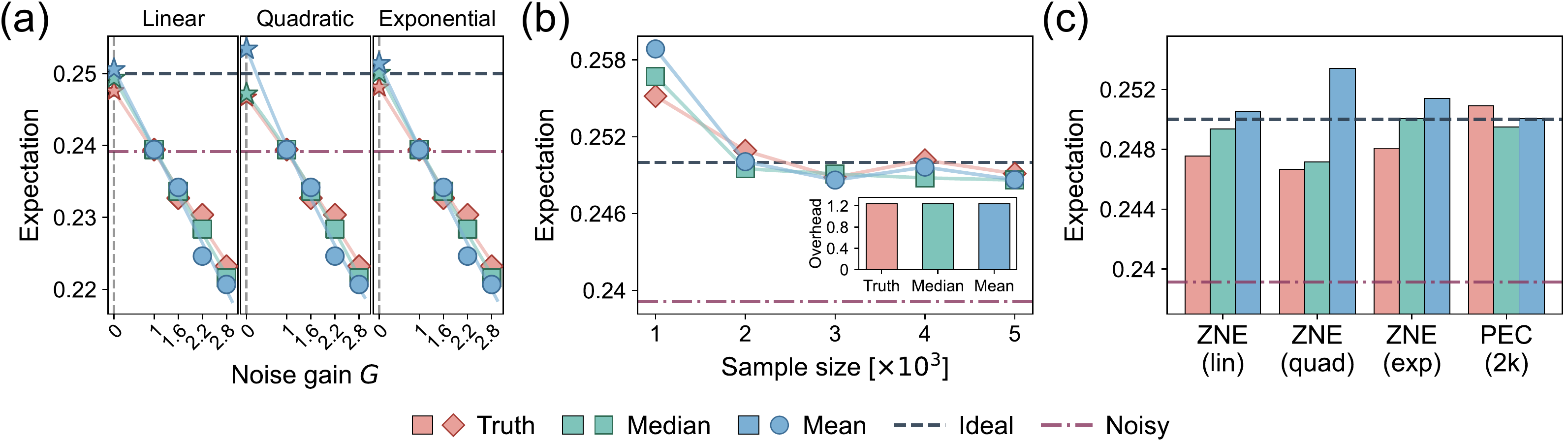}
    \caption{{ZNE and PEC based on the SBI noise posterior.} All protocols are parameterized by the ground-truth noise (red), the SBI posterior median (green), and the posterior mean (blue). 
    (a) Zero-noise extrapolation (ZNE) of the expectation value $\langle Z_5Z_6Z_7Z_8\rangle$ for a 15-qubit probe state. Mitigated estimates at the zero-noise limit ($G=0$, star) are obtained via linear (left), quadratic (middle), and exponential (right) extrapolation from unmitigated ($G=1$) and probabilistic-error-amplified ($G\in\{1.6,2.2,2.8\}$) results, with each expectation value estimated using 2,000 random circuit samples per gain level. 
    (b) Convergence of the PEC estimate for the same observable as a function of the total quasi-probability samples (in thousands). The inset shows the sampling overhead of PEC under each noise model.
    (c) Comparison of the mitigated results from panels (a) and (b): ZNE with different extrapolation fits, and PEC evaluated at a sample size of 2,000 (2k). The dashed black and dash-dotted red lines indicate the ideal and unmitigated values, respectively.} 
    \label{fig:zne_pec}
\end{figure*}

\section{Posterior-driven quantum error mitigation \label{sec:qem}}
A central application of the inferred noise posterior is QEM, inferring ideal expectation values from noisy measurements through additional circuit executions and classical post-processing~\cite{RevModPhys.95.045005}. One class of QEM methods uses noise models to guide circuit modifications and sampling, including ZNE and PEC, as we explore in \secref{sec:model_based_qem}. ZNE uses probabilistic noise amplification to evaluate expectation values at increased noise levels and extrapolate to the zero-noise limit~\cite{kim2023evidence}. PEC represents inverse noise channels as quasiprobability distributions over implementable operations and cancels noise by reweighting measurement outcomes~\cite{temme2017error,van2023probabilistic}. Achieving a given precision with these methods requires substantial sampling, whose cost can grow exponentially with circuit size~\cite{PhysRevLett.131.210601,PhysRevLett.131.210602}. An alternative class of methods estimates corrections from classically tractable reference circuits~\cite{czarnik2021error,liao2024machine, lolur2023reference,zou2025multireference}. In \secref{sec:dt_mlqem}, we leverage the classical tractability of noisy circuits in our setting to propose a strategy for reducing the data collection costs of these methods. 

\subsection{Sampling-based QEM with inferred noise \label{sec:model_based_qem}}

We assess the effectiveness of the learned posterior by using its mean and median to parameterize two standard QEM protocols: determining the quasiprobabilities for PEC (see \appref{PEC}) and constructing the amplification channels for ZNE (see \appref{ZNE}). All noisy circuit simulations use the ground-truth noise model, while the inferred models are used only to sample the random circuits employed in mitigation. Expectation values entering the mitigation procedures are estimated from 256 shots per random circuit instance.

Figure~\ref{fig:zne_pec} presents the mitigated expectation value $\langle Z_5 Z_6 Z_7 Z_8 \rangle$ of the 15-qubit probe state $|\psi\rangle = {U}_{\text{BW}} \bigotimes_{q=1}^{15} \mleft[R_z^{(q)}(\pi/4) R_y^{(q)}(\pi/4) |0\rangle \mright]$, where ${U}_{\text{BW}}$ is the brickwall CNOT layer. \figurepanel{fig:zne_pec}{a} displays the ZNE extrapolation curves under linear, quadratic, and exponential models. Across all three fits, the posterior mean and median yield mitigated estimates close to those obtained using the ground-truth noise, with all results showing clear improvement over the unmitigated baseline. \figurepanel{fig:zne_pec}{b} presents the convergence of PEC as a function of the quasiprobability sample budget. The results for all three noise parameterizations converge closely toward the ideal value, with nearly identical sampling overheads (inset). \looseness=-1

\figurepanel{fig:zne_pec}{c} summarizes the comparison across all methods. In several cases, protocols based on posterior noise estimates yield mitigated values closer to the ideal than those using the ground-truth parameters. These improvements may reflect finite-sampling fluctuations, and do not establish a systematic advantage. The comparable mitigation performance across the tested parameterizations suggests that QEM remains effective despite the non-identifiability of gauge parameters, provided the noise models represent the same effective circuit channel. A similar conclusion was reported in Ref.~\cite{69wc-gzl6}. This robustness to gauge freedom supports the use of the inferred noise models in the QEM protocols.

\subsection{Digital-twin ML-QEM \label{sec:dt_mlqem}}
\begin{figure*}[t]
    \centering
    \includegraphics[width=\linewidth]{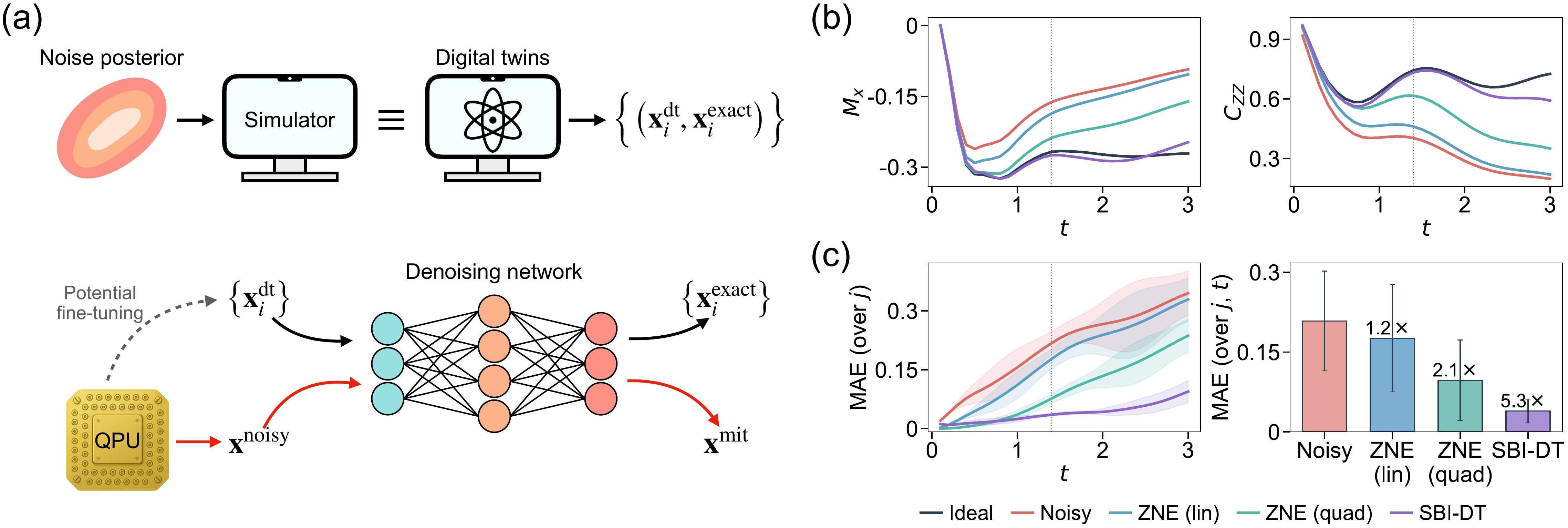} 
    \caption{Digital-twin ML-QEM with the SBI noise posterior. (a) Workflow. Training phase: the noise posterior drives a classical simulator to produce paired data $\{(\mathbf{x}_i^{\text{dt}},\mathbf{x}_i^{\text{exact}})\}$ for training a denoising network (black arrows), with optional fine-tuning on real data (gray arrows). Deployment phase: the pre-trained network maps noisy outputs $\mathbf{x}^{\text{noisy}}$ to mitigated estimates $\mathbf{x}^{\text{mit}}$ (red arrows).
    (b, c) Application to Trotterized dynamics of 12-qubit Ising chains, comparing ideal, raw noisy, linear and quadratic ZNE (gate folding at noise gains $G\in\{1,3,5\}$; see \appref{ZNE}), and SBI-DT. (b) Time-evolved transverse magnetization $M_x$ (left) and nearest-neighbor correlations $C_{ZZ}$ (right), averaged over 50 test coupling strengths $j\sim\mathcal{U}[-\pi,0]$. The vertical dotted line at $t=1.4$ separates the training and extrapolation regions. (c) Time-resolved $\mathrm{MAE}(t) = \frac{1}{|\mathcal{S}|}\sum_{O\in\mathcal{S}} |\langle O\rangle(t) - \langle O\rangle_{\text{exact}}(t)|$ across all observables $\mathcal{S} = \{X_i\}_i \cup \{Z_iZ_{i+1}\}_i$, with shaded $\pm 1$ standard deviation over test $j$ (left), and MAE aggregated over all $j$ and $t$, with improvement factors relative to the noisy baseline (right).
    } 
    \label{fig:digital_twin}
\end{figure*}

Since both ZNE and PEC require extensive sampling per observable, we turn to ML-QEM, which enables zero-shot correction at much lower runtime overhead~\cite{liao2024machine}, but typically demands large volumes of hardware data. We propose a digital-twin strategy to overcome this [\figpanel{fig:digital_twin}{a}]: noise parameters drawn from the SBI posterior parameterize a classical simulator that emulates the noisy device, producing paired noisy--ideal data $\{(\mathbf{x}^{\mathrm{dt}}_i, \mathbf{x}^{\mathrm{exact}}_i)\}$ from classically tractable circuits~\cite{czarnik2021error,liao2024machine, lh6x-7rc3,jozsa2008matchgates}. A denoising model trained on these synthetic pairs learns the correction map, with potential fine-tuning on real data to absorb model mismatch.

Cast in SBI terms, mitigation is itself an inference problem. The latent parameter is the observable correction $\boldsymbol{\theta}\equiv\Delta\mathbf{x}=\mathbf{x}^{\mathrm{exact}}-\mathbf{x}^{\mathrm{dt}}$ (or equivalently $\mathbf{x}^{\mathrm{exact}}$), and the observation $\mathbf{x}$ is the noisy data ($\mathbf{x}^{\mathrm{dt}}$ in training, $\mathbf{x}^{\mathrm{noisy}}$ at deployment). The simulator $\mathcal{S}$ is the digital twin, and the prior $\pi$ is left implicit here, induced by the training circuit ensemble rather than specified. Here we instantiate the twin $S$ as an efficient Clifford (stabilizer) simulator: the training circuits retain the CNOT gates of the target circuit and sample each Pauli rotation angle from $\{0,\pi/2,\pi,3\pi/2\}$, following Clifford data regression (CDR)~\cite{czarnik2021error}.

We depart from CDR in two respects: the training data come from $\mathcal{S}$ rather than hardware, and CDR's linear fit is replaced by a conditional network $f_{\boldsymbol{\tau}}$ that captures nonlinear dependence on circuit depth. Since this categorical prior is ill-suited to normalizing flows and mitigation requires only the point correction, we directly train a regression network $f_{\boldsymbol{\tau}}$ rather than learning the full posterior:
\begin{equation}
    \min_{\boldsymbol{\tau}}\mathbb{E}_{c,(\mathbf{x}^{\mathrm{dt}},\mathbf{x}^{\mathrm{exact}})}
    \mleft[\mleft\| f_{\boldsymbol{\tau}}(\mathbf{x}^{\mathrm{dt}},c)
    -\mleft(\mathbf{x}^{\mathrm{exact}}-\mathbf{x}^{\mathrm{dt}}\mright)
    \mright\|_2^2 \mright].
\end{equation} 
where $c$ encodes relevant circuit features. At deployment, the network corrects noisy data in a single forward pass via $\mathbf{x}^{\mathrm{mit}} = \mathbf{x}^{\mathrm{noisy}} + f_{\boldsymbol{\tau}}(\mathbf{x}^{\mathrm{noisy}}, c)$. In our simulation, $\mathbf{x}^{\mathrm{noisy}}$ is generated by exact density-matrix simulation under the ground-truth noise.

We evaluate this approach on the 1D transverse-field Ising model (TFIM), 
\begin{equation}
 H = -j\sum_i Z_i Z_{i+1} + h\sum_i X_i,
 \end{equation} 
whose time evolution is approximated by first-order Trotterization with step size $\delta t$,
\begin{equation}
    U(t) \approx \mleft[ \mleft( \prod_{i} e^{i j \delta t Z_i Z_{i+1}} \mright) \mleft( \prod_{i} e^{-i h \delta t X_i} \mright) \mright]^{t/\delta t}.
\end{equation} 
Starting from $|0\rangle^{\otimes N}$, we evolve under $U(t)$. The observation $\mathbf{x}\in\mathbb{R}^{2N-1}$ ($=\mathbb{R}^{23}$ for $N=12$) collects the single-site expectation values $\langle X_i\rangle$ and nearest-neighbor correlators $\langle Z_iZ_{i+1}\rangle$ at each time; their spatial averages---the transverse magnetization $M_x(t)=\frac{1}{N}\sum_i\langle X_i\rangle(t)$ and nearest-neighbor correlation $C_{ZZ}(t)=\frac{1}{N-1}\sum_i\langle Z_iZ_{i+1}\rangle(t)$---are reported only for visualization.
The digital twin is parameterized by the posterior mean inferred from a 12-qubit brickwall circuit, whose CNOT connectivity matches that of the Trotterized Ising circuit. Under the Markovian assumption, the noise parameters are applied identically to each repeated CNOT layer throughout the Trotterized evolution, for both the ground-truth and the SBI models. The denoising network is conditioned on Trotter steps via $c=\log (t/\delta t+1)$ and trained on 500 random Clifford variants per step, spanning 14 Trotter steps up to $t=1.4$ ($\delta t=0.1$). 

\begin{figure*}[t]
    \centering
    \includegraphics[width=1\linewidth]{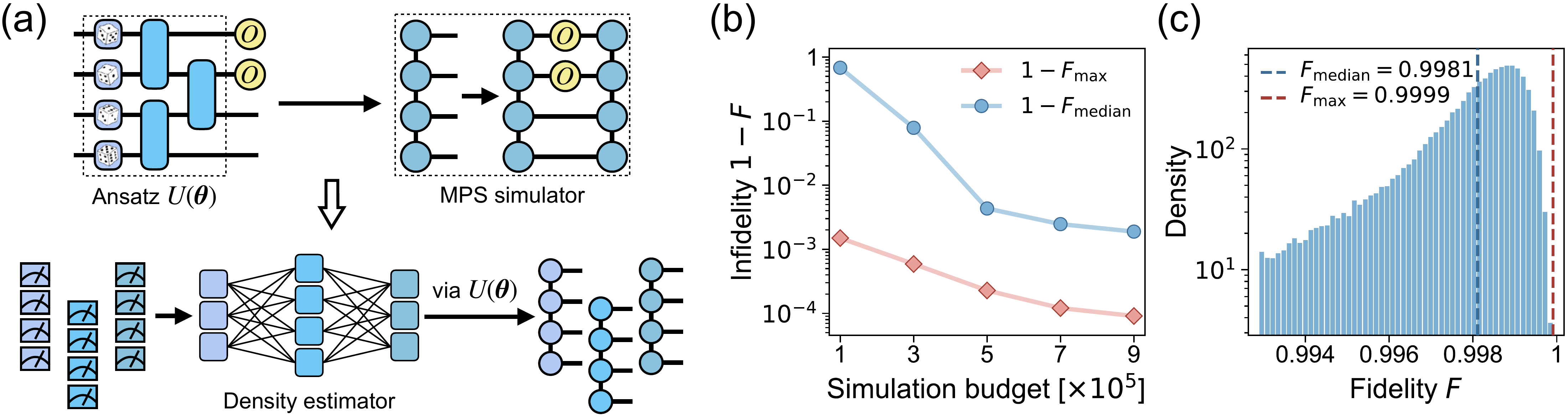}
    \caption{Quantum state tomography with NPE. (a) Workflow, illustrated using a PQC ansatz and an MPS simulator. Randomly sampled circuit parameters define states whose observables are evaluated by MPS contraction to train a neural posterior estimator. The trained estimator infers posterior distributions over circuit parameters from measurement data of new experimental instances, and posterior samples are mapped through the ansatz to reconstructed MPSs.
    (b) Reconstruction infidelity of simulated 12-qubit pure states versus simulation budget: $1-F_{\mathrm{median}}$ (blue) and $1-F_{\mathrm{max}}$ (red), computed at each budget from 500 posterior samples for each of 200 test states ($10^{5}$ samples in total). 
    (c) Probability density of the fidelity $F$ at the largest budget ($9 \times 10^{5}$), excluding outliers beyond $1.5\times$ the interquartile range. Dashed lines mark the median (blue) and maximum (red) over all samples.}
    \label{fig:pqc_qst}
\end{figure*}

We compare the results of four cases---raw noisy simulation, linear ZNE, quadratic ZNE, and the SBI-derived digital-twin denoiser (SBI-DT). As shown in \figpanel{fig:digital_twin}{b}, the SBI-DT produces the most accurate observable estimates at most evolution times, except at short times, where the ZNE-based methods are more accurate. Notably, it generalizes beyond the training boundary ($t>1.4$), where the ZNE-based methods degrade significantly. In \figpanel{fig:digital_twin}{c}, the aggregated MAE confirms improvement factors of $5.3$-fold for SBI-DT, $2.1$-fold for quadratic ZNE, and $1.2$-fold for linear ZNE, all relative to the noisy baseline. The digital-twin method thus achieves over twofold improvement compared to the quadratic ZNE baseline, while requiring no additional quantum samples at deployment. A complementary SBI instantiation targets the correction density $p(\boldsymbol{\theta}|\mathbf{x})$ instead, modestly outperforming quadratic ZNE but trailing direct regression for pointwise mitigation, as shown in \appref{app:dt_npe}. Overall, these results demonstrate that the SBI noise posteriors not only characterize the simulated device but also serve as a practical resource for error mitigation, closing the loop from error inference to mitigation.


\section{Quantum state tomography with parameterized quantum circuits \label{sec:qst}} 

Another application for our inference framework is QST. Quantum state tomography reconstructs an unknown quantum state from measurement data and is central to validating state preparation and quantum devices~\cite{PRXQuantum.6.030202,gebhart2023learning}. 
Because the full description of an $n$-qubit state grows exponentially with $n$, scalable QST generally requires structural priors that restrict the reconstruction to a tractable family of states. Such priors can be encoded through structured ans\"{a}tze, including matrix product states (MPSs)~\cite{RevModPhys.77.259, PRXQuantum.4.040345}, neural quantum states~\cite{carleo2017solving,lange2024architectures}, and parameterized quantum circuits (PQCs)~\cite{PhysRevA.101.052316,PRXQuantum.6.020346}.

Our framework provides an amortized QST approach that learns from simulated observations of many random states rather than measurement data of a single unknown state, so that one estimator can infer states throughout the ansatz family without retraining. \figurepanel{fig:pqc_qst}{a} illustrates this workflow using a PQC ansatz and an MPS simulator, with posterior parameter samples mapped through the ansatz to reconstructed states.
The PQC prepares states $\lvert\psi(\boldsymbol{\theta})\rangle =U(\boldsymbol{\theta})\lvert0\rangle^{\otimes n}$, whose parameters $\boldsymbol{\theta}$ serve as the latent variables. The parameter prior $\pi(\boldsymbol{\theta})$ induces a distribution over states supported on the ansatz manifold $\mathcal{M}_{U} = \left\{\lvert\psi(\boldsymbol{\theta})\rangle : \boldsymbol{\theta} \in\operatorname{supp}\pi \right\}$.
The circuit expressivity and the support of $\pi(\boldsymbol{\theta})$ therefore jointly determine the family of states accessible to inference. 

As a demonstration, we use the single-layer ansatz, $U(\boldsymbol{\theta})= U_{\mathrm{BW}} \mleft[\textstyle\bigotimes_{q=1}^{n} R_{x}(\theta_{q})\,R_{y}(\theta_{q})\mright]$, where the $R_x$ and $R_y$ rotations on each qubit share a single angle $\theta_q$.
The latent $\boldsymbol{\theta}\in\mathbb{R}^{n}$ collects all rotation angles, with independent uniform priors $\theta_{q}\sim\mathcal{U}[-\pi,\pi)$. The simulator $\mathcal{S}$ returns an observation $\mathbf{x}\in\mathbb{R}^{12n-9}$ of the expectation values over the Pauli ensemble $\mathcal{M}$ (weight-1 and nearest-neighbor weight-2). Each expectation value is subject to binomial shot noise with $N_{\mathrm{shots}}=10^{4}$ to mimic finite measurement statistics. These expectation values are evaluated by MPS simulation with bond dimension $\chi$, at a cost polynomial in $n$ and $\chi$ (see Appendix~\ref{app:contraction}). At fixed $\chi$, the simulator represents exactly only the states in $\mathcal{M}_{U}\cap\mathcal{M}^{\mathrm{MPS}}_{\chi}$, where $\mathcal{M}^{\mathrm{MPS}}_{\chi}$ denotes the set of pure states with Schmidt rank at most $\chi$ across every contiguous cut in the chosen qubit ordering~\cite{PhysRevLett.91.147902}. For the present ansatz, each cut is crossed by at most one CNOT gate, so $\mathcal{M}_{U}\subseteq\mathcal{M}^{\mathrm{MPS}}_{\chi=2}$ and the MPS simulation is exact.

We evaluate our estimator at $n=12$ on 200 independently sampled 12-qubit test states not used during training. For each test state, we draw 500 posterior parameter samples and evaluate the fidelity $F=\lvert\langle\psi(\boldsymbol{\theta}_{\mathrm{true}})|\psi(\boldsymbol{\theta})\rangle\rvert^{2}$ by MPS contractions. \figurepanel{fig:pqc_qst}{b} shows decreasing reconstruction
infidelity as the simulation budget grows.  The median infidelity decreases by more than two orders of magnitude, reaching approximately $2\times10^{-3}$ at the largest budget,  while the best-case infidelity remains below $1.5\times10^{-3}$ throughout. \figurepanel{fig:pqc_qst}{c} shows the fidelity distribution across all test states at the largest budget. The distribution is concentrated near unity, with $86\%$ of all posterior samples achieving fidelity above $0.99$. Reconstruction accuracy can be further improved by increasing the training budget or using the inferred $\boldsymbol{\theta}$ to initialize gradient-descent QST~\cite{gaikwad2025gradient}.
As a complementary instantiation, 4-qubit pure-state QST with a dense Cholesky parameterization under an unstructured prior is presented in Appendix~\ref{app:cholesky}. This example demonstrates amortized QST with an unrestricted pure-state ansatz, achieving median reconstruction fidelities above $0.98$ for most tested states.

\begin{figure*}[t]
    \centering
    \includegraphics[width=\linewidth]{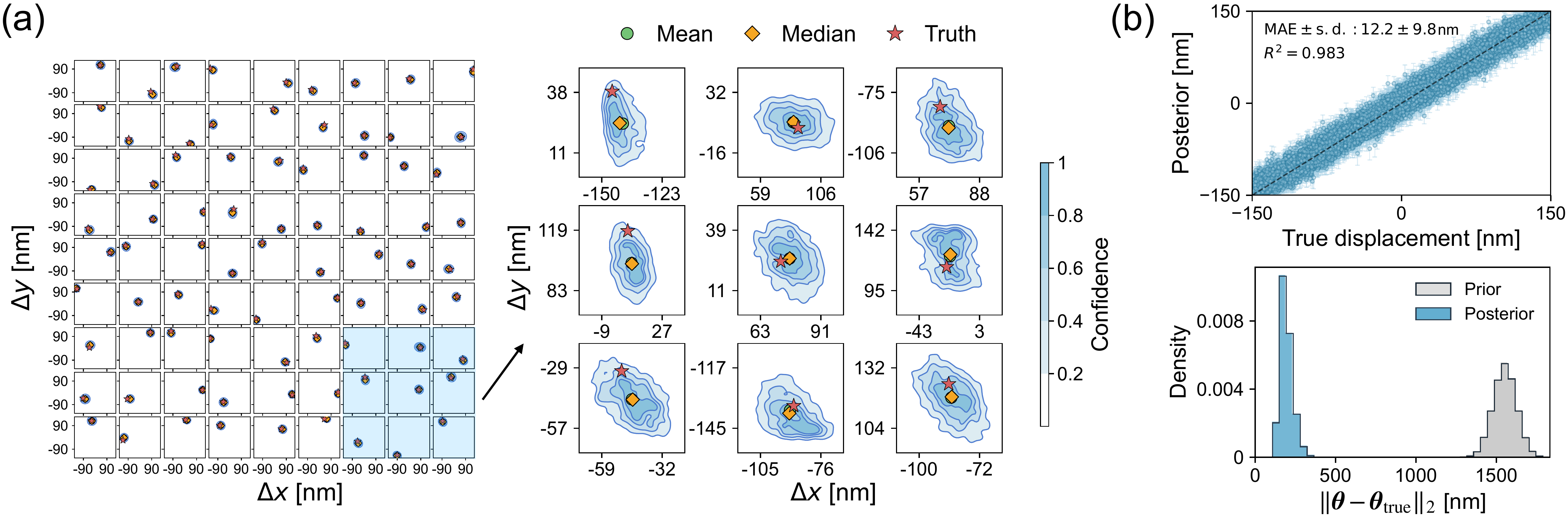}
    \caption{{Posterior localization of atomic displacements in simulated $9 \times 9$ Rydberg arrays.} (a) Posterior distribution of in-plane  displacements, $\Delta x$ and $\Delta y$ (in nm), for individual atoms within a single array instance. Each square represents a single array site, with filled contours indicating confidence intervals (color bar). The ground-truth displacements (red stars), posterior means (green circles), and posterior medians (orange diamonds) are indicated. An enlarged $3\times 3$ view highlights a representative patch. 
     (b) Amortized posterior precision evaluated across 100 independent displacements. Top, parameter-wise regression for individual displacement components; vertical bars denote the posterior standard deviation. Bottom, probability density of the parameter vector error norms, $||\boldsymbol{\theta}-\boldsymbol{\theta}_{\text{true}}||_2$, contrasting the broad prior (grey) with the localized posterior (blue).}
    \label{fig:rydberg}
\end{figure*}

\section{Hamiltonian learning of Rydberg atom arrays \label{sec:rydberg}}   
Our inference framework not only enables us to learn noise and states, but also Hamiltonians. Learning the realized Hamiltonian of a quantum computer or quantum simulator is essential for verifying that the implemented interactions match the intended model and identifying deviations that require correction~\cite{PhysRevLett.112.190501,wang2017experimental,gentile2021learning}. As a final demonstration, we turn to Hamiltonian learning of Rydberg atom arrays, a leading platform for large-scale qubit reconfiguration and control~\cite{bluvstein2024logical,bluvstein2026fault}. In these arrays, atomic positions set by optical tweezers determine the interatomic couplings, making positional deviations a source of Hamiltonian errors. Rapid inference of atomic positions could therefore inform feedback adjustments to restore the target interactions~\cite{f58h-zxs3,2ym8-vs82}. Here, we demonstrate this capability by inferring atomic displacements from simulated measurements of the system’s time evolution.

We consider a two-dimensional square array of $N = n_x \times n_y$ neutral atoms trapped at sites $\mathbf{r}_i = \mathbf{r}_i^{(0)} + \boldsymbol{\delta}_i$, where $\mathbf{r}_i^{(0)}$ is the ideal grid position and $\boldsymbol{\delta}_i =(\Delta x_i, \Delta y_i)\in \mathbb{R}^2$ is a small in-plane displacement. The system is described by the 2D TFIM 
\begin{equation}
    H = \Omega\sum_iX_i + \Delta\sum_i Z_i + \sum_{i<j} J_{ij} Z_i Z_j ,  
\end{equation}                                                         
where $\Omega$ is the Rabi frequency and $\Delta$ represents the detuning field. The coupling $J_{ij} ={C_6}/{|\mathbf{r}_i - \mathbf{r}_j|^6}$, with $C_6$ the van der Waals coefficient, encodes the pairwise geometry, so that any displacement $\boldsymbol{\delta}_i$ modifies the interactions between atom $i$ and its neighbors.  We set $\Omega = \Delta = \SI[per-mode = symbol]{15}{\radian \per \micro\second}$, lattice constant $a = \SI{8}{\micro\meter}$ (the nominal nearest-neighbor spacing), and $C_6= \SI[per-mode = symbol]{5.42e6}{ \micro\meter^6 \radian \per \micro\second}$, corresponding to $^{87}$Rb atoms in the $70S_{1/2}$ Rydberg state~\cite{f58h-zxs3}.

We consider $9 \times 9$ Rydberg arrays. The latent vector $\boldsymbol{\theta}\in\mathbb{R}^{2N}$ ($=\mathbb{R}^{162}$ for $N=81$)  encodes the in-plane displacements of all $N$ atoms, with a uniform prior ${\theta}_i\sim\mathcal{U}[-150,150]$~nm reflecting realistic positional fluctuations. Since the coupling $J_{ij}$ is effectively short-ranged, we retain only the nearest- and next-nearest-neighbor pairs, $n_{\mathrm{NN}}=2(N-\sqrt{N})$ and $n_{\mathrm{NNN}}=2(\sqrt{N}-1)^2$. The observation $\mathbf{x}\in\mathbb{R}^{N + n_{\mathrm{NN}} + n_{\mathrm{NNN}}} =\mathbb{R}^{5N-6\sqrt{N}+2}$ ($=\mathbb{R}^{353}$ for $N=81$) collects $N$ single-site magnetizations $\langle Z_i\rangle$ and $n_{\mathrm{NN}} + n_{\mathrm{NNN}}$ two-point correlators $\langle Z_i Z_j\rangle$, measured on the probe state 
\begin{equation}
    |\psi\rangle = e^{-i\Omega\delta t \sum_i X_i} e^{-i\Delta\delta t \sum_i Z_i} e^{-i\delta t \sum_{i<j} J_{ij} Z_i Z_j}|{+}\rangle^{\otimes N},
\end{equation}
obtained from a single short-time Trotter step ($\delta t = \SI{0.01}{\micro\second}$).
This short-time evolution suppresses Trotter discretization error~\cite{PhysRevX.11.011020} while keeping the circuit shallow enough for efficient classical simulation. The simulator $\mathcal{S}$ evaluates these observables at polynomial cost via truncated Pauli propagation, retaining only terms with Pauli weight $\leq 5$ and coefficient magnitude $\geq 10^{-6}$ (see more details in Appendices~\ref{pauli_propagation} and \ref{rydberg_pp}). All training and evaluation data are generated by this simulator. 
 
\figurepanel{fig:rydberg}{a} displays the posterior distributions of in-plane displacements for all 81 atoms in a single ground-truth instance. The amortized estimator, trained on $10^5$ simulations, produces well-localized posteriors whose means and medians cluster tightly around the true displacements across the entire array. The expanded $3\times3$ inset shows the posterior confidence regions in more detail. For most sites, the prescribed ground-truth displacement lies within the high-confidence contours; in the few cases where it lies near the boundary, the point estimates nonetheless remain accurate at the nanometer scale. 

\figurepanel{fig:rydberg}{b} quantifies this performance over 100 independent ground-truth configurations, with 500 posterior samples drawn for each. The parameter-wise regression (top) compares each scalar displacement component $(\Delta x \text{ or } \Delta y)$ independently, yielding $\mathrm{MAE}= 12.2 \pm \SI{9.8}{\nano\meter}$ and $R^2 = 0.983$. The distribution of the full parameter-vector error norm $\|\boldsymbol{\theta} - \boldsymbol{\theta}_{\mathrm{true}}\|_2$ (bottom) demonstrates an order-of-magnitude reduction relative to the broad prior. For context, fluorescence imaging in a tweezer-assisted positioning experiment with $^{87}\mathrm{Rb}$ atoms yields positional uncertainties on the scale of tens of nanometers~\cite{PRXQuantum.6.010322}. This suggests that the reconstruction accuracy achieved here in simulation is practically meaningful, although its performance on experimental data remains to be validated.

\section{Discussion and conclusion\label{sec:discussion}} 

We have presented a unified framework for simulation-based quantum system inference and demonstrated its efficacy through comprehensive numerical quantum simulations. Our central message is that advances in the classical simulation of quantum systems can be translated directly into advances in their characterization: any regime a classical simulator can reach becomes accessible to inference from experimental data. This turns quantum parameter inference from a bespoke, per-experiment effort into a reusable capability that grows as simulators and generative models improve.

This capability rests on four features of our approach. First, it is general: diverse problems conventionally addressed by task-specific algorithms are unified under a common SBI formulation. Second, it is scalable: training-data generation relies on simulators such as Pauli propagation and tensor networks rather than density-matrix evolution, and thus inherits their polynomial complexity. 
Third, it is amortized: an initial training investment produces a neural estimator that maps new measurement data directly to posteriors. Fourth, it is informative: beyond point estimates, posterior uncertainty provides a built-in diagnostic that may guide optimal experimental design~\cite{huan2024optimal}. \looseness=-1

Several directions for further research and development follow naturally. Currently, our Pauli noise learning assumes Markovian dynamics and idealizes SPAM. Moving beyond these simplifying assumptions to incorporate non-Markovianity~\cite{WhiteNonMarkovian,PhysRevA.97.012127,1ncg-11hz} and gate set tomography~\cite{Nielsen2021gatesettomography, Chenselfconsistent} would advance the framework toward self-consistent characterization of realistic hardware noise. Beyond refining the noise model, a crucial step is to address model misspecification: since the estimator is trained entirely on simulations, any simulator–device discrepancy can bias posteriors on experimental data. Detecting and quantifying such misspecification to calibrate posterior uncertainties~\cite{PRXQuantum.6.010354,wehenkel2025addressing,nott2023bayesian,PhysRevD.111.083013} will be essential for reliable deployment on hardware. \looseness=-1

Even when the full circuit is too complex to simulate, our framework can still act on classically tractable substructures. For instance, noise learned on repeated layers can be reused across deep circuits under stationarity~\cite{van2023probabilistic,kim2023evidence}, as demonstrated in \secref{sec:dt_mlqem}; partial tomography estimates only selected density-matrix elements~\cite{hynl-kxl2} or local reduced states~\cite{PhysRevLett.124.100401}; and circuit cutting decomposes large circuits into simulable fragments~\cite{PhysRevLett.125.150504,liu2022classicalsimu}. In addition, just as the randomized measurement toolbox~\cite{elben2023randomized} exploits randomness on the measurement side, our framework introduces a new randomized tool on the simulation side, and combining the two offers a promising direction for future work.

On the architectural side, size-agnostic models such as graph neural networks could enable estimators trained on small systems to generalize directly to larger ones~\cite{sanchez2020learning,f58h-zxs3,diez2026transferable,wang2026scalablequantumerrormitigation}, removing the fixed-size restriction of models used here. In addition, gauge equivariant or invariant architectures could help handle gauge redundancies~\cite{cohen2019gauge}, such as those in Pauli noise models and MPSs. On the training side, flow matching further offers a scalable route to enhance NPE training in high-dimensional parameter spaces~\cite{10.5555/3666122.3666859}. \looseness=-1

Looking further ahead, both components of the framework can evolve. Hybrid quantum–classical infrastructure~\cite{robledo2025chemistry,seelam2026referencearch} and eventually early fault-tolerant processors~\cite{PRXQuantum.5.020101} could generate training data beyond classical reach, extending inference to strongly entangled states and long-time dynamics, and allowing trusted devices to characterize less reliable ones~\cite{PhysRevLett.112.190501,PhysRevA.89.042314}. Likewise, quantum generative models could serve as density estimators for posteriors that are hard to represent classically~\cite{PhysRevApplied.16.044057,coyle2020born}.  Beyond inference, the principle underlying Hamiltonian learning could be extended to inverse Hamiltonian design, identifying candidate Hamiltonians predicted to realize desired properties~\cite{PhysRevX.8.031029,PhysRevResearch.6.033080, kokail2026inversequan}.
Overall, we envisage that this paradigm will mature into a versatile tool for characterizing, optimizing, and ultimately engineering complex quantum systems.

We note a recent work by Belliardo et al.~\cite{belliardo2025multi}, which also uses normalizing flows to estimate dozens of quantum parameters. A key distinction lies in the inference paradigm: their method relies on variational Bayesian inference with an explicit likelihood model and optimizes the evidence lower bound anew for each observation. In contrast, our SBI approach is likelihood-free and eliminates online optimization across varying quantum settings, at the expense of upfront simulation and training. The two frameworks are complementary, representing important steps in scaling statistical inference to high-dimensional quantum domains.\looseness=-1

\begin{acknowledgments}
The authors thank Akshay Gaikwad for valuable discussions.
This work was partially supported by the Wallenberg AI, Autonomous Systems and Software Program (WASP) funded by the Knut and Alice Wallenberg Foundation. The work was further supported by a joint project between WASP and the Wallenberg Center for Quantum Technology (WACQT), funded by the Knut and Alice Wallenberg Foundation. Preliminary results were enabled by resources provided by the National Academic Infrastructure for Supercomputing in Sweden (NAISS) at Alvis and Arrhenius (project: NAISS 2026/4-684).
AFK further acknowledges support from the Swedish Foundation for Strategic Research (grant numbers FFL21-0279 and FUS21-0063) and from the Norwegian Research Council through the Norwegian Quantum Software Center (NorQSoft, project number 361350).
AFK and MR also acknowledge support from the Horizon Europe program HORIZON-CL4-2022-QUANTUM-01-SGA via the project 101113946 OpenSuperQPlus100.

\paragraph*{Use of AI tools.} The conception, methodology, design of numerical experiments, and analysis of results were carried out entirely by the authors. Claude Sonnet 4.6 and Claude Opus 4.6 implemented most of the code following the authors' supervision, and all code was checked by the authors. Claude Opus 4.8 and ChatGPT 5.6 Sol were used to improve the readability of the manuscript and to suggest supplementary references, all of which were reviewed and verified by the authors. The authors take full responsibility for the content of this article.
\end{acknowledgments}

\section*{Data availability}
All data and models supporting the results of this study are available at \url{pending_publication}.

The source code is available at \url{pending_publication}. The software packages used to reproduce the data include \texttt{zuko}~\cite{rozet2022zuko},  \texttt{sbi}~\cite{BoeltsDeistler_sbi_2025},  \texttt{stim}~\cite{gidney2021stim},  \texttt{Qiskit}~\cite{qiskit2024}, and  \texttt{PyTorch}~\cite{NEURIPS2019_bdbca288}.
 
\appendix

 

\section{Neural spline flows \label{spline_flows}}
We adopt a neural spline flow~\cite{10.5555/3454287.3454962} as the density estimator, combining autoregressive transformations with rational-quadratic spline functions to achieve flexible density modeling.

\begin{figure*}[t]
    \centering
    \includegraphics[width=0.8\linewidth]{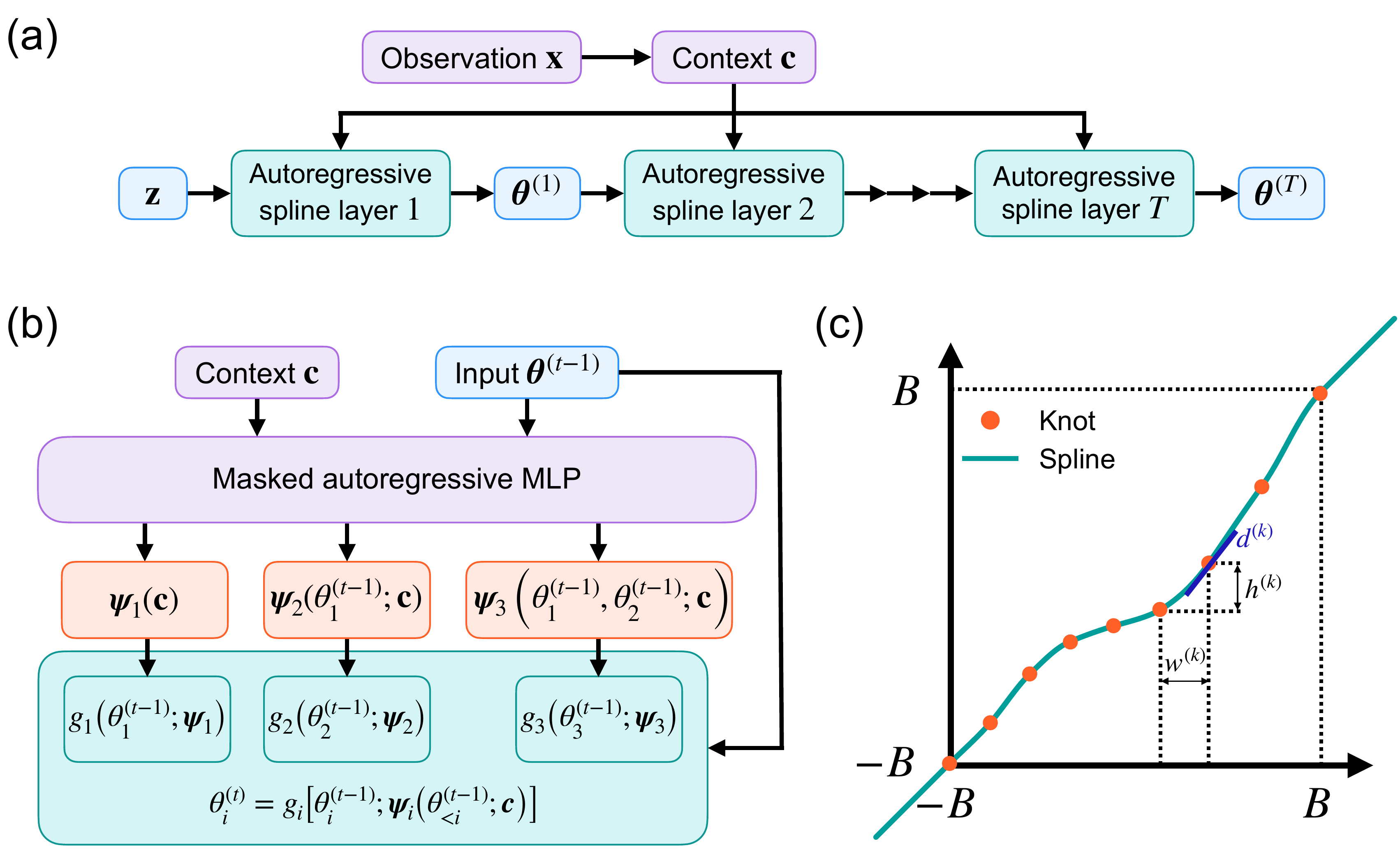}
    \caption{Neural spline flow architecture.
    (a) A base sample $\mathbf{z}$ is transformed through $T$ autoregressive spline layers into a posterior parameter sample $\boldsymbol{\theta}^{(T)}$, with each layer conditioned on the context $\mathbf{c}$ derived from the observation $\mathbf{x}$. (b) Structure of an autoregressive spline layer, illustrated for three coordinates. A masked autoregressive MLP generates the spline parameters $\boldsymbol{\psi}_i$ from the preceding coordinates $\boldsymbol{\theta}_{<i}^{(t-1)}$ and the context $\mathbf{c}$. The scalar transforms $g_i$ then map each input coordinate to $\theta_i^{(t)}$. (c) A monotonic rational-quadratic spline illustrated with $K+1=9$ knots, specified by bin widths $w^{(k)}$, heights $h^{(k)}$, and knot derivatives $d^{(k)}$, with identity tails outside $[-B,B]$.} 
    \label{fig:nsf}
\end{figure*}

Let $\boldsymbol{\theta}\in\mathbb{R}^d$ denote the parameter vector to be inferred and $\mathbf{c}\in\mathbb{R}^{d_c}$ the context vector obtained by embedding the observation $\mathbf{x}$. The flow is composed of $T$ autoregressive layers, $f_{\boldsymbol{\phi}}=f^{(T)}\circ\cdots\circ f^{(1)}$, as illustrated in \figpanel{fig:nsf}{a}. Within each layer $f^{(t)}$, the transformation acts coordinate-wise [\figpanel{fig:nsf}{b}]:  
\begin{equation}
    \theta_i^{(t)} = g_i\mleft[\theta_i^{(t-1)}; \boldsymbol\psi_i\mleft(\boldsymbol{\theta}_{<i}^{(t-1)};\boldsymbol{c}\mright)\mright], \quad i = 1, \ldots, d,
\end{equation}
where $g_i(\cdot;\boldsymbol\psi_i)$ is a monotone scalar spline transform. The spline parameters $\boldsymbol\psi_i$ are produced by a masked autoregressive network~\cite{10.5555/3045118.3045213,10.5555/3294771.3294994}, implemented as a multi-layer perceptron (MLP) with binary-masked weight matrices. This masking ensures that each $\boldsymbol\psi_i$ depends only on the preceding dimensions $\boldsymbol{\theta}_{<i}^{(t-1)}$ and the context $\boldsymbol{c}$, thereby enforcing a lower-triangular Jacobian. The log-determinant therefore decomposes exactly as:
\begin{equation}
\log\mleft|\det \frac{\partial f_{\boldsymbol{\phi}}^{-1}}{\partial \boldsymbol{\theta}}\mright| = -\sum_{t=1}^{T}\sum_{i=1}^{d} \log\mleft|\frac{\partial g_i\mleft(\theta_i^{(t-1)};\,\psi_i\mright)}{\partial \theta_i^{(t-1)}}\mright|,
 \end{equation}
requiring only $d$ scalar derivative evaluations per layer. Consecutive layers use alternating variable orderings: if layer $f^{(t)}$ uses the identity permutation $(1, 2, \ldots, d)$, then $f^{(t+1)}$ uses the reverse permutation $(d, d-1, \ldots, 1)$, ensuring full coupling among all parameters.

The spline transform $g_i$ in each coordinate is constructed by stitching together $K$ rational-quadratic segments on $[-B,B]$, with the identity applied outside this interval [\figpanel{fig:nsf}{c}]. The interval is partitioned into $K$ bins by $K+1$ knot points $(x^{(k)},y^{(k)})$, $k=0,\ldots,K$. Starting from $x^{(0)}=y^{(0)}=-B$, successive knots are placed according to $x^{(k)} = x^{(k-1)} + w^{(k)}$ and $y^{(k)} = y^{(k-1)} + h^{(k)}$, where $w^{(k)}$ and $h^{(k)}$ are the bin widths and heights, respectively. At each knot, the spline slope is specified by a derivative $d^{(k)}$. \looseness=-1

The network outputs $3K-1$ quantities per coordinate, $\boldsymbol{\psi}_i\in\mathbb R^{3K-1}$: $K$ widths, $K$ heights, and $K-1$ interior-knot derivatives. Widths and heights are normalized via softmax and accumulated to span $[-B, B]$. The $K-1$ interior derivatives are exponentiated to enforce strict positivity; the two boundary derivatives $d^{(0)}$ and $d^{(K)}$ are fixed to $1$, ensuring a smooth transition to the identity tails.
Within the $k$-th bin, let $\xi = \frac{\theta - x^{(k)}}{x^{(k+1)} - x^{(k)}} \in [0,1]$ denote the normalized position and $s_k = \frac{y^{(k+1)} - y^{(k)}}{x^{(k+1)} - x^{(k)}}$ the bin slope. The transform and its derivative are obtained by~\cite{10.5555/3454287.3454962}:
\begin{equation}
    \begin{aligned}
        g(\theta;\psi) &= y^{(k)} + \frac{(y^{(k+1)} - y^{(k)})[s_k\,\xi^2 + d^{(k)}\,\xi(1-\xi)]}{s_k + [d^{(k+1)} + d^{(k)} - 2s_k]\,\xi(1-\xi)}, \\
g'(\theta;\psi) &= \frac{s_k^2[d^{(k+1)}\,\xi^2 + 2s_k\,\xi(1-\xi) + d^{(k)}(1-\xi)^2]}{[s_k + (d^{(k+1)} + d^{(k)} - 2s_k)\,\xi(1-\xi)]^2}.
    \end{aligned}
\end{equation} 
The transform is twice continuously differentiable at each knot and analytically invertible in closed form.

The model hyperparameters used in this work are presented in \appref{hyperparameters}.

\section{Pauli propagation \label{pauli_propagation}}
The generation of extensive datasets for SBI requires the repeated evaluation of noisy quantum expectation values. To bypass the $\mathcal{O}(4^n)$ memory requirement of density-matrix simulators, we employ Pauli propagation---a classical framework that evolves observables in the Heisenberg picture~\cite{PhysRevA.99.062337,rudolph2025pauli}.

Given an $n$-qubit initial state $\rho$ and a quantum channel $\mathcal{E}$ composed of layers $\{\mathcal{E}_l\}$, the expectation value of an observable $O$ is given by the dual action:
\begin{equation}
    \langle O \rangle = \text{Tr}[\mathcal{E}(\rho) O] = \text{Tr}[\rho \mathcal{E}^\dagger(O)]  
\end{equation}
where $\mathcal{E}^\dagger = \mathcal{E}_1^\dagger \circ \mathcal{E}_2^\dagger \circ \dots \circ \mathcal{E}_L^\dagger$ denotes the adjoint channel. We decompose $O$ into a sum of Pauli strings, $O = \sum_{\alpha=1}^k c_\alpha P_\alpha $ with $P_\alpha \in \{I, X, Y, Z\}^{\otimes n}$. The propagation of $O$ is then formalized by tracking the evolution of the coefficients $c_\alpha$ and their corresponding strings $P_\alpha$ via the Pauli transfer matrix (PTM) formalism. 
Specifically, any $n$-qubit quantum operation $\mathcal{E}_l$ is completely characterized by a $4^n \times 4^n$ transformation matrix with elements $[\mathcal{E}_l]_{ij} = \text{Tr}[P_i \mathcal{E}_l(P_j)]$. Operating in the Heisenberg picture requires the dual map $\mathcal{E}_l^\dagger$, which is represented by the transpose of the PTM, $[\mathcal{E}_l^\dagger]_{ij} = [\mathcal{E}_l]_{ji}$. Thus, the backward evolution of a single Pauli string $P_\alpha$ through layer $l$ dictates a mapping to a new linear combination of strings: \begin{equation}\label{eq:backpro}
    \mathcal{E}_l^\dagger[P_\alpha] = \sum_{\beta} [\mathcal{E}_l^\dagger]_{\beta \alpha} P_\beta.
\end{equation}
By linearity, the entire observable is updated at each layer $\mathcal{E}_l$, mapping the coefficient distribution to $c_\beta^l = \sum_\alpha c_\alpha [\mathcal{E}_l^\dagger]_{\beta \alpha}$, yielding the updated observable $\mathcal{E}_l^\dagger(O) = \sum_\beta c_\beta^l P_\beta$. 

Although the PTM formally acts on a $4^n$-dimensional vector space, evaluating \eqref{eq:backpro} does not incur the prohibitive $\mathcal{O}(16^n)$ cost of dense matrix-vector multiplication.
This global $4^n \times 4^n$ matrix is never explicitly constructed in memory. Instead, the framework utilizes the backward causal cone of each observable---the subset of earlier circuit gates that can affect it under operator propagation---together with the sparsity of the active operator subspace. Gates outside this cone trivially commute with the active Pauli strings and can therefore be skipped. For a $k$-local operation within the lightcone, the algebraic transformation reduces to a localized $4^k \times 4^k$ sub-block. By dynamically tracking only the non-zero coefficients via sparse data structures, the prohibitive bottleneck is avoided.

Consequently, the computational complexity is governed by the \textit{branching factor} of the propagation path:
\paragraph{Clifford propagation.} All Clifford gates are \textit{1-branching}, mapping each Pauli string to a single Pauli string ($\mathcal{C}^\dagger[P_\alpha] \propto P_\beta$), giving an $\mathcal{O}(N_\mathcal{C})$ cost per propagating Pauli string, where $N_\mathcal{C}$ is the number of Clifford gates. However, entangling Clifford gates (e.g., CNOT) can alter the Pauli weight of propagating strings, which tends to indirectly amplify the branching induced by subsequent non-Clifford gates. 

\paragraph{Pauli noise as eigenvalue scaling.} In the Pauli basis, Pauli channels $\Lambda_p$ act as \textit{1-branching} eigenvalue contractions ($\Lambda_p^\dagger[P_\alpha] = \lambda_{P_\alpha} P_\alpha$). Incorporating complex noise profiles merely rescales existing coefficients, incurring an $\mathcal{O}(N_{\text{noise}})$ cost per propagating Pauli string, where $N_{\text{noise}}$ is the number of local Pauli noise channels.

\paragraph{Non-Clifford propagation.} Non-Clifford Pauli rotations $R_{\sigma}(\theta)=e^{-i\theta \sigma/2}$ are either \textit{1-branching} or \textit{2-branching}. Specifically, the evolution of a Pauli string $P$ follows:
  \begin{equation}
      R_{\sigma}(\theta)[P]=\begin{cases} P & \text{if } [\sigma,P]=0 \\ \cos(\theta)P + \sin(\theta)P' & \text{otherwise} \end{cases},
  \end{equation}
where $P' = i[\sigma,P]/2$ is a newly generated Pauli string. 

When the expanding operator lightcone encounters an increasing number of branching gates, the number of Pauli strings may grow exponentially. To maintain computational tractability, this proliferation necessitates heuristic truncation strategies. These typically include \textit{coefficient truncation}, which discards Pauli paths whose accumulated amplitude falls below a precision threshold ($|c_\alpha| < \epsilon$), and \textit{weight truncation}, which eliminates strings exceeding a maximum Pauli weight ($w(P) > w_{\text{max}}$)~\cite{rudolph2025pauli}. For current hardware, the latter is also motivated by the physical reality that high-weight paths are typically exponentially suppressed by noise, contributing negligibly to local observables.

\section{Benchmarking Pauli-propagation simulators}

\subsection{Simulator for Pauli noise learning \label{pauli_noise_pp}}

In our special cases of Pauli noise learning (see Fig.~\ref{fig:circuit_layout}), we restrict all non-Clifford operations to the initial state preparation layer. Instead of explicitly propagating the Pauli strings backward through this densely branching initial layer, we terminate the Heisenberg evolution after traversing the Clifford and noise layers. At this terminal step, the non-Clifford operations simply define an unentangled product state, $|\psi_{\text{init}}\rangle = \bigotimes_{i=1}^n U_i |0\rangle$. The expectation value of any fully propagated Pauli string $P = \bigotimes_{i=1}^n P_i$ can therefore be evaluated analytically by factorizing it over the individual qubits:
\begin{equation}
    \langle P \rangle = \prod_{i=1}^n \langle 0 | U_i^\dagger P_i U_i | 0 \rangle 
\end{equation} 
For instance, when applying $U_i = R_z(\pi/4)R_y(\pi/4)$ to each qubit identically, these local expectation values reduce to fixed analytical constants: 
\begin{equation}
    \langle 0 | U_i^\dagger P_i U_i | 0 \rangle = \begin{cases} 1, & \text{if } P_i = I \\ 1/2, & \text{if } P_i = X \text{ or } Y \\ 1/\sqrt{2}, & \text{if } P_i = Z \end{cases}.
\end{equation}
This local evaluation requires merely $\mathcal{O}(n)$ operations per string. This architectural advantage eliminates the nominal $\mathcal{O}(2^n)$ branching overhead associated with the initial $n$ non-Clifford gates. This approach guarantees exact, non-truncated noisy simulation with an efficient complexity of $\mathcal{O}(n k)$, representing a bilinear scaling with respect to both the system size $n$ and the number of Pauli strings $k$ in the target observable. Such an efficiency profile theoretically enables the use of SBI to characterize sparse Markovian Pauli noise in large-scale circuits exceeding 100 qubits.

\begin{figure}[t] 
    \centering
    \includegraphics[width=\linewidth]{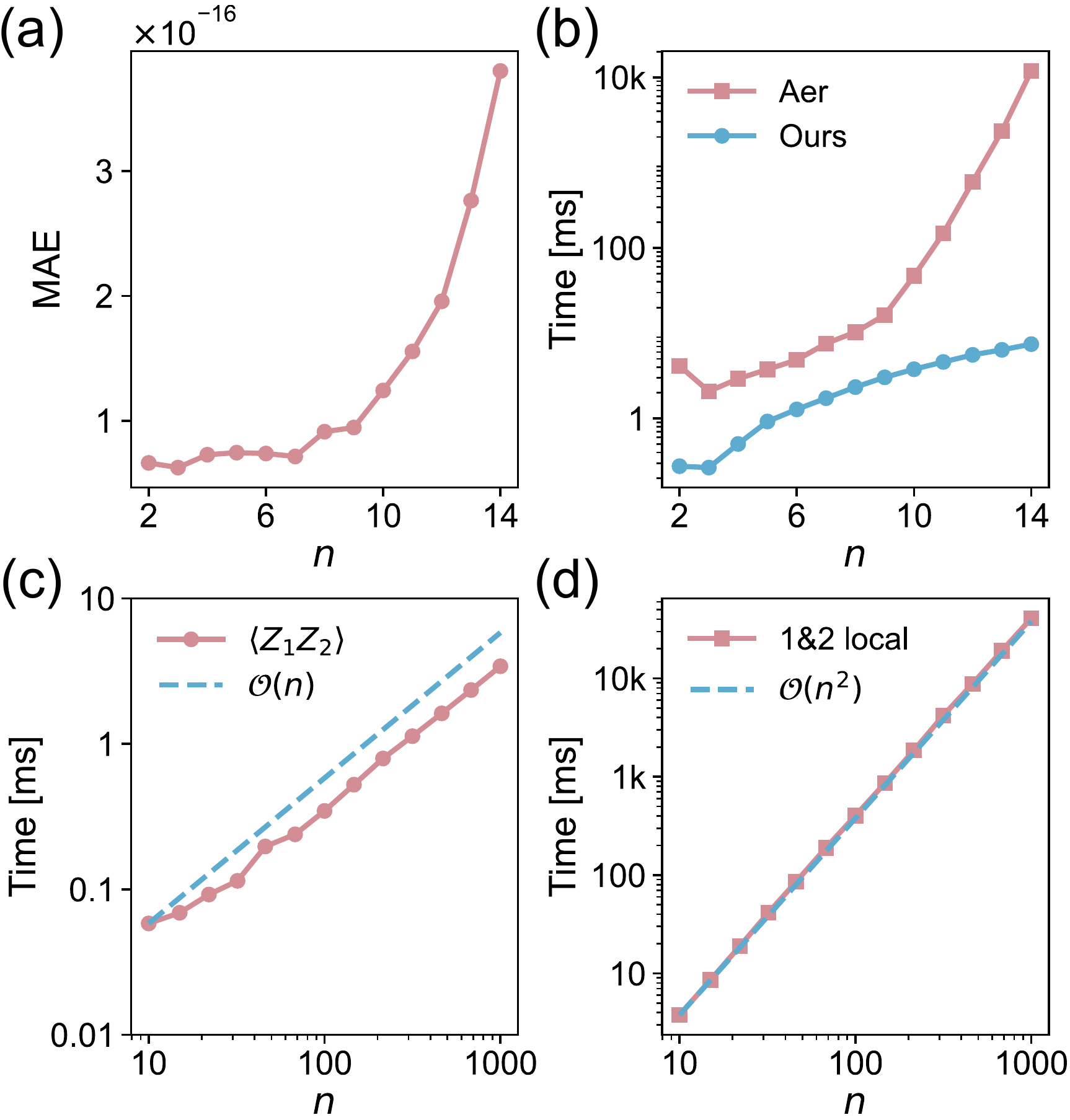}
    \caption{{Benchmarking the simulator for Pauli noise learning.} (a) Mean absolute error (MAE) between our method and \texttt{Qiskit Aer} density-matrix simulator over all $k=12n-9$ Pauli observables (system size $n=2$--$14$).
    (b) Wall time comparison for both methods over the same setting as in panel (a). 
    (c) Wall time of our method for a single observable $\langle Z_1 Z_2\rangle$ as a function of $n=10$--$1000$, with an $\mathcal{O}(n)$ reference.
    (d) Total wall time for all $k$ observables, with an $\mathcal{O}(n^2)$ reference. Reference curves are initialized at the leftmost data point. All quantities are averaged over five independent random seeds (Pauli rotation angles and Pauli noise coefficients).}
    \label{fig:pauli_noise_pp}
\end{figure}

We benchmark our simulation approach against the \texttt{Qiskit Aer} density-matrix simulator, as illustrated in Fig.~\ref{fig:pauli_noise_pp}. The benchmark evaluates the full set of $k = 12n - 9$ (1- and 2-local) Pauli observables. All tests were run on a single thread of an Apple M4 Pro CPU, using a linear circuit template in which $n-1$ CNOT gates, each followed by a two-qubit Pauli noise channel, are applied sequentially between nearest-neighbor qubits. Both methods demonstrate numerical agreement within floating-point precision [\figpanel{fig:pauli_noise_pp}{a}]. While the  \texttt{Qiskit Aer} runtime grows exponentially with $n$ and becomes computationally prohibitive beyond $n \sim 14$ [\figpanel{fig:pauli_noise_pp}{b}], our approach reduces the cost per observable to $\mathcal{O}(n)$ [\figpanel{fig:pauli_noise_pp}{c}]. This efficiency allows a single CPU thread to process an observable for $n = 1000$ in under \SI{4}{\milli\second}. Furthermore, the overall complexity for all $k$ observables scales as $\mathcal{O}(nk)\sim\mathcal{O}(n^2)$ [\figpanel{fig:pauli_noise_pp}{d}], enabling simulations of $n = 100$ in about 0.4~s and $n = 1000$ in roughly \SI{40}{\second}. 

\subsection{Simulator for Rydberg Hamiltonian learning \label{rydberg_pp}}

We now benchmark the weight-truncated Pauli-propagation simulator used for Rydberg Hamiltonian learning. The circuit structure is identical to that of \secref{sec:rydberg} in the main text (a single short-time Trotter step applied to $\ket{+}^{\otimes N}$). We take $\langle Z_1 Z_2\rangle$ as a representative observable and average over twenty random displacement samples. We sweep the weight-truncation order from $w \leq 1$ to $w \leq 5$.

\begin{figure}[t]
    \centering
    \includegraphics[width=\linewidth]{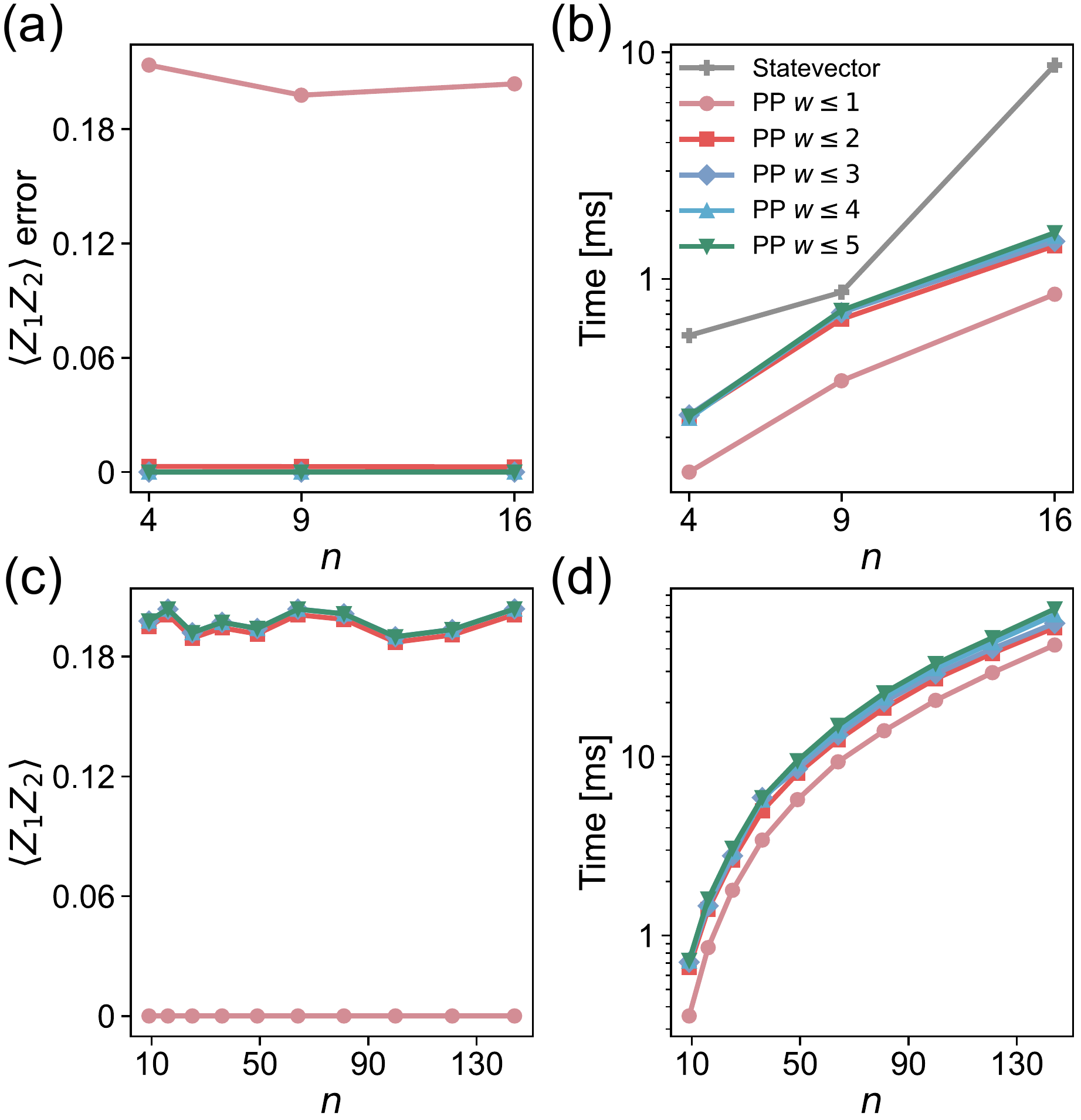}
    \caption{{Benchmarking the simulator for Rydberg Hamiltonian learning.}
    (a) Absolute error between the weight-truncated Pauli propagation and the \texttt{Qiskit Aer} statevector simulator for the observable $\langle Z_1Z_2 \rangle$, shown for truncation orders $w \leq 1$--$5$ on $2\times2$, $3\times3$, and $4\times4$ lattices ($n = 4, 9, 16$).
    (b) Wall time comparison between the statevector reference and  Pauli propagation over the same setting as in panel (a).
    (c) Estimated $\langle Z_1Z_2 \rangle$ as a function of system size up to the $12\times 12$ lattice ($n=144$), across the same truncation orders.
    (d) Wall time for the same scaling sweep as in panel (c). All quantities are averaged over twenty random displacement samples. \looseness=-1 }
    \label{fig:pp_rydberg}
\end{figure}

On $2\times 2,~ 3\times 3$, and $4\times 4$ lattices [\figpanel{fig:pp_rydberg}{a}], the truncation error converges rapidly and uniformly in $n$: $w\leq 1$ is clearly insufficient, $w\leq 2$ already reduces it by more than an order of magnitude,  and from $w\leq 3$ onward the error is negligible. The statevector runtime rises steeply with system size while Pauli propagation stays modest, making propagation up to nearly an order of magnitude cheaper at $n=16$ [\figpanel{fig:pp_rydberg}{b}]. Extending to square lattices up to $n=144$, the estimated $\langle Z_1Z_2\rangle$ maintains convergence for $w \geq 3$ [\figpanel{fig:pp_rydberg}{c}], while the propagation wall time follows a clean polynomial trend within tens of milliseconds [\figpanel{fig:pp_rydberg}{d}]. These results confirm that, for the (at most) 2-local observables entering the observation, the weight-5 truncation used in the main text incurs negligible error while preserving polynomial cost.

\section{Local gauge ambiguity \label{gauge}}               
The unlearnable degrees of freedom observed in the posterior distributions arise from a fundamental gauge symmetry inherent to composite gate benchmarking. For a composition of two gates $\mathcal{G}_2 \circ \mathcal{G}_1$, a gauge transformation $\mathcal{V}$ satisfies $\tilde{\mathcal{G}}_2 \circ \tilde{\mathcal{G}}_1 = \bigl(\mathcal{G}_2 \circ \mathcal{V}^{-1}\bigr)\circ \bigl(\mathcal{V} \circ \mathcal{G}_1\bigr)$, leaving the composite channel invariant while redistributing errors between the two gates. 

\begin{figure*}[t]
    \centering
    \includegraphics[width=\linewidth]{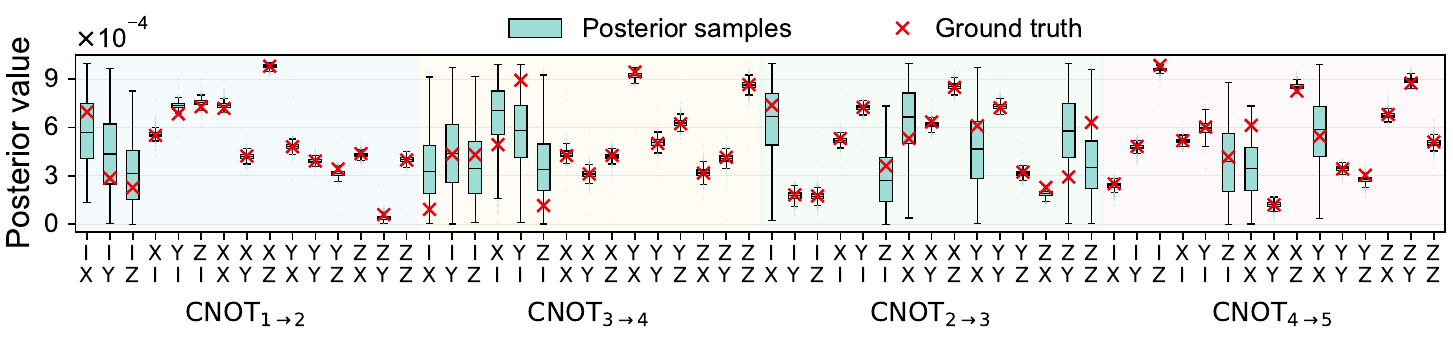}
    \caption{{Posterior Pauli noise distributions of a five-qubit circuit with brickwall connectivity.} Layer one consists of $\text{CNOT}_{1\to 2}$ and $\text{CNOT}_{3\to 4}$; layer two consists of $\text{CNOT}_{2\to 3}$ and $\text{CNOT}_{4\to 5}$. }
    \label{fig:5qubitnosie}
\end{figure*}

We build intuition on an illustrative five-qubit brickwall circuit (compare \figref{fig:circuit_layout} in the main text), whose posteriors are shown in Fig.~\ref{fig:5qubitnosie}. The first layer applies $\mathrm{CNOT}_{1,2}$ and $\mathrm{CNOT}_{3,4}$, and the second layer applies $\mathrm{CNOT}_{2,3}$ and $\mathrm{CNOT}_{4,5}$; each CNOT is followed by its own two-qubit Pauli noise channel. Two-qubit Pauli labels refer to the qubit pair of the corresponding gate, ordered by qubit index; for example, $IX$ for $\mathrm{CNOT}_{1,2}$ denotes an $X$ error on $q_2$.  Qubits $q_2$, $q_3$, and $q_4$ are each shared by a first-layer (upstream) gate and a second-layer (downstream) gate; we refer to them as bridge qubits.

Consider a single-qubit Pauli error $E \in \{X, Y, Z\}$ on a bridge qubit, generated by the noise channel of the upstream gate. Before reaching the measurement, this error passes through the downstream gate $C$ (a CNOT), which maps it to $E^\prime=CEC^\dagger$. Since $CE=E^\prime C$, an error $E$ occurring before $C$ acts identically to the error $E^\prime$ occurring after $C$. Because $E^\prime$ is also one of the Pauli terms of the downstream noise channel, the upstream probability of $E$ and the downstream probability of $E^\prime$ enter the composite channel only through their combination. Equivalently, a single-qubit Pauli channel on a bridge qubit can be shifted from one gate to the other without changing the composite channel, which gives three gauge directions $\{X, Y, Z\}$ per bridge qubit.

Using the CNOT conjugation rules, written in (control, target) order,
\begin{equation}
\begin{gathered}
     XI \to XX,\quad YI \to YX,\quad ZI \to ZI,\\
     IX \to IX,\quad IY \to ZY,\quad IZ \to ZZ,
\end{gathered}
\end{equation}
the coupled pairs (upstream $\leftrightarrow$ downstream) are:
\begin{enumerate}
    \item Bridge $q_2$ (control of $\text{CNOT}_{2,3}$): $\{IX, IY, IZ\}$ of $\text{CNOT}_{1,2}$ $\leftrightarrow$ $\{XX, YX, ZI\}$ of $\text{CNOT}_{2,3}$, since
    \begin{align*}
        I_1X_2I_3 &\xrightarrow{\text{CNOT}_{2,3}} I_1X_2X_3,\\
        I_1Y_2I_3 &\xrightarrow{\text{CNOT}_{2,3}} I_1Y_2X_3,\\
        I_1Z_2I_3 &\xrightarrow{\text{CNOT}_{2,3}} I_1Z_2I_3.
    \end{align*}
    \item Bridge $q_3$ (target of $\text{CNOT}_{2,3}$): $\{XI, YI, ZI\}$ of $\text{CNOT}_{3,4}$ $\leftrightarrow$ $\{IX, ZY, ZZ\}$ of $\text{CNOT}_{2,3}$, since
    \begin{align*}
        I_2X_3I_4 &\xrightarrow{\text{CNOT}_{2,3}} I_2X_3I_4,\\
        I_2Y_3I_4 &\xrightarrow{\text{CNOT}_{2,3}} Z_2Y_3I_4,\\
        I_2Z_3I_4 &\xrightarrow{\text{CNOT}_{2,3}} Z_2Z_3I_4.
    \end{align*}
    \item Bridge $q_4$ (control of $\text{CNOT}_{4,5}$): $\{IX, IY, IZ\}$ of $\text{CNOT}_{3,4}$ $\leftrightarrow$ $\{XX, YX, ZI\}$ of $\text{CNOT}_{4,5}$, since
    \begin{align*}
        I_3X_4I_5 &\xrightarrow{\text{CNOT}_{4,5}} I_3X_4X_5,\\
        I_3Y_4I_5 &\xrightarrow{\text{CNOT}_{4,5}} I_3Y_4X_5,\\
        I_3Z_4I_5 &\xrightarrow{\text{CNOT}_{4,5}} I_3Z_4I_5.
    \end{align*}
\end{enumerate}

Each of these nine pairs carries one gauge degree of freedom, so neither parameter in a pair can be determined individually. Collecting the pairs by gate gives the unlearnable parameters of each gate position:
\begin{enumerate}
    \item Odd-layer boundary ($\text{CNOT}_{1,2}$; bridge $q_2$): 3 parameters, $\{IX, IY, IZ\}$.
    \item Odd-layer interior ($\text{CNOT}_{3,4}$; bridges $q_3$, $q_4$): 6 parameters, $\{XI, YI, ZI\}$ via $q_3$ and $\{IX, IY, IZ\}$ via $q_4$.
    \item Even-layer interior ($\text{CNOT}_{2,3}$; bridges $q_2$, $q_3$): 6 parameters, $\{XX, YX, ZI\}$ via $q_2$ and $\{IX, ZY, ZZ\}$ via $q_3$.
    \item Even-layer boundary ($\text{CNOT}_{4,5}$; bridge $q_4$): 3 parameters, $\{XX, YX, ZI\}$ via $q_4$.
\end{enumerate}
In total, the three bridge qubits yield $18$ unlearnable parameters, consistent with the broad posteriors in Fig.~\ref{fig:5qubitnosie}.  

Generalizing to $n$ qubits, the brickwall circuit considered here contains $n-2$ bridge qubits, each shared by two adjacent gates and contributing three gauge directions, i.e., three coupled pairs. This yields $6(n-2)$ unlearnable parameters in total. The specific error coefficients that become coupled are determined by how each bridge qubit enters the adjacent gates (as control or target), and thus vary with the circuit topology.
  
Without external constraints---such as independent single-gate calibrations or gauge-fixing circuit elements---these gauge parameters are fundamentally unresolvable from composite circuit measurements alone. This is not a limitation of the SBI framework but an inherent consequence of the objective itself: resolving \emph{individual} gate-level noise from \emph{composite} measurements. Standard quantum process tomography of the entire gate sequence would yield only the aggregate process matrix, providing no additional power to decompose noise attribution among constituent gates. Crucially, even tomographically complete input states and measurements cannot resolve this ambiguity, as it arises from local algebraic equivalences at the bridge qubits that leave the composite channel itself invariant. The inclusion of SPAM noise would further exacerbate the issue by introducing additional gauge freedom~\cite{chen2023learnability, Nielsen2021gatesettomography}.

\section{Complete posterior distributions for 50-qubit Pauli noise \label{complete_posterior}}

\begin{figure}[t] 
    \centering
    \includegraphics[width=\linewidth]{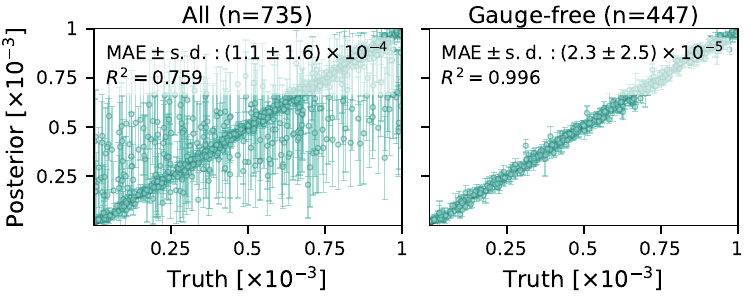}
    \caption{{Parameter-wise regression for 50-qubit Pauli noise learning.} Posterior means versus ground-truth values for all 735 parameters (left) and the 447-parameter gauge-free subset (right). Vertical bars denote $\pm 1$ posterior standard deviation among 2,000 samples.} 
    \label{fig:regression_50q}
\end{figure}

As a supplement to \figpanel{fig:noise}{a} in the main text, \figref{fig:regression_50q} presents the parameter-wise regression for the 50-qubit noise model. Over all 735 parameters (left), the posterior means show considerable scatter ($R^2 = 0.759$, MAE $= 1.1 \times 10^{-4}$) due to the gauge parameters. Restricting to the 447 gauge-free parameters (right) yields tight agreement with the ground truth ($R^2 = 0.996$, MAE $= 2.3 \times 10^{-5}$), confirming that the identifiable noise structure is accurately resolved despite the high dimensionality.

\begin{figure*}[tp] 
    \centering
    \includegraphics[width=\linewidth]{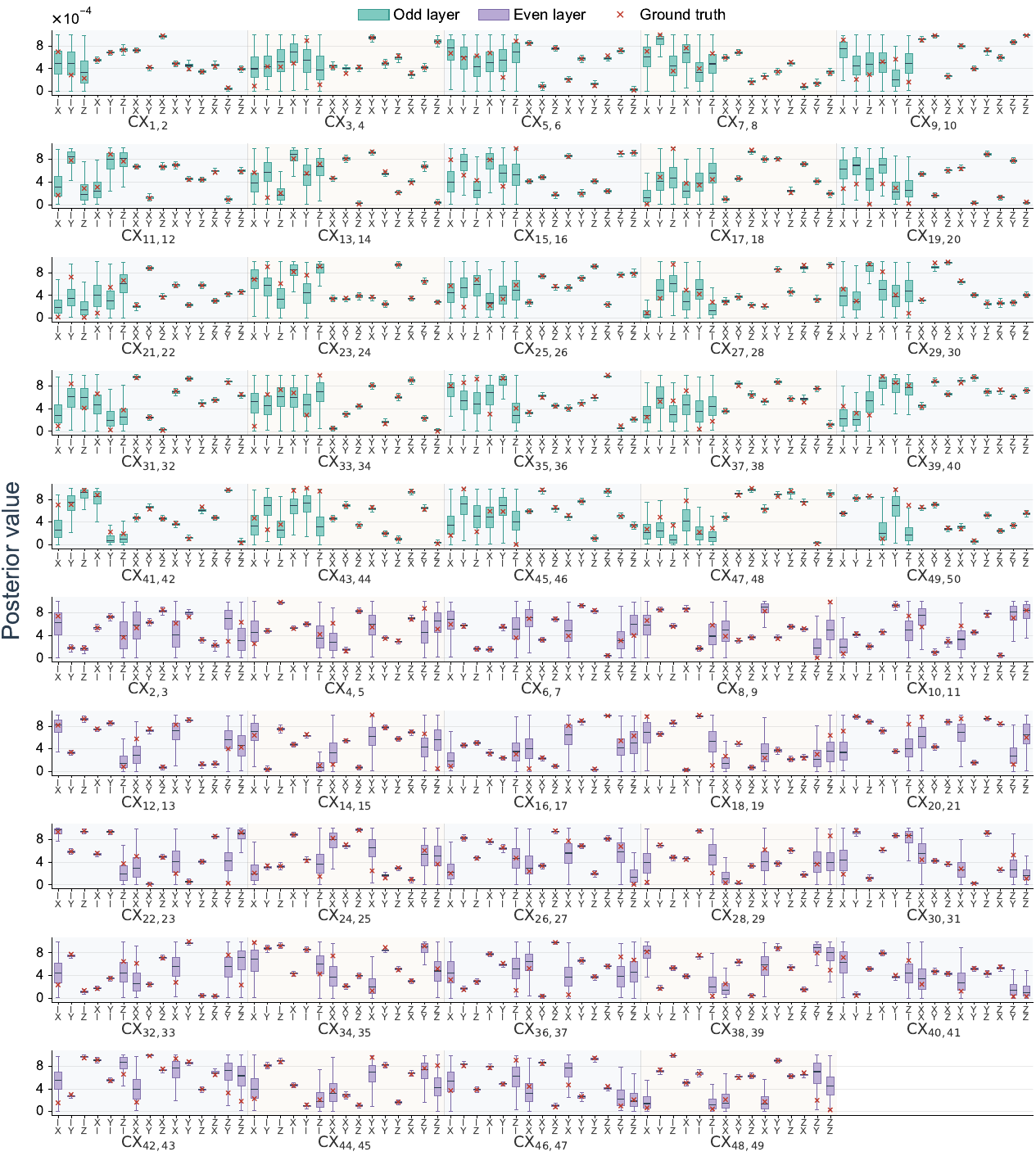}
    \caption{{Complete posterior distributions for 50-qubit Pauli noise learning.} Posterior samples (box plots) and ground-truth values (red crosses) for all 49 CNOT gates in the brickwall circuit. For each box, the central line marks the median, the box spans the interquartile range (25th--75th percentiles), and the whiskers extend to $1.5\times$ the interquartile range; outliers beyond the whiskers are omitted. Each box summarizes 2,000 posterior samples. Gates are grouped by sublayer: odd-control (top five rows) and even-control (bottom five rows).}
    \label{fig:complete_distribution_50q}
\end{figure*}

Figure~\ref{fig:complete_distribution_50q} displays the complete posterior distributions of the 50-qubit noise model. These results are obtained from the amortized NPE model trained on 30,000 simulations. The gauge-freedom patterns exhibit four distinct gate categories: odd-layer top boundary (CNOT$_{1,2}$), odd-layer interior (CNOT$_{3,4}$--CNOT$_{47,48}$), odd-layer bottom boundary (CNOT$_{49,50}$), and even-layer interior (CNOT$_{2,3}$--CNOT$_{48,49}$). Compared with the 5-qubit case (Fig.~\ref{fig:5qubitnosie}), the larger circuit gains an odd-layer bottom boundary but loses the even-layer boundary.

\section{Standard quantum error mitigation}
\subsection{Probabilistic error cancellation \label{PEC}}

PEC cancels noise statistically by sampling operations from a quasiprobability representation of the inverse noise channel~\cite{van2023probabilistic,temme2017error}. Each Pauli noise channel $\Lambda_k$ is diagonal in the Pauli basis with eigenvalues $\lambda_{k,Q}=\mathrm{Tr}[Q^\dagger\Lambda_k(Q)]/2^n$ for each Pauli string $Q$. The inverse channel is decomposed as  $\Lambda_k^{-1}(\cdot)=\sum_{P} q_{k,P}\, P(\cdot)P,$ 
where the quasiprobability coefficients are recovered via $q_{k,P} = \frac{1}{4^n}\sum_{Q} \lambda_{k,Q}^{-1}(-1)^{[P,Q]},$
with $(-1)^{[P,Q]}$ equal to $+1$ if $[P,Q]=0$ and $-1$ otherwise, and $n=2$ for the two-qubit channels considered here. The coefficients $q_{k,P}$ are real but not necessarily positive, reflecting the non-physicality of the inverse channel.

The sampling distribution at each gate $k$ is $\pi_{k,P}= {|q_{k,P}|}/{\gamma_k}$ with the per-gate sampling overhead  $\gamma_k=\sum_P |q_{k,P}|$. For each circuit instance $C_s$, random Pauli corrections $P_k^{(s)}$ are independently drawn from $\pi_k$ for every gate, and the mitigated expectation value is estimated via the Monte Carlo average:
\begin{equation}
    \langle O \rangle_{\mathrm{PEC}} = \frac{\gamma}{N}\sum_{s=1}^{N} \mathrm{sgn}(w_s)\, o_s , 
\end{equation}
where $o_s$ is the measurement outcome for instance $C_s$, $w_s=\prod_k q_{k,P_k^{(s)}}$ accumulates the sign into the estimator, and $\gamma=\prod_k \gamma_k$ is the total sampling overhead.  
\subsection{Zero-noise extrapolation \label{ZNE}}

ZNE recovers ideal expectation values by extrapolating measurements at amplified noise levels to the zero-noise limit~\cite{temme2017error,li2017efficient,kim2023evidence,giurgica2020digital}. We implement ZNE using two complementary noise-amplification strategies, depending on whether a noise model is available.

\paragraph{Probabilistic error amplification (PEA).} When a Pauli noise model is available, here supplied by the SBI posterior, we amplify noise by inserting additional random Pauli operations~\cite{kim2023evidence}. For a noise gain $G\geq1$, we define the supplementary error probabilities as 
\begin{equation}\begin{aligned}
      p_{k,P}^{\mathrm{add}}(G)&=\frac{1-(1-2p_{k,P})^{G-1}}{2}\\
      &=(G-1)p_{k,P}+O(p_{k,P}^{2}),
\end{aligned}
\end{equation} 
where the expansion holds for fixed $G$ in the weak-noise regime. The factor $G-1$ accounts for the native noise already present, so composing the supplementary channel with the native noise amplifies the noise strength by $G$ to first order.
At each gate $k$, a supplementary Pauli operator $P$ is sampled and inserted with probability $p_{k,P}^{\mathrm{add}}(G)$. The amplified expectation value $\langle O \rangle^{(G)}$ is estimated at each gain over an ensemble of independently sampled circuits, and the zero-noise limit ($G=0$) is extracted by extrapolation using linear, quadratic, or exponential fits. PEA can realize non-integer gains, and we use it for the posterior-driven ZNE in \figpanel{fig:zne_pec}{a} of the main text $(G \in \{1.6, 2.2, 2.8\})$.

\paragraph{Gate folding.} When no noise model is assumed, noise can be amplified by gate folding~\cite{giurgica2020digital}. Each folded gate $U$ is replaced by $U(U^\dagger U)^n$, preserving the ideal action while amplifying the effective noise by $G=2n+1$. Unlike PEA, which rescales the noise channel in place, folding lengthens the circuit, incurring a higher coherence-time overhead, while restricting the accessible gains to odd integers. We use it for the model-agnostic ZNE baselines in \figref{fig:digital_twin} of the main text ($G \in \{1, 3, 5\}$), which amplify the ground-truth Markovian noise carried by the CNOT gates of the Trotterized circuit.

\section{NPE-based digital-twin ML-QEM \label{app:dt_npe}}

The digital-twin denoiser, mentioned in the main text (Sec.~\ref{sec:dt_mlqem}), is trained by direct regression of the correction $\boldsymbol{\theta}\equiv\Delta\mathbf{x}=\mathbf{x}^{\mathrm{exact}}-\mathbf{x}^{\mathrm{dt}}$. Here we detail an alternative variant that instead learns the full posterior $p(\boldsymbol{\theta}|\mathbf{x}^{\mathrm{dt}})$ via NPE.
While the posterior-driven twin $\mathcal{S}$, the observable set, the noise model, the Trotter schedule (14 steps up to $t=1.4$), and the training-data budget (500 circuits per Trotter step) remain identical to the main implementation, this variant introduces key differences in its prior specification and training data generation.

The categorical Clifford sampling leaves the prior $\pi(\boldsymbol{\theta})$ analytically unspecified, which inherently precludes standard NPE training. To recover the standard NPE approach, we construct a data-driven empirical prior defined as a diagonal Gaussian, $\pi(\boldsymbol{\theta})=\prod_i \mathcal{N}\mleft(\theta_i;\mu_i,\sigma_i^2\mright)$ with $\mu_i$ and $\sigma_i$ the empirical mean and standard deviation of the correction $\theta_i$ across the training circuit ensemble. Under this prior, we train a conditional normalizing flow $q_{\boldsymbol{\phi}}(\boldsymbol{\theta}|\mathbf{x}^{\mathrm{dt}},c)\approx p(\boldsymbol{\theta}|\mathbf{x}^{\mathrm{dt}},c)$, and at deployment correct with the posterior mean:
\begin{equation}
  \mathbf{x}^{\mathrm{mit}}=\mathbf{x}^{\mathrm{noisy}}
  +\mathbb{E}_{q_{\boldsymbol{\phi}}}\mleft[\boldsymbol{\theta}|\mathbf{x}^{\mathrm{noisy}},c\mright],
  \label{eq:dt_npe_mit}
\end{equation}
substituting the direct regression estimate used in the main text. 

Although the introduction of the empirical prior renders the flow theoretically trainable on the original Clifford ensemble, direct application to these circuits failed to yield any meaningful correction signal, a failure fundamentally rooted in the lack of continuous support in $\boldsymbol{\theta}$ required for flow-based density estimation. To circumvent this limitation, we employ a data augmentation scheme, prepending each circuit with a random single-qubit product-state layer $\bigotimes_q R_y(\alpha_q)R_x(\beta_q)$ ($\alpha_q,\beta_q\sim\mathcal{U}[0,2\pi)$) prior to the Clifford Trotter blocks. This injection of randomness smoothens the training distribution, which adequately diversifies the data for the flow model to learn.

\begin{figure}[t]
  \centering
  \includegraphics[width=\columnwidth]{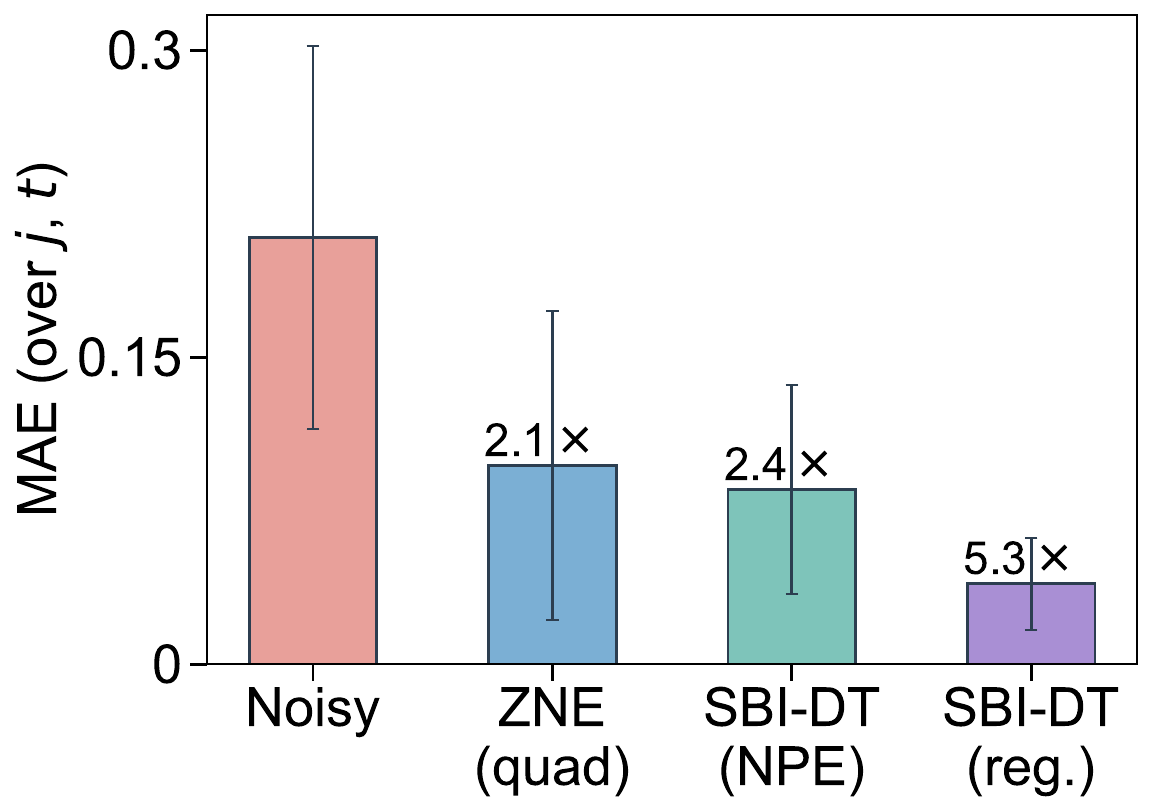}
  \caption{Overall MAE of the NPE-based SBI-DT, shown in direct comparison with the results in Fig.~\ref{fig:digital_twin}. The plot contrasts this variant against the noisy baseline, quadratic ZNE, and the regression-based SBI-DT. Improvement factors are relative to the noisy baseline.}
  \label{fig:dt_npe}
\end{figure}

As demonstrated in Fig.~\ref{fig:dt_npe} (presented in comparison with the primary results in Fig.~\ref{fig:digital_twin}), the NPE variant reduces the aggregated MAE by $2.4\times$ relative to the noisy baseline, compared to a $2.1\times$ reduction achieved by quadratic ZNE and a $5.3\times$ reduction by direct regression. It thus trails the regression-based approach for pointwise mitigation. This performance gap is expected: direct regression merely learns a point estimate (a deterministic map $\mathbf{x}\mapsto\boldsymbol{\theta}$) and does not model the distribution of $\boldsymbol{\theta}$, rendering it structurally immune to the training failures that discrete support imposes on flow-based density estimation. In contrast, NPE is forced to resolve the far more challenging task of learning the full density. Nevertheless, the NPE variant still yields a marginal improvement over the standard ZNE baseline.

\section{Matrix product states \label{app:contraction}} 
 
The QST simulator in \secref{sec:qst} represents the variational-circuit state as a bond-$\chi$ MPS~\cite{RevModPhys.77.259,cramer2010efficient,PRXQuantum.4.040345},
\begin{equation}
  \ket\psi 
  = \sum_{s_{1},\dots,s_{n}} A^{[1]}_{s_{1}} A^{[2]}_{s_{2}} \cdots A^{[n]}_{s_{n}}\, \ket{s_{1}\cdots s_{n}},
  \label{eq:mps}
\end{equation}
with each $A^{[i]}_{s_i}\in\mathbb{C}^{D_{i-1}\times D_{i}}$ a complex matrix for the physical leg $s_{i}\in\{0,1\}$, open boundaries $D_{0}=D_{n}=1$, and ramped bond dimensions $D_{i}=\min(2^{i},2^{\,n-i},\chi)$ with maximum bond dimension $\chi$.
The MPS stores $4\sum_{i}D_{i-1}D_{i}$ real parameters; for $\chi=2^{c}$ ($c\in\mathbb{Z}_{\ge 0}$) and $n\ge 2c$, this is $4(n-2c)\chi^{2}+\tfrac{16}{3}(\chi^{2}-1)=\mathcal{O}(n\chi^{2})$. The simulator evaluates the MPS's Pauli expectation values by transfer-matrix contraction, without ever forming the dense $2^{n}$ statevector. For an observable $P=\bigotimes_{i=1}^{n}P_i$, we define at each site $i$ the doubled-layer transfer matrix on the squared bond space $\mathbb{C}^{D_{i-1}}\!\otimes\mathbb{C}^{D_{i-1}}\to \mathbb{C}^{D_{i}}\!\otimes\mathbb{C}^{D_{i}}$,
\begin{equation}
  T^{i}[P_i]=\sum_{s_i,s_i'}(P_i)_{s_i s_i'}\,\overline{A^{[i]}_{s_i}}\otimes A^{[i]}_{s_i'}, 
\end{equation}
where $(P_i)_{s_i s_i'}=\langle s_i\rvert P_i\lvert s_i'\rangle$ is a matrix element of the single-qubit operator $P_i$,  $\overline{A^{[i]}_{s_i}}$ is the complex-conjugate (bra) copy of the MPS matrix and $A^{[i]}_{s_i'}$ the original (ket) copy, with the Kronecker product $\otimes$ pairing their bond indices on the doubled (bra--ket) space. The expectation then factorizes along the chain,
\begin{equation}
  \langle P\rangle=\frac{T^{1}[P_1]\,T^{2}[P_2]\cdots T^{n}[P_n]}{Z},  
    \label{eq:mps-expval}
\end{equation}
where $Z=\langle\psi|\psi\rangle=E^{1}E^{2}\cdots E^{n}$ is the normalization factor and $E^{i}\equiv T^{i}[I]=\sum_{s_i}\overline{A^{[i]}_{s_i}}\otimes A^{[i]}_{s_i}$. 
The product in \eqref{eq:mps-expval} is evaluated by a single left-to-right sweep that carries an \emph{environment}---a running $\chi\times\chi$ matrix holding the partial contraction of all sites traversed so far, with one index on the bra layer and one on the ket. At each site the environment is updated by the action of $T^i[P_i]$, computed as a few $\chi\times\chi$ matrix products at $\mathcal{O}(\chi^{3})$ cost rather than by assembling the doubled-layer matrix $T^i$, which would cost $\mathcal{O}(\chi^{4})$. 
Each observable thus costs $\mathcal{O}(n\chi^{3})$, and the full set $\mathcal{M}$ of $12n-9$ Pauli operators costs $\mathcal{O}(n^{2}\chi^{3})$. 
 
\section{Quantum state tomography with Cholesky parameterization \label{app:cholesky}}

In the main text (Sec.~\ref{sec:qst}), the state is parameterized by a structured variational quantum circuit, whose parameter count scales as $\mathrm{poly}(n)$. Here we describe the alternative dense Cholesky parameterization used for small systems. In contrast to the VQC ansatz, it can represent arbitrary states (up to full-rank mixed states) without structural restriction, at $\exp (n)$ cost.  

The Cholesky decomposition writes $\rho = TT^{\dagger}/\mathrm{Tr}(TT^{\dagger})$ with an unconstrained complex matrix $T\in\mathbb{C}^{2^{n}\times r}$ of rank $r$, ensuring the inferred state is automatically Hermitian, positive semidefinite, and trace-normalized~\cite{PhysRevA.64.052312}.  
Here we restrict the model to $n$-qubit pure states ($r=1$), for which $T$ reduces to a single complex vector $\mathbf{z} \in \mathbb{C}^{2^n}$, yielding $\rho = {\mathbf{z}\mathbf{z}^\dagger}/{\text{Tr}(\mathbf{z}\mathbf{z}^\dagger)} = |\psi\rangle\langle\psi|$. 

For a four-qubit pure state, the latent vector $\boldsymbol{\theta} \in \mathbb{R}^{2\cdot 2^n}$ ($=\mathbb{R}^{32}$)  comprises the real and imaginary components of $\mathbf{z}$, providing a real parameter space for inference. The prior is chosen as a standard normal, $\theta_i\sim \mathcal{N}(0,1)$, which after normalization induces the Haar measure over the pure-state manifold---an unstructured choice in contrast to the structured prior of the main text. The observation $\mathbf{x}\in\mathbb{R}^{180}$ collects expectation values over 180 randomly selected Pauli observables out of the $4^n-1=255$ basis elements. The simulator $\mathcal{S}$ evaluates exact expectation values via $\mathrm{Tr}[\rho(\boldsymbol{\theta})P]$ and adds binomial shot noise with $N_{\mathrm{shots}} = 10^4$. 

\begin{figure}[t]
    \centering
    \includegraphics[width=1\linewidth]{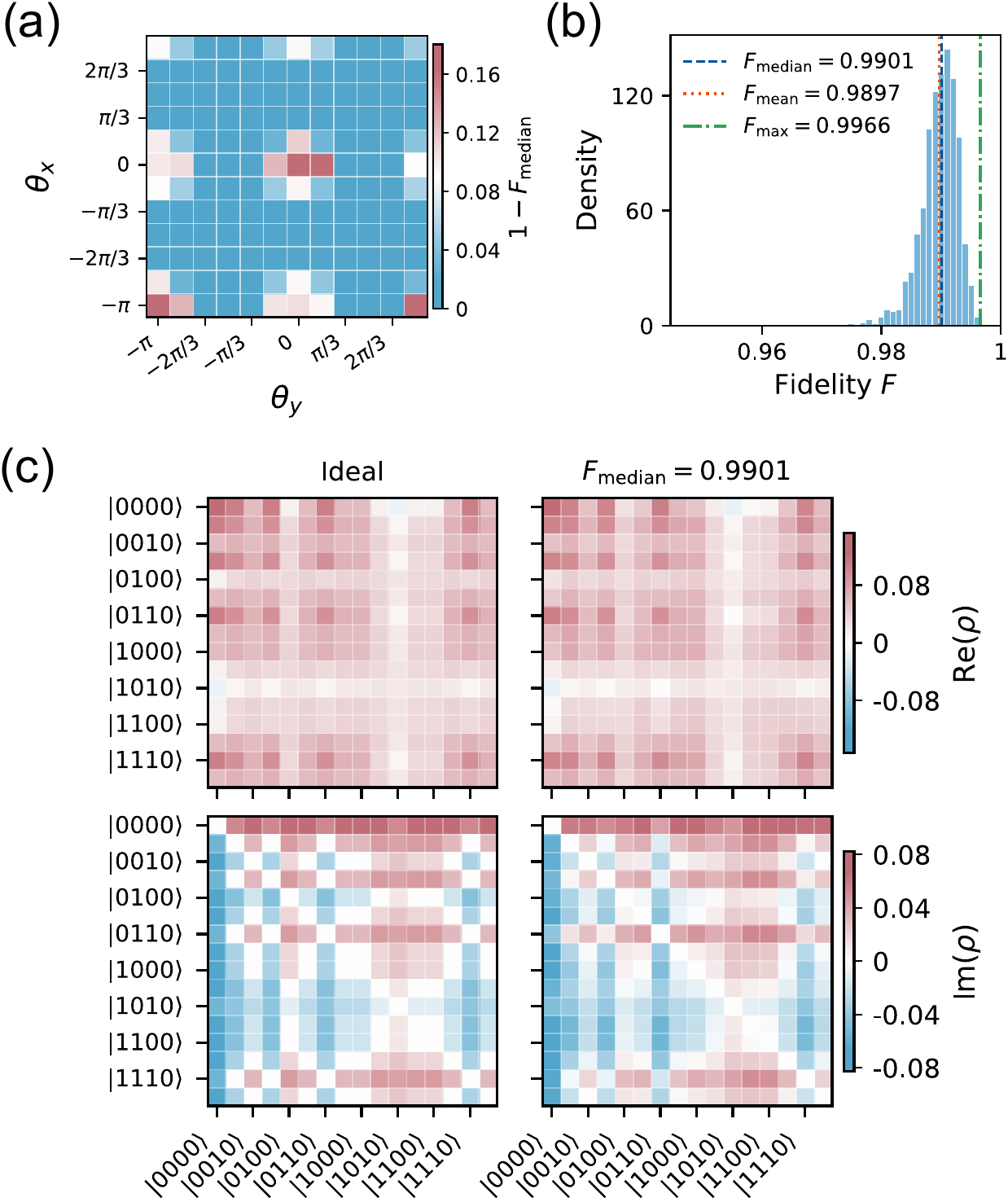}
    \caption{{Amortized quantum state tomography with the Cholesky parameterization.}
    (a) Heatmap of the median reconstruction infidelity over 2,000 posterior samples, where each cell $(\theta_x, \theta_y)$ represents a specific $4$-qubit pure state. 
    (b) Posterior fidelity distribution for a representative state prepared at $(\theta_x, \theta_y) = (\pi/3, \pi/3)$.
    (c) Real (top) and imaginary (bottom) components of the reconstructed density matrix at the median fidelity in panel (b).
} \label{fig:cd_qst}
\end{figure}

We evaluate the amortized estimator, trained on $10^6$ simulations, across a family of states parameterized by two rotation angles $(\theta_x,\theta_y)$, $|\psi(\theta_x,\theta_y)\rangle = U_{\text{BW}}\Bigl(\bigotimes_{q=1}^{4} R_x(\theta_x)\,R_y(\theta_y)\Bigr)|0\rangle^{\otimes 4}$. \figurepanel{fig:cd_qst}{a} maps the median reconstruction infidelity $1-F_{\text{median}}$ across the $(\theta_x,\theta_y)$ landscape. The majority of cells achieve $1-F_{\text{median}}<0.04$ (blue region), demonstrating high-fidelity reconstruction over a broad range of states. Elevated infidelity appears at cells where $\theta_x$ or $\theta_y$ approach the periodic boundaries (0 or $\pm \pi$). Periodically equivalent rotation settings can produce the same prepared state, and at these cells state reconstruction therefore becomes ambiguous.

\figurepanel{fig:cd_qst}{b} examines a representative state at $(\theta_x,\theta_y)=(\pi/3,\pi/3)$ in detail. The posterior fidelity distribution concentrates near unity, with $F_{\text{median}}=0.9901$, $F_{\text{mean}}=0.9897$, and $F_{\text{max}}=0.9966$. \figurepanel{fig:cd_qst}{c} visualizes the real and imaginary components of the density matrix reconstructed from the median-fidelity sample, showing close element-wise agreement with the ideal target. These fidelities make the amortized estimator directly applicable to benchmarking workflows requiring routine small-qubit tomography.

\section{Model and training details \label{hyperparameters}}

\subsection{Neural posterior estimation} 
The NPE hyperparameters are summarized in Table~\ref{tab:npe}; settings are shared across tasks except for the batch size, training budget, and observation-vector preprocessing (standardization and embedding). All flows are trained with the Adam optimizer, using PyTorch defaults aside from the learning rate. For the Rydberg task, a two-layer MLP ($\text{dim}_{\text{obs}}\!\to\!512\!\to\!256$, SiLU activations) embeds the observation vector before conditioning the flow. Listed training budgets are the maxima per task; smaller budgets appear in the scaling and sweep analyses.

\begin{table}[H]
\centering
\caption{Neural spline flow architecture and training hyperparameters. Upper block: settings shared across all models; lower block: model-specific.}
\label{tab:npe}
\begin{tabular}{lcccc}
\hline\hline
 & Pauli noise & QST & Rydberg & SBI-DT \\
\hline
Flow transforms   & \multicolumn{4}{c}{ 3 } \\
Hidden features &  \multicolumn{4}{c}{ 2048 }   \\
Spline bins     & \multicolumn{4}{c}{16 per dimension} \\
Learning rate   & \multicolumn{4}{c}{$10^{-4}$} \\
Early stopping  & \multicolumn{4}{c}{100 epochs without improvement} \\ 
\hline
Standardize $\mathbf{x}$ & yes & yes & yes & no \\
Embedding $\mathbf{x}$ & no & no & yes & no \\
Batch size      & $2048$ & $25{,}600$ & $4096$ & $500$ \\
Training budget & $30{,}000$ & $900{,}000$ & $100{,}000$ & $7{,}000$ \\ 
Figure  & \figref{fig:noise} & \figref{fig:pqc_qst} & \figref{fig:rydberg} & \figref{fig:dt_npe} \\
\hline\hline
\end{tabular}
\end{table}  

\subsection{Digital-twin denoising}
The denoising network (used in \figref{fig:digital_twin}) employs an MLP with two hidden layers (each of width 756), trained for 200 epochs using the Adam optimizer with a learning rate of $10^{-3}$ and a batch size of 256, on a total of $7{,}000$ training samples.

All models were trained on a single NVIDIA A100 GPU with \SI{40}{\giga\byte} of memory. 

\bibliography{apssamp.bib} 

\begin{thebibliography}{133}%
\makeatletter
\providecommand \@ifxundefined [1]{%
 \@ifx{#1\undefined}
}%
\providecommand \@ifnum [1]{%
 \ifnum #1\expandafter \@firstoftwo
 \else \expandafter \@secondoftwo
 \fi
}%
\providecommand \@ifx [1]{%
 \ifx #1\expandafter \@firstoftwo
 \else \expandafter \@secondoftwo
 \fi
}%
\providecommand \natexlab [1]{#1}%
\providecommand \enquote  [1]{``#1''}%
\providecommand \bibnamefont  [1]{#1}%
\providecommand \bibfnamefont [1]{#1}%
\providecommand \citenamefont [1]{#1}%
\providecommand \href@noop [0]{\@secondoftwo}%
\providecommand \href [0]{\begingroup \@sanitize@url \@href}%
\providecommand \@href[1]{\@@startlink{#1}\@@href}%
\providecommand \@@href[1]{\endgroup#1\@@endlink}%
\providecommand \@sanitize@url [0]{\catcode `\\12\catcode `\$12\catcode `\&12\catcode `\#12\catcode `\^12\catcode `\_12\catcode `\%12\relax}%
\providecommand \@@startlink[1]{}%
\providecommand \@@endlink[0]{}%
\providecommand \url  [0]{\begingroup\@sanitize@url \@url }%
\providecommand \@url [1]{\endgroup\@href {#1}{\urlprefix }}%
\providecommand \urlprefix  [0]{URL }%
\providecommand \Eprint [0]{\href }%
\providecommand \doibase [0]{https://doi.org/}%
\providecommand \selectlanguage [0]{\@gobble}%
\providecommand \bibinfo  [0]{\@secondoftwo}%
\providecommand \bibfield  [0]{\@secondoftwo}%
\providecommand \translation [1]{[#1]}%
\providecommand \BibitemOpen [0]{}%
\providecommand \bibitemStop [0]{}%
\providecommand \bibitemNoStop [0]{.\EOS\space}%
\providecommand \EOS [0]{\spacefactor3000\relax}%
\providecommand \BibitemShut  [1]{\csname bibitem#1\endcsname}%
\let\auto@bib@innerbib\@empty
\bibitem [{\citenamefont {Cramer}\ \emph {et~al.}(2010)\citenamefont {Cramer}, \citenamefont {Plenio}, \citenamefont {Flammia}, \citenamefont {Somma}, \citenamefont {Gross}, \citenamefont {Bartlett}, \citenamefont {Landon-Cardinal}, \citenamefont {Poulin},\ and\ \citenamefont {Liu}}]{cramer2010efficient}%
  \BibitemOpen
  \bibfield  {author} {\bibinfo {author} {\bibfnamefont {M.}~\bibnamefont {Cramer}}, \bibinfo {author} {\bibfnamefont {M.~B.}\ \bibnamefont {Plenio}}, \bibinfo {author} {\bibfnamefont {S.~T.}\ \bibnamefont {Flammia}}, \bibinfo {author} {\bibfnamefont {R.}~\bibnamefont {Somma}}, \bibinfo {author} {\bibfnamefont {D.}~\bibnamefont {Gross}}, \bibinfo {author} {\bibfnamefont {S.~D.}\ \bibnamefont {Bartlett}}, \bibinfo {author} {\bibfnamefont {O.}~\bibnamefont {Landon-Cardinal}}, \bibinfo {author} {\bibfnamefont {D.}~\bibnamefont {Poulin}},\ and\ \bibinfo {author} {\bibfnamefont {Y.-K.}\ \bibnamefont {Liu}},\ }\bibfield  {title} {\bibinfo {title} {Efficient quantum state tomography},\ }\href {https://doi.org/https://doi.org/10.1038/ncomms1147} {\bibfield  {journal} {\bibinfo  {journal} {Nature Communications}\ }\textbf {\bibinfo {volume} {1}},\ \bibinfo {pages} {149} (\bibinfo {year} {2010})}\BibitemShut {NoStop}%
\bibitem [{\citenamefont {Torlai}\ \emph {et~al.}(2018)\citenamefont {Torlai}, \citenamefont {Mazzola}, \citenamefont {Carrasquilla}, \citenamefont {Troyer}, \citenamefont {Melko},\ and\ \citenamefont {Carleo}}]{torlai2018neural}%
  \BibitemOpen
  \bibfield  {author} {\bibinfo {author} {\bibfnamefont {G.}~\bibnamefont {Torlai}}, \bibinfo {author} {\bibfnamefont {G.}~\bibnamefont {Mazzola}}, \bibinfo {author} {\bibfnamefont {J.}~\bibnamefont {Carrasquilla}}, \bibinfo {author} {\bibfnamefont {M.}~\bibnamefont {Troyer}}, \bibinfo {author} {\bibfnamefont {R.}~\bibnamefont {Melko}},\ and\ \bibinfo {author} {\bibfnamefont {G.}~\bibnamefont {Carleo}},\ }\bibfield  {title} {\bibinfo {title} {Neural-network quantum state tomography},\ }\href {https://doi.org/https://doi.org/10.1038/s41567-018-0048-5} {\bibfield  {journal} {\bibinfo  {journal} {Nature Physics}\ }\textbf {\bibinfo {volume} {14}},\ \bibinfo {pages} {447} (\bibinfo {year} {2018})}\BibitemShut {NoStop}%
\bibitem [{\citenamefont {Ahmed}\ \emph {et~al.}(2021)\citenamefont {Ahmed}, \citenamefont {S\'anchez Mu\~noz}, \citenamefont {Nori},\ and\ \citenamefont {Kockum}}]{PhysRevLett.127.140502}%
  \BibitemOpen
  \bibfield  {author} {\bibinfo {author} {\bibfnamefont {S.}~\bibnamefont {Ahmed}}, \bibinfo {author} {\bibfnamefont {C.}~\bibnamefont {S\'anchez Mu\~noz}}, \bibinfo {author} {\bibfnamefont {F.}~\bibnamefont {Nori}},\ and\ \bibinfo {author} {\bibfnamefont {A.~F.}\ \bibnamefont {Kockum}},\ }\bibfield  {title} {\bibinfo {title} {Quantum state tomography with conditional generative adversarial networks},\ }\href {https://doi.org/10.1103/PhysRevLett.127.140502} {\bibfield  {journal} {\bibinfo  {journal} {Physical Review Letters}\ }\textbf {\bibinfo {volume} {127}},\ \bibinfo {pages} {140502} (\bibinfo {year} {2021})}\BibitemShut {NoStop}%
\bibitem [{\citenamefont {Hashim}\ \emph {et~al.}(2025)\citenamefont {Hashim}, \citenamefont {Nguyen}, \citenamefont {Goss}, \citenamefont {Marinelli}, \citenamefont {Naik}, \citenamefont {Chistolini}, \citenamefont {Hines}, \citenamefont {Marceaux}, \citenamefont {Kim}, \citenamefont {Gokhale}, \citenamefont {Tomesh}, \citenamefont {Chen}, \citenamefont {Jiang}, \citenamefont {Ferracin}, \citenamefont {Rudinger}, \citenamefont {Proctor}, \citenamefont {Young}, \citenamefont {Siddiqi},\ and\ \citenamefont {Blume-Kohout}}]{PRXQuantum.6.030202}%
  \BibitemOpen
  \bibfield  {author} {\bibinfo {author} {\bibfnamefont {A.}~\bibnamefont {Hashim}}, \bibinfo {author} {\bibfnamefont {L.~B.}\ \bibnamefont {Nguyen}}, \bibinfo {author} {\bibfnamefont {N.}~\bibnamefont {Goss}}, \bibinfo {author} {\bibfnamefont {B.}~\bibnamefont {Marinelli}}, \bibinfo {author} {\bibfnamefont {R.~K.}\ \bibnamefont {Naik}}, \bibinfo {author} {\bibfnamefont {T.}~\bibnamefont {Chistolini}}, \bibinfo {author} {\bibfnamefont {J.}~\bibnamefont {Hines}}, \bibinfo {author} {\bibfnamefont {J.}~\bibnamefont {Marceaux}}, \bibinfo {author} {\bibfnamefont {Y.}~\bibnamefont {Kim}}, \bibinfo {author} {\bibfnamefont {P.}~\bibnamefont {Gokhale}}, \bibinfo {author} {\bibfnamefont {T.}~\bibnamefont {Tomesh}}, \bibinfo {author} {\bibfnamefont {S.}~\bibnamefont {Chen}}, \bibinfo {author} {\bibfnamefont {L.}~\bibnamefont {Jiang}}, \bibinfo {author} {\bibfnamefont {S.}~\bibnamefont {Ferracin}}, \bibinfo {author} {\bibfnamefont {K.}~\bibnamefont {Rudinger}}, \bibinfo {author} {\bibfnamefont {T.}~\bibnamefont
  {Proctor}}, \bibinfo {author} {\bibfnamefont {K.~C.}\ \bibnamefont {Young}}, \bibinfo {author} {\bibfnamefont {I.}~\bibnamefont {Siddiqi}},\ and\ \bibinfo {author} {\bibfnamefont {R.}~\bibnamefont {Blume-Kohout}},\ }\bibfield  {title} {\bibinfo {title} {Practical introduction to benchmarking and characterization of quantum computers},\ }\href {https://doi.org/10.1103/PRXQuantum.6.030202} {\bibfield  {journal} {\bibinfo  {journal} {PRX Quantum}\ }\textbf {\bibinfo {volume} {6}},\ \bibinfo {pages} {030202} (\bibinfo {year} {2025})}\BibitemShut {NoStop}%
\bibitem [{\citenamefont {Gebhart}\ \emph {et~al.}(2023)\citenamefont {Gebhart}, \citenamefont {Santagati}, \citenamefont {Gentile}, \citenamefont {Gauger}, \citenamefont {Craig}, \citenamefont {Ares}, \citenamefont {Banchi}, \citenamefont {Marquardt}, \citenamefont {Pezze},\ and\ \citenamefont {Bonato}}]{gebhart2023learning}%
  \BibitemOpen
  \bibfield  {author} {\bibinfo {author} {\bibfnamefont {V.}~\bibnamefont {Gebhart}}, \bibinfo {author} {\bibfnamefont {R.}~\bibnamefont {Santagati}}, \bibinfo {author} {\bibfnamefont {A.~A.}\ \bibnamefont {Gentile}}, \bibinfo {author} {\bibfnamefont {E.~M.}\ \bibnamefont {Gauger}}, \bibinfo {author} {\bibfnamefont {D.}~\bibnamefont {Craig}}, \bibinfo {author} {\bibfnamefont {N.}~\bibnamefont {Ares}}, \bibinfo {author} {\bibfnamefont {L.}~\bibnamefont {Banchi}}, \bibinfo {author} {\bibfnamefont {F.}~\bibnamefont {Marquardt}}, \bibinfo {author} {\bibfnamefont {L.}~\bibnamefont {Pezze}},\ and\ \bibinfo {author} {\bibfnamefont {C.}~\bibnamefont {Bonato}},\ }\bibfield  {title} {\bibinfo {title} {Learning quantum systems},\ }\href {https://doi.org/https://doi.org/10.1038/s42254-022-00552-1} {\bibfield  {journal} {\bibinfo  {journal} {Nature Reviews Physics}\ }\textbf {\bibinfo {volume} {5}},\ \bibinfo {pages} {141} (\bibinfo {year} {2023})}\BibitemShut {NoStop}%
\bibitem [{\citenamefont {Proctor}\ \emph {et~al.}(2025)\citenamefont {Proctor}, \citenamefont {Young}, \citenamefont {Baczewski},\ and\ \citenamefont {Blume-Kohout}}]{proctor2025benchmarking}%
  \BibitemOpen
  \bibfield  {author} {\bibinfo {author} {\bibfnamefont {T.}~\bibnamefont {Proctor}}, \bibinfo {author} {\bibfnamefont {K.}~\bibnamefont {Young}}, \bibinfo {author} {\bibfnamefont {A.~D.}\ \bibnamefont {Baczewski}},\ and\ \bibinfo {author} {\bibfnamefont {R.}~\bibnamefont {Blume-Kohout}},\ }\bibfield  {title} {\bibinfo {title} {Benchmarking quantum computers},\ }\href {https://doi.org/10.1038/s42254-024-00796-z} {\bibfield  {journal} {\bibinfo  {journal} {Nature Reviews Physics}\ }\textbf {\bibinfo {volume} {7}},\ \bibinfo {pages} {105} (\bibinfo {year} {2025})}\BibitemShut {NoStop}%
\bibitem [{\citenamefont {Nielsen}\ \emph {et~al.}(2021)\citenamefont {Nielsen}, \citenamefont {Gamble}, \citenamefont {Rudinger}, \citenamefont {Scholten}, \citenamefont {Young},\ and\ \citenamefont {Blume-Kohout}}]{Nielsen2021gatesettomography}%
  \BibitemOpen
  \bibfield  {author} {\bibinfo {author} {\bibfnamefont {E.}~\bibnamefont {Nielsen}}, \bibinfo {author} {\bibfnamefont {J.~K.}\ \bibnamefont {Gamble}}, \bibinfo {author} {\bibfnamefont {K.}~\bibnamefont {Rudinger}}, \bibinfo {author} {\bibfnamefont {T.}~\bibnamefont {Scholten}}, \bibinfo {author} {\bibfnamefont {K.}~\bibnamefont {Young}},\ and\ \bibinfo {author} {\bibfnamefont {R.}~\bibnamefont {Blume-Kohout}},\ }\bibfield  {title} {\bibinfo {title} {Gate set tomography},\ }\href {https://doi.org/10.22331/q-2021-10-05-557} {\bibfield  {journal} {\bibinfo  {journal} {{Quantum}}\ }\textbf {\bibinfo {volume} {5}},\ \bibinfo {pages} {557} (\bibinfo {year} {2021})}\BibitemShut {NoStop}%
\bibitem [{\citenamefont {Wiebe}\ \emph {et~al.}(2014{\natexlab{a}})\citenamefont {Wiebe}, \citenamefont {Granade}, \citenamefont {Ferrie},\ and\ \citenamefont {Cory}}]{PhysRevLett.112.190501}%
  \BibitemOpen
  \bibfield  {author} {\bibinfo {author} {\bibfnamefont {N.}~\bibnamefont {Wiebe}}, \bibinfo {author} {\bibfnamefont {C.}~\bibnamefont {Granade}}, \bibinfo {author} {\bibfnamefont {C.}~\bibnamefont {Ferrie}},\ and\ \bibinfo {author} {\bibfnamefont {D.~G.}\ \bibnamefont {Cory}},\ }\bibfield  {title} {\bibinfo {title} {Hamiltonian learning and certification using quantum resources},\ }\href {https://doi.org/10.1103/PhysRevLett.112.190501} {\bibfield  {journal} {\bibinfo  {journal} {Physical Review Letters}\ }\textbf {\bibinfo {volume} {112}},\ \bibinfo {pages} {190501} (\bibinfo {year} {2014}{\natexlab{a}})}\BibitemShut {NoStop}%
\bibitem [{\citenamefont {Wang}\ \emph {et~al.}(2017)\citenamefont {Wang}, \citenamefont {Paesani}, \citenamefont {Santagati}, \citenamefont {Knauer}, \citenamefont {Gentile}, \citenamefont {Wiebe}, \citenamefont {Petruzzella}, \citenamefont {O’Brien}, \citenamefont {Rarity}, \citenamefont {Laing} \emph {et~al.}}]{wang2017experimental}%
  \BibitemOpen
  \bibfield  {author} {\bibinfo {author} {\bibfnamefont {J.}~\bibnamefont {Wang}}, \bibinfo {author} {\bibfnamefont {S.}~\bibnamefont {Paesani}}, \bibinfo {author} {\bibfnamefont {R.}~\bibnamefont {Santagati}}, \bibinfo {author} {\bibfnamefont {S.}~\bibnamefont {Knauer}}, \bibinfo {author} {\bibfnamefont {A.~A.}\ \bibnamefont {Gentile}}, \bibinfo {author} {\bibfnamefont {N.}~\bibnamefont {Wiebe}}, \bibinfo {author} {\bibfnamefont {M.}~\bibnamefont {Petruzzella}}, \bibinfo {author} {\bibfnamefont {J.~L.}\ \bibnamefont {O’Brien}}, \bibinfo {author} {\bibfnamefont {J.~G.}\ \bibnamefont {Rarity}}, \bibinfo {author} {\bibfnamefont {A.}~\bibnamefont {Laing}}, \emph {et~al.},\ }\bibfield  {title} {\bibinfo {title} {{Experimental quantum Hamiltonian learning}},\ }\href {https://doi.org/https://doi.org/10.1038/nphys4074} {\bibfield  {journal} {\bibinfo  {journal} {Nature Physics}\ }\textbf {\bibinfo {volume} {13}},\ \bibinfo {pages} {551} (\bibinfo {year} {2017})}\BibitemShut {NoStop}%
\bibitem [{\citenamefont {Gentile}\ \emph {et~al.}(2021)\citenamefont {Gentile}, \citenamefont {Flynn}, \citenamefont {Knauer}, \citenamefont {Wiebe}, \citenamefont {Paesani}, \citenamefont {Granade}, \citenamefont {Rarity}, \citenamefont {Santagati},\ and\ \citenamefont {Laing}}]{gentile2021learning}%
  \BibitemOpen
  \bibfield  {author} {\bibinfo {author} {\bibfnamefont {A.~A.}\ \bibnamefont {Gentile}}, \bibinfo {author} {\bibfnamefont {B.}~\bibnamefont {Flynn}}, \bibinfo {author} {\bibfnamefont {S.}~\bibnamefont {Knauer}}, \bibinfo {author} {\bibfnamefont {N.}~\bibnamefont {Wiebe}}, \bibinfo {author} {\bibfnamefont {S.}~\bibnamefont {Paesani}}, \bibinfo {author} {\bibfnamefont {C.~E.}\ \bibnamefont {Granade}}, \bibinfo {author} {\bibfnamefont {J.~G.}\ \bibnamefont {Rarity}}, \bibinfo {author} {\bibfnamefont {R.}~\bibnamefont {Santagati}},\ and\ \bibinfo {author} {\bibfnamefont {A.}~\bibnamefont {Laing}},\ }\bibfield  {title} {\bibinfo {title} {Learning models of quantum systems from experiments},\ }\href {https://doi.org/https://doi.org/10.1038/s41567-021-01201-7} {\bibfield  {journal} {\bibinfo  {journal} {Nature Physics}\ }\textbf {\bibinfo {volume} {17}},\ \bibinfo {pages} {837} (\bibinfo {year} {2021})}\BibitemShut {NoStop}%
\bibitem [{\citenamefont {Simard}\ \emph {et~al.}(2025)\citenamefont {Simard}, \citenamefont {Dawid}, \citenamefont {Tindall}, \citenamefont {Ferrero}, \citenamefont {Sengupta},\ and\ \citenamefont {Georges}}]{f58h-zxs3}%
  \BibitemOpen
  \bibfield  {author} {\bibinfo {author} {\bibfnamefont {O.}~\bibnamefont {Simard}}, \bibinfo {author} {\bibfnamefont {A.}~\bibnamefont {Dawid}}, \bibinfo {author} {\bibfnamefont {J.}~\bibnamefont {Tindall}}, \bibinfo {author} {\bibfnamefont {M.}~\bibnamefont {Ferrero}}, \bibinfo {author} {\bibfnamefont {A.~M.}\ \bibnamefont {Sengupta}},\ and\ \bibinfo {author} {\bibfnamefont {A.}~\bibnamefont {Georges}},\ }\bibfield  {title} {\bibinfo {title} {{Learning interactions between Rydberg atoms}},\ }\href {https://doi.org/10.1103/f58h-zxs3} {\bibfield  {journal} {\bibinfo  {journal} {PRX Quantum}\ }\textbf {\bibinfo {volume} {6}},\ \bibinfo {pages} {030324} (\bibinfo {year} {2025})}\BibitemShut {NoStop}%
\bibitem [{\citenamefont {Chertkov}\ and\ \citenamefont {Clark}(2018)}]{PhysRevX.8.031029}%
  \BibitemOpen
  \bibfield  {author} {\bibinfo {author} {\bibfnamefont {E.}~\bibnamefont {Chertkov}}\ and\ \bibinfo {author} {\bibfnamefont {B.~K.}\ \bibnamefont {Clark}},\ }\bibfield  {title} {\bibinfo {title} {Computational inverse method for constructing spaces of quantum models from wave functions},\ }\href {https://doi.org/10.1103/PhysRevX.8.031029} {\bibfield  {journal} {\bibinfo  {journal} {Physical Review X}\ }\textbf {\bibinfo {volume} {8}},\ \bibinfo {pages} {031029} (\bibinfo {year} {2018})}\BibitemShut {NoStop}%
\bibitem [{\citenamefont {Inui}\ and\ \citenamefont {Motome}(2024)}]{PhysRevResearch.6.033080}%
  \BibitemOpen
  \bibfield  {author} {\bibinfo {author} {\bibfnamefont {K.}~\bibnamefont {Inui}}\ and\ \bibinfo {author} {\bibfnamefont {Y.}~\bibnamefont {Motome}},\ }\bibfield  {title} {\bibinfo {title} {{Inverse Hamiltonian design of highly entangled quantum systems}},\ }\href {https://doi.org/10.1103/PhysRevResearch.6.033080} {\bibfield  {journal} {\bibinfo  {journal} {Physical Review Research}\ }\textbf {\bibinfo {volume} {6}},\ \bibinfo {pages} {033080} (\bibinfo {year} {2024})}\BibitemShut {NoStop}%
\bibitem [{\citenamefont {Kokail}\ \emph {et~al.}(2026)\citenamefont {Kokail}, \citenamefont {Dolgirev}, \citenamefont {van Bijnen}, \citenamefont {Gonzalez-Cuadra}, \citenamefont {Lukin},\ and\ \citenamefont {Zoller}}]{kokail2026inversequan}%
  \BibitemOpen
  \bibfield  {author} {\bibinfo {author} {\bibfnamefont {C.}~\bibnamefont {Kokail}}, \bibinfo {author} {\bibfnamefont {P.~E.}\ \bibnamefont {Dolgirev}}, \bibinfo {author} {\bibfnamefont {R.}~\bibnamefont {van Bijnen}}, \bibinfo {author} {\bibfnamefont {D.}~\bibnamefont {Gonzalez-Cuadra}}, \bibinfo {author} {\bibfnamefont {M.~D.}\ \bibnamefont {Lukin}},\ and\ \bibinfo {author} {\bibfnamefont {P.}~\bibnamefont {Zoller}},\ }\href@noop {} {\bibinfo {title} {Inverse quantum simulation for quantum material design}} (\bibinfo {year} {2026}),\ \Eprint {https://arxiv.org/abs/2601.12239} {arXiv:2601.12239} \BibitemShut {NoStop}%
\bibitem [{\citenamefont {Ferrie}\ and\ \citenamefont {Granade}(2014)}]{PhysRevLett.112.130402}%
  \BibitemOpen
  \bibfield  {author} {\bibinfo {author} {\bibfnamefont {C.}~\bibnamefont {Ferrie}}\ and\ \bibinfo {author} {\bibfnamefont {C.~E.}\ \bibnamefont {Granade}},\ }\bibfield  {title} {\bibinfo {title} {Likelihood-free methods for quantum parameter estimation},\ }\href {https://doi.org/10.1103/PhysRevLett.112.130402} {\bibfield  {journal} {\bibinfo  {journal} {Physical Review Letters}\ }\textbf {\bibinfo {volume} {112}},\ \bibinfo {pages} {130402} (\bibinfo {year} {2014})}\BibitemShut {NoStop}%
\bibitem [{\citenamefont {Catana}\ \emph {et~al.}(2014)\citenamefont {Catana}, \citenamefont {Kypraios},\ and\ \citenamefont {Gu{\c{t}}{\u{a}}}}]{catana2014maximum}%
  \BibitemOpen
  \bibfield  {author} {\bibinfo {author} {\bibfnamefont {C.}~\bibnamefont {Catana}}, \bibinfo {author} {\bibfnamefont {T.}~\bibnamefont {Kypraios}},\ and\ \bibinfo {author} {\bibfnamefont {M.}~\bibnamefont {Gu{\c{t}}{\u{a}}}},\ }\bibfield  {title} {\bibinfo {title} {Maximum likelihood versus likelihood-free quantum system identification in the atom maser},\ }\href {https://doi.org/10.1088/1751-8113/47/41/415302} {\bibfield  {journal} {\bibinfo  {journal} {Journal of Physics A: Mathematical and Theoretical}\ }\textbf {\bibinfo {volume} {47}},\ \bibinfo {pages} {415302} (\bibinfo {year} {2014})}\BibitemShut {NoStop}%
\bibitem [{\citenamefont {Clark}\ and\ \citenamefont {Ko\l{}ody\ifmmode~\acute{n}\else \'{n}\fi{}ski}(2025)}]{PhysRevApplied.23.044040}%
  \BibitemOpen
  \bibfield  {author} {\bibinfo {author} {\bibfnamefont {L.~A.}\ \bibnamefont {Clark}}\ and\ \bibinfo {author} {\bibfnamefont {J.}~\bibnamefont {Ko\l{}ody\ifmmode~\acute{n}\else \'{n}\fi{}ski}},\ }\bibfield  {title} {\bibinfo {title} {{Efficient inference of quantum system parameters by approximate Bayesian computation}},\ }\href {https://doi.org/10.1103/PhysRevApplied.23.044040} {\bibfield  {journal} {\bibinfo  {journal} {Physical Review Applied}\ }\textbf {\bibinfo {volume} {23}},\ \bibinfo {pages} {044040} (\bibinfo {year} {2025})}\BibitemShut {NoStop}%
\bibitem [{\citenamefont {Cranmer}\ \emph {et~al.}(2020)\citenamefont {Cranmer}, \citenamefont {Brehmer},\ and\ \citenamefont {Louppe}}]{cranmer2020frontier}%
  \BibitemOpen
  \bibfield  {author} {\bibinfo {author} {\bibfnamefont {K.}~\bibnamefont {Cranmer}}, \bibinfo {author} {\bibfnamefont {J.}~\bibnamefont {Brehmer}},\ and\ \bibinfo {author} {\bibfnamefont {G.}~\bibnamefont {Louppe}},\ }\bibfield  {title} {\bibinfo {title} {The frontier of simulation-based inference},\ }\href {https://doi.org/https://doi.org/10.1073/pnas.1912789117} {\bibfield  {journal} {\bibinfo  {journal} {Proceedings of the National Academy of Sciences}\ }\textbf {\bibinfo {volume} {117}},\ \bibinfo {pages} {30055} (\bibinfo {year} {2020})}\BibitemShut {NoStop}%
\bibitem [{\citenamefont {Deistler}\ \emph {et~al.}(2025)\citenamefont {Deistler}, \citenamefont {Boelts}, \citenamefont {Steinbach}, \citenamefont {Moss}, \citenamefont {Moreau}, \citenamefont {Gloeckler}, \citenamefont {Rodrigues}, \citenamefont {Linhart}, \citenamefont {Lappalainen}, \citenamefont {Miller}, \citenamefont {Gonçalves}, \citenamefont {Lueckmann}, \citenamefont {Schröder},\ and\ \citenamefont {Macke}}]{deistler2025simul}%
  \BibitemOpen
  \bibfield  {author} {\bibinfo {author} {\bibfnamefont {M.}~\bibnamefont {Deistler}}, \bibinfo {author} {\bibfnamefont {J.}~\bibnamefont {Boelts}}, \bibinfo {author} {\bibfnamefont {P.}~\bibnamefont {Steinbach}}, \bibinfo {author} {\bibfnamefont {G.}~\bibnamefont {Moss}}, \bibinfo {author} {\bibfnamefont {T.}~\bibnamefont {Moreau}}, \bibinfo {author} {\bibfnamefont {M.}~\bibnamefont {Gloeckler}}, \bibinfo {author} {\bibfnamefont {P.~L.~C.}\ \bibnamefont {Rodrigues}}, \bibinfo {author} {\bibfnamefont {J.}~\bibnamefont {Linhart}}, \bibinfo {author} {\bibfnamefont {J.~K.}\ \bibnamefont {Lappalainen}}, \bibinfo {author} {\bibfnamefont {B.~K.}\ \bibnamefont {Miller}}, \bibinfo {author} {\bibfnamefont {P.~J.}\ \bibnamefont {Gonçalves}}, \bibinfo {author} {\bibfnamefont {J.-M.}\ \bibnamefont {Lueckmann}}, \bibinfo {author} {\bibfnamefont {C.}~\bibnamefont {Schröder}},\ and\ \bibinfo {author} {\bibfnamefont {J.~H.}\ \bibnamefont {Macke}},\ }\href@noop {} {\bibinfo {title} {{Simulation-based inference: A practical
  guide}}} (\bibinfo {year} {2025}),\ \Eprint {https://arxiv.org/abs/2508.12939} {arXiv:2508.12939} \BibitemShut {NoStop}%
\bibitem [{\citenamefont {Dax}\ \emph {et~al.}(2025)\citenamefont {Dax}, \citenamefont {Green}, \citenamefont {Gair}, \citenamefont {Gupte}, \citenamefont {P{\"u}rrer}, \citenamefont {Raymond}, \citenamefont {Wildberger}, \citenamefont {Macke}, \citenamefont {Buonanno},\ and\ \citenamefont {Sch{\"o}lkopf}}]{dax2025real}%
  \BibitemOpen
  \bibfield  {author} {\bibinfo {author} {\bibfnamefont {M.}~\bibnamefont {Dax}}, \bibinfo {author} {\bibfnamefont {S.~R.}\ \bibnamefont {Green}}, \bibinfo {author} {\bibfnamefont {J.}~\bibnamefont {Gair}}, \bibinfo {author} {\bibfnamefont {N.}~\bibnamefont {Gupte}}, \bibinfo {author} {\bibfnamefont {M.}~\bibnamefont {P{\"u}rrer}}, \bibinfo {author} {\bibfnamefont {V.}~\bibnamefont {Raymond}}, \bibinfo {author} {\bibfnamefont {J.}~\bibnamefont {Wildberger}}, \bibinfo {author} {\bibfnamefont {J.~H.}\ \bibnamefont {Macke}}, \bibinfo {author} {\bibfnamefont {A.}~\bibnamefont {Buonanno}},\ and\ \bibinfo {author} {\bibfnamefont {B.}~\bibnamefont {Sch{\"o}lkopf}},\ }\bibfield  {title} {\bibinfo {title} {Real-time inference for binary neutron star mergers using machine learning},\ }\href {https://doi.org/https://doi.org/10.1038/s41586-025-08593-z} {\bibfield  {journal} {\bibinfo  {journal} {Nature}\ }\textbf {\bibinfo {volume} {639}},\ \bibinfo {pages} {49} (\bibinfo {year} {2025})}\BibitemShut {NoStop}%
\bibitem [{\citenamefont {Dingeldein}\ \emph {et~al.}(2025{\natexlab{a}})\citenamefont {Dingeldein}, \citenamefont {Cossio},\ and\ \citenamefont {Covino}}]{DINGELDEIN2025102988}%
  \BibitemOpen
  \bibfield  {author} {\bibinfo {author} {\bibfnamefont {L.}~\bibnamefont {Dingeldein}}, \bibinfo {author} {\bibfnamefont {P.}~\bibnamefont {Cossio}},\ and\ \bibinfo {author} {\bibfnamefont {R.}~\bibnamefont {Covino}},\ }\bibfield  {title} {\bibinfo {title} {Simulation-based inference of single-molecule experiments},\ }\href {https://doi.org/https://doi.org/10.1016/j.sbi.2025.102988} {\bibfield  {journal} {\bibinfo  {journal} {Current Opinion in Structural Biology}\ }\textbf {\bibinfo {volume} {91}},\ \bibinfo {pages} {102988} (\bibinfo {year} {2025}{\natexlab{a}})}\BibitemShut {NoStop}%
\bibitem [{\citenamefont {{The ATLAS Collaboration}}(2025)}]{atlas2025implementation}%
  \BibitemOpen
  \bibfield  {author} {\bibinfo {author} {\bibnamefont {{The ATLAS Collaboration}}},\ }\bibfield  {title} {\bibinfo {title} {{An implementation of neural simulation-based inference for parameter estimation in ATLAS}},\ }\href {https://doi.org/10.1088/1361-6633/add370} {\bibfield  {journal} {\bibinfo  {journal} {Reports on Progress in Physics}\ }\textbf {\bibinfo {volume} {88}},\ \bibinfo {pages} {067801} (\bibinfo {year} {2025})}\BibitemShut {NoStop}%
\bibitem [{\citenamefont {Zhang}\ \emph {et~al.}(2026)\citenamefont {Zhang}, \citenamefont {Zhang}, \citenamefont {Yi}, \citenamefont {Ren}, \citenamefont {Jiao}, \citenamefont {Bai}, \citenamefont {Jiang},\ and\ \citenamefont {Song}}]{zhang2026discovery}%
  \BibitemOpen
  \bibfield  {author} {\bibinfo {author} {\bibfnamefont {J.}~\bibnamefont {Zhang}}, \bibinfo {author} {\bibfnamefont {Y.}~\bibnamefont {Zhang}}, \bibinfo {author} {\bibfnamefont {B.}~\bibnamefont {Yi}}, \bibinfo {author} {\bibfnamefont {Y.}~\bibnamefont {Ren}}, \bibinfo {author} {\bibfnamefont {Q.}~\bibnamefont {Jiao}}, \bibinfo {author} {\bibfnamefont {H.}~\bibnamefont {Bai}}, \bibinfo {author} {\bibfnamefont {W.}~\bibnamefont {Jiang}},\ and\ \bibinfo {author} {\bibfnamefont {Z.}~\bibnamefont {Song}},\ }\bibfield  {title} {\bibinfo {title} {Discovery learning predicts battery cycle life from minimal experiments},\ }\href {https://doi.org/10.1038/s41586-025-09951-7} {\bibfield  {journal} {\bibinfo  {journal} {Nature}\ }\textbf {\bibinfo {volume} {650}},\ \bibinfo {pages} {110} (\bibinfo {year} {2026})}\BibitemShut {NoStop}%
\bibitem [{\citenamefont {Dingeldein}\ \emph {et~al.}(2025{\natexlab{b}})\citenamefont {Dingeldein}, \citenamefont {Silva-Sánchez}, \citenamefont {Evans}, \citenamefont {D’Imprima}, \citenamefont {Grigorieff}, \citenamefont {Covino},\ and\ \citenamefont {Cossio}}]{doi:10.1073/pnas.2420158122}%
  \BibitemOpen
  \bibfield  {author} {\bibinfo {author} {\bibfnamefont {L.}~\bibnamefont {Dingeldein}}, \bibinfo {author} {\bibfnamefont {D.}~\bibnamefont {Silva-Sánchez}}, \bibinfo {author} {\bibfnamefont {L.}~\bibnamefont {Evans}}, \bibinfo {author} {\bibfnamefont {E.}~\bibnamefont {D’Imprima}}, \bibinfo {author} {\bibfnamefont {N.}~\bibnamefont {Grigorieff}}, \bibinfo {author} {\bibfnamefont {R.}~\bibnamefont {Covino}},\ and\ \bibinfo {author} {\bibfnamefont {P.}~\bibnamefont {Cossio}},\ }\bibfield  {title} {\bibinfo {title} {Amortized template matching of molecular conformations from cryoelectron microscopy images using simulation-based inference},\ }\href {https://doi.org/10.1073/pnas.2420158122} {\bibfield  {journal} {\bibinfo  {journal} {Proceedings of the National Academy of Sciences}\ }\textbf {\bibinfo {volume} {122}},\ \bibinfo {pages} {e2420158122} (\bibinfo {year} {2025}{\natexlab{b}})}\BibitemShut {NoStop}%
\bibitem [{\citenamefont {Bond-Taylor}\ \emph {et~al.}(2022)\citenamefont {Bond-Taylor}, \citenamefont {Leach}, \citenamefont {Long},\ and\ \citenamefont {Willcocks}}]{9555209}%
  \BibitemOpen
  \bibfield  {author} {\bibinfo {author} {\bibfnamefont {S.}~\bibnamefont {Bond-Taylor}}, \bibinfo {author} {\bibfnamefont {A.}~\bibnamefont {Leach}}, \bibinfo {author} {\bibfnamefont {Y.}~\bibnamefont {Long}},\ and\ \bibinfo {author} {\bibfnamefont {C.~G.}\ \bibnamefont {Willcocks}},\ }\bibfield  {title} {\bibinfo {title} {{Deep generative modelling: A comparative review of VAEs, GANs, normalizing flows, energy-based and autoregressive models}},\ }\href {https://doi.org/10.1109/TPAMI.2021.3116668} {\bibfield  {journal} {\bibinfo  {journal} {IEEE Transactions on Pattern Analysis and Machine Intelligence}\ }\textbf {\bibinfo {volume} {44}},\ \bibinfo {pages} {7327} (\bibinfo {year} {2022})}\BibitemShut {NoStop}%
\bibitem [{\citenamefont {Kobyzev}\ \emph {et~al.}(2021)\citenamefont {Kobyzev}, \citenamefont {Prince},\ and\ \citenamefont {Brubaker}}]{9089305}%
  \BibitemOpen
  \bibfield  {author} {\bibinfo {author} {\bibfnamefont {I.}~\bibnamefont {Kobyzev}}, \bibinfo {author} {\bibfnamefont {S.~J.}\ \bibnamefont {Prince}},\ and\ \bibinfo {author} {\bibfnamefont {M.~A.}\ \bibnamefont {Brubaker}},\ }\bibfield  {title} {\bibinfo {title} {Normalizing flows: An introduction and review of current methods},\ }\href {https://doi.org/10.1109/TPAMI.2020.2992934} {\bibfield  {journal} {\bibinfo  {journal} {IEEE Transactions on Pattern Analysis and Machine Intelligence}\ }\textbf {\bibinfo {volume} {43}},\ \bibinfo {pages} {3964} (\bibinfo {year} {2021})}\BibitemShut {NoStop}%
\bibitem [{\citenamefont {Lueckmann}\ \emph {et~al.}(2017)\citenamefont {Lueckmann}, \citenamefont {Goncalves}, \citenamefont {Bassetto}, \citenamefont {{\"O}cal}, \citenamefont {Nonnenmacher},\ and\ \citenamefont {Macke}}]{10.5555/3294771.3294894}%
  \BibitemOpen
  \bibfield  {author} {\bibinfo {author} {\bibfnamefont {J.-M.}\ \bibnamefont {Lueckmann}}, \bibinfo {author} {\bibfnamefont {P.~J.}\ \bibnamefont {Goncalves}}, \bibinfo {author} {\bibfnamefont {G.}~\bibnamefont {Bassetto}}, \bibinfo {author} {\bibfnamefont {K.}~\bibnamefont {{\"O}cal}}, \bibinfo {author} {\bibfnamefont {M.}~\bibnamefont {Nonnenmacher}},\ and\ \bibinfo {author} {\bibfnamefont {J.~H.}\ \bibnamefont {Macke}},\ }\bibfield  {title} {\bibinfo {title} {Flexible statistical inference for mechanistic models of neural dynamics},\ }in\ \href@noop {} {\emph {\bibinfo {booktitle} {Advances in Neural Information Processing Systems}}},\ Vol.~\bibinfo {volume} {30}\ (\bibinfo {year} {2017})\BibitemShut {NoStop}%
\bibitem [{\citenamefont {Papamakarios}\ and\ \citenamefont {Murray}(2016)}]{10.5555/3157096.3157212}%
  \BibitemOpen
  \bibfield  {author} {\bibinfo {author} {\bibfnamefont {G.}~\bibnamefont {Papamakarios}}\ and\ \bibinfo {author} {\bibfnamefont {I.}~\bibnamefont {Murray}},\ }\bibfield  {title} {\bibinfo {title} {{Fast $\varepsilon$-free inference of simulation models with Bayesian conditional density estimation}},\ }in\ \href@noop {} {\emph {\bibinfo {booktitle} {Advances in Neural Information Processing Systems}}},\ Vol.~\bibinfo {volume} {29}\ (\bibinfo {year} {2016})\BibitemShut {NoStop}%
\bibitem [{\citenamefont {Greenberg}\ \emph {et~al.}(2019)\citenamefont {Greenberg}, \citenamefont {Nonnenmacher},\ and\ \citenamefont {Macke}}]{greenberg2019automatic}%
  \BibitemOpen
  \bibfield  {author} {\bibinfo {author} {\bibfnamefont {D.}~\bibnamefont {Greenberg}}, \bibinfo {author} {\bibfnamefont {M.}~\bibnamefont {Nonnenmacher}},\ and\ \bibinfo {author} {\bibfnamefont {J.}~\bibnamefont {Macke}},\ }\bibfield  {title} {\bibinfo {title} {Automatic posterior transformation for likelihood-free inference},\ }in\ \href@noop {} {\emph {\bibinfo {booktitle} {International Conference on Machine Learning}}}\ (\bibinfo {year} {2019})\ pp.\ \bibinfo {pages} {2404--2414}\BibitemShut {NoStop}%
\bibitem [{\citenamefont {Wildberger}\ \emph {et~al.}(2023)\citenamefont {Wildberger}, \citenamefont {Dax}, \citenamefont {Buchholz}, \citenamefont {Green}, \citenamefont {Macke},\ and\ \citenamefont {Sch\"{o}lkopf}}]{10.5555/3666122.3666859}%
  \BibitemOpen
  \bibfield  {author} {\bibinfo {author} {\bibfnamefont {J.}~\bibnamefont {Wildberger}}, \bibinfo {author} {\bibfnamefont {M.}~\bibnamefont {Dax}}, \bibinfo {author} {\bibfnamefont {S.}~\bibnamefont {Buchholz}}, \bibinfo {author} {\bibfnamefont {S.~R.}\ \bibnamefont {Green}}, \bibinfo {author} {\bibfnamefont {J.~H.}\ \bibnamefont {Macke}},\ and\ \bibinfo {author} {\bibfnamefont {B.}~\bibnamefont {Sch\"{o}lkopf}},\ }\bibfield  {title} {\bibinfo {title} {Flow matching for scalable simulation-based inference},\ }in\ \href@noop {} {\emph {\bibinfo {booktitle} {Advances in Neural Information Processing Systems}}},\ Vol.~\bibinfo {volume} {36}\ (\bibinfo {year} {2023})\BibitemShut {NoStop}%
\bibitem [{\citenamefont {Zammit-Mangion}\ \emph {et~al.}(2025)\citenamefont {Zammit-Mangion}, \citenamefont {Sainsbury-Dale},\ and\ \citenamefont {Huser}}]{zammit2025neural}%
  \BibitemOpen
  \bibfield  {author} {\bibinfo {author} {\bibfnamefont {A.}~\bibnamefont {Zammit-Mangion}}, \bibinfo {author} {\bibfnamefont {M.}~\bibnamefont {Sainsbury-Dale}},\ and\ \bibinfo {author} {\bibfnamefont {R.}~\bibnamefont {Huser}},\ }\bibfield  {title} {\bibinfo {title} {Neural methods for amortized inference},\ }\href {https://doi.org/10.1146/annurev-statistics-112723-034123} {\bibfield  {journal} {\bibinfo  {journal} {Annual Review of Statistics and Its Application}\ }\textbf {\bibinfo {volume} {12}},\ \bibinfo {pages} {311} (\bibinfo {year} {2025})}\BibitemShut {NoStop}%
\bibitem [{\citenamefont {Proctor}\ \emph {et~al.}(2020)\citenamefont {Proctor}, \citenamefont {Revelle}, \citenamefont {Nielsen}, \citenamefont {Rudinger}, \citenamefont {Lobser}, \citenamefont {Maunz}, \citenamefont {Blume-Kohout},\ and\ \citenamefont {Young}}]{proctor2020detecting}%
  \BibitemOpen
  \bibfield  {author} {\bibinfo {author} {\bibfnamefont {T.}~\bibnamefont {Proctor}}, \bibinfo {author} {\bibfnamefont {M.}~\bibnamefont {Revelle}}, \bibinfo {author} {\bibfnamefont {E.}~\bibnamefont {Nielsen}}, \bibinfo {author} {\bibfnamefont {K.}~\bibnamefont {Rudinger}}, \bibinfo {author} {\bibfnamefont {D.}~\bibnamefont {Lobser}}, \bibinfo {author} {\bibfnamefont {P.}~\bibnamefont {Maunz}}, \bibinfo {author} {\bibfnamefont {R.}~\bibnamefont {Blume-Kohout}},\ and\ \bibinfo {author} {\bibfnamefont {K.}~\bibnamefont {Young}},\ }\bibfield  {title} {\bibinfo {title} {Detecting and tracking drift in quantum information processors},\ }\href {https://doi.org/10.1038/s41467-020-19074-4} {\bibfield  {journal} {\bibinfo  {journal} {Nature Communications}\ }\textbf {\bibinfo {volume} {11}},\ \bibinfo {pages} {5396} (\bibinfo {year} {2020})}\BibitemShut {NoStop}%
\bibitem [{\citenamefont {Kim}\ \emph {et~al.}(2025)\citenamefont {Kim}, \citenamefont {Govia}, \citenamefont {Dane}, \citenamefont {van~den Berg}, \citenamefont {Zajac}, \citenamefont {Mitchell}, \citenamefont {Liu}, \citenamefont {Balakrishnan}, \citenamefont {Keefe}, \citenamefont {Stabile} \emph {et~al.}}]{kim2025error}%
  \BibitemOpen
  \bibfield  {author} {\bibinfo {author} {\bibfnamefont {Y.}~\bibnamefont {Kim}}, \bibinfo {author} {\bibfnamefont {L.~C.}\ \bibnamefont {Govia}}, \bibinfo {author} {\bibfnamefont {A.}~\bibnamefont {Dane}}, \bibinfo {author} {\bibfnamefont {E.}~\bibnamefont {van~den Berg}}, \bibinfo {author} {\bibfnamefont {D.~M.}\ \bibnamefont {Zajac}}, \bibinfo {author} {\bibfnamefont {B.}~\bibnamefont {Mitchell}}, \bibinfo {author} {\bibfnamefont {Y.}~\bibnamefont {Liu}}, \bibinfo {author} {\bibfnamefont {K.}~\bibnamefont {Balakrishnan}}, \bibinfo {author} {\bibfnamefont {G.}~\bibnamefont {Keefe}}, \bibinfo {author} {\bibfnamefont {A.}~\bibnamefont {Stabile}}, \emph {et~al.},\ }\bibfield  {title} {\bibinfo {title} {Error mitigation with stabilized noise in superconducting quantum processors},\ }\href {https://doi.org/https://doi.org/10.1038/s41467-025-62820-9} {\bibfield  {journal} {\bibinfo  {journal} {Nature Communications}\ }\textbf {\bibinfo {volume} {16}},\ \bibinfo {pages} {8439} (\bibinfo {year} {2025})}\BibitemShut
  {NoStop}%
\bibitem [{\citenamefont {Zhu}\ \emph {et~al.}(2022)\citenamefont {Zhu}, \citenamefont {Cian}, \citenamefont {Noel}, \citenamefont {Risinger}, \citenamefont {Biswas}, \citenamefont {Egan}, \citenamefont {Zhu}, \citenamefont {Green}, \citenamefont {Alderete}, \citenamefont {Nguyen} \emph {et~al.}}]{zhu2022cross}%
  \BibitemOpen
  \bibfield  {author} {\bibinfo {author} {\bibfnamefont {D.}~\bibnamefont {Zhu}}, \bibinfo {author} {\bibfnamefont {Z.-P.}\ \bibnamefont {Cian}}, \bibinfo {author} {\bibfnamefont {C.}~\bibnamefont {Noel}}, \bibinfo {author} {\bibfnamefont {A.}~\bibnamefont {Risinger}}, \bibinfo {author} {\bibfnamefont {D.}~\bibnamefont {Biswas}}, \bibinfo {author} {\bibfnamefont {L.}~\bibnamefont {Egan}}, \bibinfo {author} {\bibfnamefont {Y.}~\bibnamefont {Zhu}}, \bibinfo {author} {\bibfnamefont {A.~M.}\ \bibnamefont {Green}}, \bibinfo {author} {\bibfnamefont {C.~H.}\ \bibnamefont {Alderete}}, \bibinfo {author} {\bibfnamefont {N.~H.}\ \bibnamefont {Nguyen}}, \emph {et~al.},\ }\bibfield  {title} {\bibinfo {title} {Cross-platform comparison of arbitrary quantum states},\ }\href {https://doi.org/https://doi.org/10.1038/s41467-022-34279-5} {\bibfield  {journal} {\bibinfo  {journal} {Nature Communications}\ }\textbf {\bibinfo {volume} {13}},\ \bibinfo {pages} {6620} (\bibinfo {year} {2022})}\BibitemShut {NoStop}%
\bibitem [{\citenamefont {Zhou}\ \emph {et~al.}(2020)\citenamefont {Zhou}, \citenamefont {Stoudenmire},\ and\ \citenamefont {Waintal}}]{PhysRevX.10.041038}%
  \BibitemOpen
  \bibfield  {author} {\bibinfo {author} {\bibfnamefont {Y.}~\bibnamefont {Zhou}}, \bibinfo {author} {\bibfnamefont {E.~M.}\ \bibnamefont {Stoudenmire}},\ and\ \bibinfo {author} {\bibfnamefont {X.}~\bibnamefont {Waintal}},\ }\bibfield  {title} {\bibinfo {title} {What limits the simulation of quantum computers?},\ }\href {https://doi.org/10.1103/PhysRevX.10.041038} {\bibfield  {journal} {\bibinfo  {journal} {Physical Review X}\ }\textbf {\bibinfo {volume} {10}},\ \bibinfo {pages} {041038} (\bibinfo {year} {2020})}\BibitemShut {NoStop}%
\bibitem [{\citenamefont {Tindall}\ \emph {et~al.}(2024)\citenamefont {Tindall}, \citenamefont {Fishman}, \citenamefont {Stoudenmire},\ and\ \citenamefont {Sels}}]{PRXQuantum.5.010308}%
  \BibitemOpen
  \bibfield  {author} {\bibinfo {author} {\bibfnamefont {J.}~\bibnamefont {Tindall}}, \bibinfo {author} {\bibfnamefont {M.}~\bibnamefont {Fishman}}, \bibinfo {author} {\bibfnamefont {E.~M.}\ \bibnamefont {Stoudenmire}},\ and\ \bibinfo {author} {\bibfnamefont {D.}~\bibnamefont {Sels}},\ }\bibfield  {title} {\bibinfo {title} {{Efficient tensor network simulation of IBM's Eagle kicked Ising experiment}},\ }\href {https://doi.org/10.1103/PRXQuantum.5.010308} {\bibfield  {journal} {\bibinfo  {journal} {PRX Quantum}\ }\textbf {\bibinfo {volume} {5}},\ \bibinfo {pages} {010308} (\bibinfo {year} {2024})}\BibitemShut {NoStop}%
\bibitem [{\citenamefont {Begušić}\ \emph {et~al.}(2024)\citenamefont {Begušić}, \citenamefont {Gray},\ and\ \citenamefont {Chan}}]{doi:10.1126/sciadv.adk4321}%
  \BibitemOpen
  \bibfield  {author} {\bibinfo {author} {\bibfnamefont {T.}~\bibnamefont {Begušić}}, \bibinfo {author} {\bibfnamefont {J.}~\bibnamefont {Gray}},\ and\ \bibinfo {author} {\bibfnamefont {G.~K.-L.}\ \bibnamefont {Chan}},\ }\bibfield  {title} {\bibinfo {title} {Fast and converged classical simulations of evidence for the utility of quantum computing before fault tolerance},\ }\href {https://doi.org/10.1126/sciadv.adk4321} {\bibfield  {journal} {\bibinfo  {journal} {Science Advances}\ }\textbf {\bibinfo {volume} {10}},\ \bibinfo {pages} {eadk4321} (\bibinfo {year} {2024})}\BibitemShut {NoStop}%
\bibitem [{\citenamefont {Shao}\ \emph {et~al.}(2024)\citenamefont {Shao}, \citenamefont {Wei}, \citenamefont {Cheng},\ and\ \citenamefont {Liu}}]{PhysRevLett.133.120603}%
  \BibitemOpen
  \bibfield  {author} {\bibinfo {author} {\bibfnamefont {Y.}~\bibnamefont {Shao}}, \bibinfo {author} {\bibfnamefont {F.}~\bibnamefont {Wei}}, \bibinfo {author} {\bibfnamefont {S.}~\bibnamefont {Cheng}},\ and\ \bibinfo {author} {\bibfnamefont {Z.}~\bibnamefont {Liu}},\ }\bibfield  {title} {\bibinfo {title} {{Simulating noisy variational quantum algorithms: A polynomial approach}},\ }\href {https://doi.org/10.1103/PhysRevLett.133.120603} {\bibfield  {journal} {\bibinfo  {journal} {Physical Review Letters}\ }\textbf {\bibinfo {volume} {133}},\ \bibinfo {pages} {120603} (\bibinfo {year} {2024})}\BibitemShut {NoStop}%
\bibitem [{\citenamefont {Aharonov}\ \emph {et~al.}(2023)\citenamefont {Aharonov}, \citenamefont {Gao}, \citenamefont {Landau}, \citenamefont {Liu},\ and\ \citenamefont {Vazirani}}]{aharonov2023polynomial}%
  \BibitemOpen
  \bibfield  {author} {\bibinfo {author} {\bibfnamefont {D.}~\bibnamefont {Aharonov}}, \bibinfo {author} {\bibfnamefont {X.}~\bibnamefont {Gao}}, \bibinfo {author} {\bibfnamefont {Z.}~\bibnamefont {Landau}}, \bibinfo {author} {\bibfnamefont {Y.}~\bibnamefont {Liu}},\ and\ \bibinfo {author} {\bibfnamefont {U.}~\bibnamefont {Vazirani}},\ }\bibfield  {title} {\bibinfo {title} {A polynomial-time classical algorithm for noisy random circuit sampling},\ }in\ \href {https://doi.org/https://doi.org/10.1145/3564246.3585234} {\emph {\bibinfo {booktitle} {55th Annual ACM Symposium on Theory of Computing}}}\ (\bibinfo {year} {2023})\BibitemShut {NoStop}%
\bibitem [{\citenamefont {Schuster}\ \emph {et~al.}(2025)\citenamefont {Schuster}, \citenamefont {Yin}, \citenamefont {Gao},\ and\ \citenamefont {Yao}}]{schuster2025polynomial}%
  \BibitemOpen
  \bibfield  {author} {\bibinfo {author} {\bibfnamefont {T.}~\bibnamefont {Schuster}}, \bibinfo {author} {\bibfnamefont {C.}~\bibnamefont {Yin}}, \bibinfo {author} {\bibfnamefont {X.}~\bibnamefont {Gao}},\ and\ \bibinfo {author} {\bibfnamefont {N.~Y.}\ \bibnamefont {Yao}},\ }\bibfield  {title} {\bibinfo {title} {A polynomial-time classical algorithm for noisy quantum circuits},\ }\href {https://doi.org/10.1103/xct1-7kf2} {\bibfield  {journal} {\bibinfo  {journal} {Physical Review X}\ }\textbf {\bibinfo {volume} {15}},\ \bibinfo {pages} {041018} (\bibinfo {year} {2025})}\BibitemShut {NoStop}%
\bibitem [{\citenamefont {Rudolph}\ \emph {et~al.}(2026)\citenamefont {Rudolph}, \citenamefont {Jones}, \citenamefont {Teng}, \citenamefont {Angrisani},\ and\ \citenamefont {Holmes}}]{rudolph2025pauli}%
  \BibitemOpen
  \bibfield  {author} {\bibinfo {author} {\bibfnamefont {M.~S.}\ \bibnamefont {Rudolph}}, \bibinfo {author} {\bibfnamefont {T.}~\bibnamefont {Jones}}, \bibinfo {author} {\bibfnamefont {Y.}~\bibnamefont {Teng}}, \bibinfo {author} {\bibfnamefont {A.}~\bibnamefont {Angrisani}},\ and\ \bibinfo {author} {\bibfnamefont {Z.}~\bibnamefont {Holmes}},\ }\bibfield  {title} {\bibinfo {title} {{Pauli Propagation: A computational framework for simulating quantum systems}},\ }\href {https://doi.org/10.1103/6vd7-l9bn} {\bibfield  {journal} {\bibinfo  {journal} {PRX Quantum}\ }\textbf {\bibinfo {volume} {7}},\ \bibinfo {pages} {032001} (\bibinfo {year} {2026})}\BibitemShut {NoStop}%
\bibitem [{\citenamefont {Anschuetz}\ \emph {et~al.}(2023)\citenamefont {Anschuetz}, \citenamefont {Bauer}, \citenamefont {Kiani},\ and\ \citenamefont {Lloyd}}]{anschuetz2023efficient}%
  \BibitemOpen
  \bibfield  {author} {\bibinfo {author} {\bibfnamefont {E.~R.}\ \bibnamefont {Anschuetz}}, \bibinfo {author} {\bibfnamefont {A.}~\bibnamefont {Bauer}}, \bibinfo {author} {\bibfnamefont {B.~T.}\ \bibnamefont {Kiani}},\ and\ \bibinfo {author} {\bibfnamefont {S.}~\bibnamefont {Lloyd}},\ }\bibfield  {title} {\bibinfo {title} {Efficient classical algorithms for simulating symmetric quantum systems},\ }\href@noop {} {\bibfield  {journal} {\bibinfo  {journal} {Quantum}\ }\textbf {\bibinfo {volume} {7}},\ \bibinfo {pages} {1189} (\bibinfo {year} {2023})}\BibitemShut {NoStop}%
\bibitem [{\citenamefont {Camillo}\ \emph {et~al.}(2026)\citenamefont {Camillo}, \citenamefont {Peres}, \citenamefont {Heinrich},\ and\ \citenamefont {Bermejo-Vega}}]{nnzz-481j}%
  \BibitemOpen
  \bibfield  {author} {\bibinfo {author} {\bibfnamefont {G.}~\bibnamefont {Camillo}}, \bibinfo {author} {\bibfnamefont {F.~C.}\ \bibnamefont {Peres}}, \bibinfo {author} {\bibfnamefont {M.}~\bibnamefont {Heinrich}},\ and\ \bibinfo {author} {\bibfnamefont {J.}~\bibnamefont {Bermejo-Vega}},\ }\bibfield  {title} {\bibinfo {title} {{Symmetry-accelerated classical simulation of Clifford-dominated circuits}},\ }\href {https://doi.org/10.1103/nnzz-481j} {\bibfield  {journal} {\bibinfo  {journal} {PRX Quantum}\ }\textbf {\bibinfo {volume} {7}},\ \bibinfo {pages} {020356} (\bibinfo {year} {2026})}\BibitemShut {NoStop}%
\bibitem [{\citenamefont {Chang}\ \emph {et~al.}(2026)\citenamefont {Chang}, \citenamefont {Larocca},\ and\ \citenamefont {Cerezo}}]{chang2026permutationequivariant}%
  \BibitemOpen
  \bibfield  {author} {\bibinfo {author} {\bibfnamefont {S.~Y.}\ \bibnamefont {Chang}}, \bibinfo {author} {\bibfnamefont {M.}~\bibnamefont {Larocca}},\ and\ \bibinfo {author} {\bibfnamefont {M.}~\bibnamefont {Cerezo}},\ }\href@noop {} {\bibinfo {title} {Practical framework for simulating permutation-equivariant quantum circuits}} (\bibinfo {year} {2026}),\ \Eprint {https://arxiv.org/abs/2603.13072} {arXiv:2603.13072} \BibitemShut {NoStop}%
\bibitem [{\citenamefont {Aaronson}\ and\ \citenamefont {Gottesman}(2004)}]{PhysRevA.70.052328}%
  \BibitemOpen
  \bibfield  {author} {\bibinfo {author} {\bibfnamefont {S.}~\bibnamefont {Aaronson}}\ and\ \bibinfo {author} {\bibfnamefont {D.}~\bibnamefont {Gottesman}},\ }\bibfield  {title} {\bibinfo {title} {Improved simulation of stabilizer circuits},\ }\href {https://doi.org/10.1103/PhysRevA.70.052328} {\bibfield  {journal} {\bibinfo  {journal} {Physical Review A}\ }\textbf {\bibinfo {volume} {70}},\ \bibinfo {pages} {052328} (\bibinfo {year} {2004})}\BibitemShut {NoStop}%
\bibitem [{\citenamefont {Jozsa}\ and\ \citenamefont {Miyake}(2008)}]{jozsa2008matchgates}%
  \BibitemOpen
  \bibfield  {author} {\bibinfo {author} {\bibfnamefont {R.}~\bibnamefont {Jozsa}}\ and\ \bibinfo {author} {\bibfnamefont {A.}~\bibnamefont {Miyake}},\ }\bibfield  {title} {\bibinfo {title} {Matchgates and classical simulation of quantum circuits},\ }\href {https://doi.org/10.1098/rspa.2008.0189} {\bibfield  {journal} {\bibinfo  {journal} {Proceedings: Mathematical, Physical and Engineering Sciences}\ }\textbf {\bibinfo {volume} {464}},\ \bibinfo {pages} {3089} (\bibinfo {year} {2008})}\BibitemShut {NoStop}%
\bibitem [{\citenamefont {Goh}\ \emph {et~al.}(2025)\citenamefont {Goh}, \citenamefont {Larocca}, \citenamefont {Cincio}, \citenamefont {Cerezo},\ and\ \citenamefont {Sauvage}}]{3y65-f5w6}%
  \BibitemOpen
  \bibfield  {author} {\bibinfo {author} {\bibfnamefont {M.~L.}\ \bibnamefont {Goh}}, \bibinfo {author} {\bibfnamefont {M.}~\bibnamefont {Larocca}}, \bibinfo {author} {\bibfnamefont {L.}~\bibnamefont {Cincio}}, \bibinfo {author} {\bibfnamefont {M.}~\bibnamefont {Cerezo}},\ and\ \bibinfo {author} {\bibfnamefont {F.}~\bibnamefont {Sauvage}},\ }\bibfield  {title} {\bibinfo {title} {Lie-algebraic classical simulations for quantum computing},\ }\href {https://doi.org/10.1103/3y65-f5w6} {\bibfield  {journal} {\bibinfo  {journal} {Physical Review Research}\ }\textbf {\bibinfo {volume} {7}},\ \bibinfo {pages} {033266} (\bibinfo {year} {2025})}\BibitemShut {NoStop}%
\bibitem [{\citenamefont {Vidal}(2003)}]{PhysRevLett.91.147902}%
  \BibitemOpen
  \bibfield  {author} {\bibinfo {author} {\bibfnamefont {G.}~\bibnamefont {Vidal}},\ }\bibfield  {title} {\bibinfo {title} {Efficient classical simulation of slightly entangled quantum computations},\ }\href {https://doi.org/10.1103/PhysRevLett.91.147902} {\bibfield  {journal} {\bibinfo  {journal} {Physical Review Letters}\ }\textbf {\bibinfo {volume} {91}},\ \bibinfo {pages} {147902} (\bibinfo {year} {2003})}\BibitemShut {NoStop}%
\bibitem [{\citenamefont {Dowling}(2026)}]{dowling2026operatorentanglement}%
  \BibitemOpen
  \bibfield  {author} {\bibinfo {author} {\bibfnamefont {N.}~\bibnamefont {Dowling}},\ }\href@noop {} {\bibinfo {title} {Classical simulability from operator entanglement scaling}} (\bibinfo {year} {2026}),\ \Eprint {https://arxiv.org/abs/2603.05656} {arXiv:2603.05656} \BibitemShut {NoStop}%
\bibitem [{\citenamefont {Schuster}\ and\ \citenamefont {Yao}(2023)}]{PhysRevLett.131.160402}%
  \BibitemOpen
  \bibfield  {author} {\bibinfo {author} {\bibfnamefont {T.}~\bibnamefont {Schuster}}\ and\ \bibinfo {author} {\bibfnamefont {N.~Y.}\ \bibnamefont {Yao}},\ }\bibfield  {title} {\bibinfo {title} {Operator growth in open quantum systems},\ }\href {https://doi.org/10.1103/PhysRevLett.131.160402} {\bibfield  {journal} {\bibinfo  {journal} {Physical Review Letters}\ }\textbf {\bibinfo {volume} {131}},\ \bibinfo {pages} {160402} (\bibinfo {year} {2023})}\BibitemShut {NoStop}%
\bibitem [{\citenamefont {Markov}\ and\ \citenamefont {Shi}(2008)}]{doi:10.1137/050644756}%
  \BibitemOpen
  \bibfield  {author} {\bibinfo {author} {\bibfnamefont {I.~L.}\ \bibnamefont {Markov}}\ and\ \bibinfo {author} {\bibfnamefont {Y.}~\bibnamefont {Shi}},\ }\bibfield  {title} {\bibinfo {title} {Simulating quantum computation by contracting tensor networks},\ }\href {https://doi.org/10.1137/050644756} {\bibfield  {journal} {\bibinfo  {journal} {SIAM Journal on Computing}\ }\textbf {\bibinfo {volume} {38}},\ \bibinfo {pages} {963} (\bibinfo {year} {2008})}\BibitemShut {NoStop}%
\bibitem [{\citenamefont {Hartnett}\ \emph {et~al.}(2026)\citenamefont {Hartnett}, \citenamefont {Najafi}, \citenamefont {Khindanov}, \citenamefont {Liao}, \citenamefont {Schutzman}, \citenamefont {Hush}, \citenamefont {Biercuk},\ and\ \citenamefont {Baum}}]{hartnett2026fastaccurate}%
  \BibitemOpen
  \bibfield  {author} {\bibinfo {author} {\bibfnamefont {G.~S.}\ \bibnamefont {Hartnett}}, \bibinfo {author} {\bibfnamefont {K.~S.}\ \bibnamefont {Najafi}}, \bibinfo {author} {\bibfnamefont {A.}~\bibnamefont {Khindanov}}, \bibinfo {author} {\bibfnamefont {H.}~\bibnamefont {Liao}}, \bibinfo {author} {\bibfnamefont {M.}~\bibnamefont {Schutzman}}, \bibinfo {author} {\bibfnamefont {M.~R.}\ \bibnamefont {Hush}}, \bibinfo {author} {\bibfnamefont {M.~J.}\ \bibnamefont {Biercuk}},\ and\ \bibinfo {author} {\bibfnamefont {Y.}~\bibnamefont {Baum}},\ }\href@noop {} {\bibinfo {title} {{Fast, accurate, high-resolution simulation of large-scale Fermi-Hubbard models on a digital quantum processor}}} (\bibinfo {year} {2026}),\ \Eprint {https://arxiv.org/abs/2605.04025} {arXiv:2605.04025} \BibitemShut {NoStop}%
\bibitem [{\citenamefont {Deger}\ \emph {et~al.}(2026)\citenamefont {Deger}, \citenamefont {Koutsioumpas}, \citenamefont {Webster}, \citenamefont {Sayginel}, \citenamefont {Roffe},\ and\ \citenamefont {Browne}}]{deger2026efficiently}%
  \BibitemOpen
  \bibfield  {author} {\bibinfo {author} {\bibfnamefont {A.}~\bibnamefont {Deger}}, \bibinfo {author} {\bibfnamefont {S.}~\bibnamefont {Koutsioumpas}}, \bibinfo {author} {\bibfnamefont {M.}~\bibnamefont {Webster}}, \bibinfo {author} {\bibfnamefont {H.}~\bibnamefont {Sayginel}}, \bibinfo {author} {\bibfnamefont {J.}~\bibnamefont {Roffe}},\ and\ \bibinfo {author} {\bibfnamefont {D.~E.}\ \bibnamefont {Browne}},\ }\href@noop {} {\bibinfo {title} {{Efficiently simulable quantum circuits with large entanglement, magic, and non-Gaussianity via code-compiled tensor networks}}} (\bibinfo {year} {2026}),\ \Eprint {https://arxiv.org/abs/2607.08396} {arXiv:2607.08396} \BibitemShut {NoStop}%
\bibitem [{\citenamefont {AI}\ and\ \citenamefont {Collaborators}(2025)}]{acharya2025willow}%
  \BibitemOpen
  \bibfield  {author} {\bibinfo {author} {\bibfnamefont {G.~Q.}\ \bibnamefont {AI}}\ and\ \bibinfo {author} {\bibnamefont {Collaborators}},\ }\bibfield  {title} {\bibinfo {title} {Quantum error correction below the surface code threshold},\ }\href {https://doi.org/https://doi.org/10.1038/s41586-024-08449-y} {\bibfield  {journal} {\bibinfo  {journal} {Nature}\ }\textbf {\bibinfo {volume} {638}},\ \bibinfo {pages} {920} (\bibinfo {year} {2025})}\BibitemShut {NoStop}%
\bibitem [{\citenamefont {McKay}\ \emph {et~al.}(2023)\citenamefont {McKay}, \citenamefont {Hincks}, \citenamefont {Pritchett}, \citenamefont {Carroll}, \citenamefont {Govia},\ and\ \citenamefont {Merkel}}]{mckay2023benchmarking}%
  \BibitemOpen
  \bibfield  {author} {\bibinfo {author} {\bibfnamefont {D.~C.}\ \bibnamefont {McKay}}, \bibinfo {author} {\bibfnamefont {I.}~\bibnamefont {Hincks}}, \bibinfo {author} {\bibfnamefont {E.~J.}\ \bibnamefont {Pritchett}}, \bibinfo {author} {\bibfnamefont {M.}~\bibnamefont {Carroll}}, \bibinfo {author} {\bibfnamefont {L.~C.~G.}\ \bibnamefont {Govia}},\ and\ \bibinfo {author} {\bibfnamefont {S.~T.}\ \bibnamefont {Merkel}},\ }\href@noop {} {\bibinfo {title} {Benchmarking quantum processor performance at scale}} (\bibinfo {year} {2023}),\ \Eprint {https://arxiv.org/abs/2311.05933} {arXiv:2311.05933} \BibitemShut {NoStop}%
\bibitem [{\citenamefont {Wurtz}\ \emph {et~al.}(2023)\citenamefont {Wurtz}, \citenamefont {Bylinskii}, \citenamefont {Braverman}, \citenamefont {Amato-Grill}, \citenamefont {Cantu}, \citenamefont {Huber}, \citenamefont {Lukin}, \citenamefont {Liu}, \citenamefont {Weinberg}, \citenamefont {Long}, \citenamefont {Wang}, \citenamefont {Gemelke},\ and\ \citenamefont {Keesling}}]{wurtz2023aquilaqueras256}%
  \BibitemOpen
  \bibfield  {author} {\bibinfo {author} {\bibfnamefont {J.}~\bibnamefont {Wurtz}}, \bibinfo {author} {\bibfnamefont {A.}~\bibnamefont {Bylinskii}}, \bibinfo {author} {\bibfnamefont {B.}~\bibnamefont {Braverman}}, \bibinfo {author} {\bibfnamefont {J.}~\bibnamefont {Amato-Grill}}, \bibinfo {author} {\bibfnamefont {S.~H.}\ \bibnamefont {Cantu}}, \bibinfo {author} {\bibfnamefont {F.}~\bibnamefont {Huber}}, \bibinfo {author} {\bibfnamefont {A.}~\bibnamefont {Lukin}}, \bibinfo {author} {\bibfnamefont {F.}~\bibnamefont {Liu}}, \bibinfo {author} {\bibfnamefont {P.}~\bibnamefont {Weinberg}}, \bibinfo {author} {\bibfnamefont {J.}~\bibnamefont {Long}}, \bibinfo {author} {\bibfnamefont {S.-T.}\ \bibnamefont {Wang}}, \bibinfo {author} {\bibfnamefont {N.}~\bibnamefont {Gemelke}},\ and\ \bibinfo {author} {\bibfnamefont {A.}~\bibnamefont {Keesling}},\ }\href@noop {} {\bibinfo {title} {{Aquila: QuEra's 256-qubit neutral-atom quantum computer}}} (\bibinfo {year} {2023}),\ \Eprint {https://arxiv.org/abs/2306.11727}
  {arXiv:2306.11727} \BibitemShut {NoStop}%
\bibitem [{\citenamefont {Bayraktar}\ \emph {et~al.}(2023)\citenamefont {Bayraktar}, \citenamefont {Charara}, \citenamefont {Clark}, \citenamefont {Cohen}, \citenamefont {Costa}, \citenamefont {Fang}, \citenamefont {Gao}, \citenamefont {Guan}, \citenamefont {Gunnels}, \citenamefont {Haidar} \emph {et~al.}}]{bayraktar2023cuquantum}%
  \BibitemOpen
  \bibfield  {author} {\bibinfo {author} {\bibfnamefont {H.}~\bibnamefont {Bayraktar}}, \bibinfo {author} {\bibfnamefont {A.}~\bibnamefont {Charara}}, \bibinfo {author} {\bibfnamefont {D.}~\bibnamefont {Clark}}, \bibinfo {author} {\bibfnamefont {S.}~\bibnamefont {Cohen}}, \bibinfo {author} {\bibfnamefont {T.}~\bibnamefont {Costa}}, \bibinfo {author} {\bibfnamefont {Y.-L.~L.}\ \bibnamefont {Fang}}, \bibinfo {author} {\bibfnamefont {Y.}~\bibnamefont {Gao}}, \bibinfo {author} {\bibfnamefont {J.}~\bibnamefont {Guan}}, \bibinfo {author} {\bibfnamefont {J.}~\bibnamefont {Gunnels}}, \bibinfo {author} {\bibfnamefont {A.}~\bibnamefont {Haidar}}, \emph {et~al.},\ }\bibfield  {title} {\bibinfo {title} {{cuQuantum SDK: A high-performance library for accelerating quantum science}},\ }in\ \href@noop {} {\emph {\bibinfo {booktitle} {IEEE International Conference on Quantum Computing and Engineering}}},\ Vol.~\bibinfo {volume} {1}\ (\bibinfo {year} {2023})\ pp.\ \bibinfo {pages} {1050--1061}\BibitemShut {NoStop}%
\bibitem [{\citenamefont {Cicero}\ \emph {et~al.}(2025)\citenamefont {Cicero}, \citenamefont {Maleki}, \citenamefont {Azhar}, \citenamefont {Kockum},\ and\ \citenamefont {Trancoso}}]{10.1145/3762672}%
  \BibitemOpen
  \bibfield  {author} {\bibinfo {author} {\bibfnamefont {A.}~\bibnamefont {Cicero}}, \bibinfo {author} {\bibfnamefont {M.~A.}\ \bibnamefont {Maleki}}, \bibinfo {author} {\bibfnamefont {M.~W.}\ \bibnamefont {Azhar}}, \bibinfo {author} {\bibfnamefont {A.~F.}\ \bibnamefont {Kockum}},\ and\ \bibinfo {author} {\bibfnamefont {P.}~\bibnamefont {Trancoso}},\ }\bibfield  {title} {\bibinfo {title} {{Simulation of quantum computers: Review and acceleration opportunities}},\ }\href {https://doi.org/10.1145/3762672} {\bibfield  {journal} {\bibinfo  {journal} {ACM Transactions on Quantum Computing}\ }\textbf {\bibinfo {volume} {7}} (\bibinfo {year} {2025})}\BibitemShut {NoStop}%
\bibitem [{\citenamefont {Morningstar}\ \emph {et~al.}(2022)\citenamefont {Morningstar}, \citenamefont {Hauru}, \citenamefont {Beall}, \citenamefont {Ganahl}, \citenamefont {Lewis}, \citenamefont {Khemani},\ and\ \citenamefont {Vidal}}]{PRXQuantum.3.020331}%
  \BibitemOpen
  \bibfield  {author} {\bibinfo {author} {\bibfnamefont {A.}~\bibnamefont {Morningstar}}, \bibinfo {author} {\bibfnamefont {M.}~\bibnamefont {Hauru}}, \bibinfo {author} {\bibfnamefont {J.}~\bibnamefont {Beall}}, \bibinfo {author} {\bibfnamefont {M.}~\bibnamefont {Ganahl}}, \bibinfo {author} {\bibfnamefont {A.~G.}\ \bibnamefont {Lewis}}, \bibinfo {author} {\bibfnamefont {V.}~\bibnamefont {Khemani}},\ and\ \bibinfo {author} {\bibfnamefont {G.}~\bibnamefont {Vidal}},\ }\bibfield  {title} {\bibinfo {title} {{Simulation of quantum many-body dynamics with tensor processing units: Floquet prethermalization}},\ }\href {https://doi.org/10.1103/PRXQuantum.3.020331} {\bibfield  {journal} {\bibinfo  {journal} {PRX Quantum}\ }\textbf {\bibinfo {volume} {3}},\ \bibinfo {pages} {020331} (\bibinfo {year} {2022})}\BibitemShut {NoStop}%
\bibitem [{\citenamefont {Chen}\ \emph {et~al.}(2023)\citenamefont {Chen}, \citenamefont {Liu}, \citenamefont {Otten}, \citenamefont {Seif}, \citenamefont {Fefferman},\ and\ \citenamefont {Jiang}}]{chen2023learnability}%
  \BibitemOpen
  \bibfield  {author} {\bibinfo {author} {\bibfnamefont {S.}~\bibnamefont {Chen}}, \bibinfo {author} {\bibfnamefont {Y.}~\bibnamefont {Liu}}, \bibinfo {author} {\bibfnamefont {M.}~\bibnamefont {Otten}}, \bibinfo {author} {\bibfnamefont {A.}~\bibnamefont {Seif}}, \bibinfo {author} {\bibfnamefont {B.}~\bibnamefont {Fefferman}},\ and\ \bibinfo {author} {\bibfnamefont {L.}~\bibnamefont {Jiang}},\ }\bibfield  {title} {\bibinfo {title} {{The learnability of Pauli noise}},\ }\href {https://doi.org/https://doi.org/10.1038/s41467-022-35759-4} {\bibfield  {journal} {\bibinfo  {journal} {Nature Communications}\ }\textbf {\bibinfo {volume} {14}},\ \bibinfo {pages} {52} (\bibinfo {year} {2023})}\BibitemShut {NoStop}%
\bibitem [{\citenamefont {Chen}\ \emph {et~al.}(2026{\natexlab{a}})\citenamefont {Chen}, \citenamefont {Chen}, \citenamefont {Fischer}, \citenamefont {Eddins}, \citenamefont {Govia}, \citenamefont {Mitchell}, \citenamefont {He}, \citenamefont {Kim}, \citenamefont {Jiang},\ and\ \citenamefont {Seif}}]{69wc-gzl6}%
  \BibitemOpen
  \bibfield  {author} {\bibinfo {author} {\bibfnamefont {E.~H.}\ \bibnamefont {Chen}}, \bibinfo {author} {\bibfnamefont {S.}~\bibnamefont {Chen}}, \bibinfo {author} {\bibfnamefont {L.~E.}\ \bibnamefont {Fischer}}, \bibinfo {author} {\bibfnamefont {A.}~\bibnamefont {Eddins}}, \bibinfo {author} {\bibfnamefont {L.~C.~G.}\ \bibnamefont {Govia}}, \bibinfo {author} {\bibfnamefont {B.}~\bibnamefont {Mitchell}}, \bibinfo {author} {\bibfnamefont {A.}~\bibnamefont {He}}, \bibinfo {author} {\bibfnamefont {Y.}~\bibnamefont {Kim}}, \bibinfo {author} {\bibfnamefont {L.}~\bibnamefont {Jiang}},\ and\ \bibinfo {author} {\bibfnamefont {A.}~\bibnamefont {Seif}},\ }\bibfield  {title} {\bibinfo {title} {{Disambiguating Pauli noise in quantum computers}},\ }\href {https://doi.org/10.1103/69wc-gzl6} {\bibfield  {journal} {\bibinfo  {journal} {PRX Quantum}\ }\textbf {\bibinfo {volume} {7}},\ \bibinfo {pages} {033045} (\bibinfo {year} {2026}{\natexlab{a}})}\BibitemShut {NoStop}%
\bibitem [{\citenamefont {Van Den~Berg}\ \emph {et~al.}(2023)\citenamefont {Van Den~Berg}, \citenamefont {Minev}, \citenamefont {Kandala},\ and\ \citenamefont {Temme}}]{van2023probabilistic}%
  \BibitemOpen
  \bibfield  {author} {\bibinfo {author} {\bibfnamefont {E.}~\bibnamefont {Van Den~Berg}}, \bibinfo {author} {\bibfnamefont {Z.~K.}\ \bibnamefont {Minev}}, \bibinfo {author} {\bibfnamefont {A.}~\bibnamefont {Kandala}},\ and\ \bibinfo {author} {\bibfnamefont {K.}~\bibnamefont {Temme}},\ }\bibfield  {title} {\bibinfo {title} {{Probabilistic error cancellation with sparse Pauli--Lindblad models on noisy quantum processors}},\ }\href {https://doi.org/10.1038/s41567-023-02042-2} {\bibfield  {journal} {\bibinfo  {journal} {Nature Physics}\ }\textbf {\bibinfo {volume} {19}},\ \bibinfo {pages} {1116} (\bibinfo {year} {2023})}\BibitemShut {NoStop}%
\bibitem [{\citenamefont {Temme}\ \emph {et~al.}(2017)\citenamefont {Temme}, \citenamefont {Bravyi},\ and\ \citenamefont {Gambetta}}]{temme2017error}%
  \BibitemOpen
  \bibfield  {author} {\bibinfo {author} {\bibfnamefont {K.}~\bibnamefont {Temme}}, \bibinfo {author} {\bibfnamefont {S.}~\bibnamefont {Bravyi}},\ and\ \bibinfo {author} {\bibfnamefont {J.~M.}\ \bibnamefont {Gambetta}},\ }\bibfield  {title} {\bibinfo {title} {Error mitigation for short-depth quantum circuits},\ }\href {https://doi.org/https://doi.org/10.1103/PhysRevLett.119.180509} {\bibfield  {journal} {\bibinfo  {journal} {Physical Review Letters}\ }\textbf {\bibinfo {volume} {119}},\ \bibinfo {pages} {180509} (\bibinfo {year} {2017})}\BibitemShut {NoStop}%
\bibitem [{\citenamefont {Li}\ and\ \citenamefont {Benjamin}(2017)}]{li2017efficient}%
  \BibitemOpen
  \bibfield  {author} {\bibinfo {author} {\bibfnamefont {Y.}~\bibnamefont {Li}}\ and\ \bibinfo {author} {\bibfnamefont {S.~C.}\ \bibnamefont {Benjamin}},\ }\bibfield  {title} {\bibinfo {title} {Efficient variational quantum simulator incorporating active error minimization},\ }\href {https://doi.org/https://doi.org/10.1103/PhysRevX.7.021050} {\bibfield  {journal} {\bibinfo  {journal} {Physical Review X}\ }\textbf {\bibinfo {volume} {7}},\ \bibinfo {pages} {021050} (\bibinfo {year} {2017})}\BibitemShut {NoStop}%
\bibitem [{\citenamefont {Kim}\ \emph {et~al.}(2023)\citenamefont {Kim}, \citenamefont {Eddins}, \citenamefont {Anand}, \citenamefont {Wei}, \citenamefont {Van Den~Berg}, \citenamefont {Rosenblatt}, \citenamefont {Nayfeh}, \citenamefont {Wu}, \citenamefont {Zaletel}, \citenamefont {Temme} \emph {et~al.}}]{kim2023evidence}%
  \BibitemOpen
  \bibfield  {author} {\bibinfo {author} {\bibfnamefont {Y.}~\bibnamefont {Kim}}, \bibinfo {author} {\bibfnamefont {A.}~\bibnamefont {Eddins}}, \bibinfo {author} {\bibfnamefont {S.}~\bibnamefont {Anand}}, \bibinfo {author} {\bibfnamefont {K.~X.}\ \bibnamefont {Wei}}, \bibinfo {author} {\bibfnamefont {E.}~\bibnamefont {Van Den~Berg}}, \bibinfo {author} {\bibfnamefont {S.}~\bibnamefont {Rosenblatt}}, \bibinfo {author} {\bibfnamefont {H.}~\bibnamefont {Nayfeh}}, \bibinfo {author} {\bibfnamefont {Y.}~\bibnamefont {Wu}}, \bibinfo {author} {\bibfnamefont {M.}~\bibnamefont {Zaletel}}, \bibinfo {author} {\bibfnamefont {K.}~\bibnamefont {Temme}}, \emph {et~al.},\ }\bibfield  {title} {\bibinfo {title} {Evidence for the utility of quantum computing before fault tolerance},\ }\href {https://doi.org/https://doi.org/10.1038/s41586-023-06096-3} {\bibfield  {journal} {\bibinfo  {journal} {Nature}\ }\textbf {\bibinfo {volume} {618}},\ \bibinfo {pages} {500} (\bibinfo {year} {2023})}\BibitemShut {NoStop}%
\bibitem [{\citenamefont {Liao}\ \emph {et~al.}(2024)\citenamefont {Liao}, \citenamefont {Wang}, \citenamefont {Sitdikov}, \citenamefont {Salcedo}, \citenamefont {Seif},\ and\ \citenamefont {Minev}}]{liao2024machine}%
  \BibitemOpen
  \bibfield  {author} {\bibinfo {author} {\bibfnamefont {H.}~\bibnamefont {Liao}}, \bibinfo {author} {\bibfnamefont {D.~S.}\ \bibnamefont {Wang}}, \bibinfo {author} {\bibfnamefont {I.}~\bibnamefont {Sitdikov}}, \bibinfo {author} {\bibfnamefont {C.}~\bibnamefont {Salcedo}}, \bibinfo {author} {\bibfnamefont {A.}~\bibnamefont {Seif}},\ and\ \bibinfo {author} {\bibfnamefont {Z.~K.}\ \bibnamefont {Minev}},\ }\bibfield  {title} {\bibinfo {title} {Machine learning for practical quantum error mitigation},\ }\href {https://doi.org/https://doi.org/10.1038/s42256-024-00927-2} {\bibfield  {journal} {\bibinfo  {journal} {Nature Machine Intelligence}\ }\textbf {\bibinfo {volume} {6}},\ \bibinfo {pages} {1478} (\bibinfo {year} {2024})}\BibitemShut {NoStop}%
\bibitem [{\citenamefont {Czarnik}\ \emph {et~al.}(2021)\citenamefont {Czarnik}, \citenamefont {Arrasmith}, \citenamefont {Coles},\ and\ \citenamefont {Cincio}}]{czarnik2021error}%
  \BibitemOpen
  \bibfield  {author} {\bibinfo {author} {\bibfnamefont {P.}~\bibnamefont {Czarnik}}, \bibinfo {author} {\bibfnamefont {A.}~\bibnamefont {Arrasmith}}, \bibinfo {author} {\bibfnamefont {P.~J.}\ \bibnamefont {Coles}},\ and\ \bibinfo {author} {\bibfnamefont {L.}~\bibnamefont {Cincio}},\ }\bibfield  {title} {\bibinfo {title} {{Error mitigation with Clifford quantum-circuit data}},\ }\href {https://doi.org/https://doi.org/10.22331/q-2021-11-26-592} {\bibfield  {journal} {\bibinfo  {journal} {Quantum}\ }\textbf {\bibinfo {volume} {5}},\ \bibinfo {pages} {592} (\bibinfo {year} {2021})}\BibitemShut {NoStop}%
\bibitem [{\citenamefont {Liu}\ \emph {et~al.}(2020)\citenamefont {Liu}, \citenamefont {Wang}, \citenamefont {Xue}, \citenamefont {Huang}, \citenamefont {Fu}, \citenamefont {Qiang}, \citenamefont {Xu}, \citenamefont {Huang}, \citenamefont {Deng}, \citenamefont {Guo}, \citenamefont {Yang},\ and\ \citenamefont {Wu}}]{PhysRevA.101.052316}%
  \BibitemOpen
  \bibfield  {author} {\bibinfo {author} {\bibfnamefont {Y.}~\bibnamefont {Liu}}, \bibinfo {author} {\bibfnamefont {D.}~\bibnamefont {Wang}}, \bibinfo {author} {\bibfnamefont {S.}~\bibnamefont {Xue}}, \bibinfo {author} {\bibfnamefont {A.}~\bibnamefont {Huang}}, \bibinfo {author} {\bibfnamefont {X.}~\bibnamefont {Fu}}, \bibinfo {author} {\bibfnamefont {X.}~\bibnamefont {Qiang}}, \bibinfo {author} {\bibfnamefont {P.}~\bibnamefont {Xu}}, \bibinfo {author} {\bibfnamefont {H.-L.}\ \bibnamefont {Huang}}, \bibinfo {author} {\bibfnamefont {M.}~\bibnamefont {Deng}}, \bibinfo {author} {\bibfnamefont {C.}~\bibnamefont {Guo}}, \bibinfo {author} {\bibfnamefont {X.}~\bibnamefont {Yang}},\ and\ \bibinfo {author} {\bibfnamefont {J.}~\bibnamefont {Wu}},\ }\bibfield  {title} {\bibinfo {title} {Variational quantum circuits for quantum state tomography},\ }\href {https://doi.org/10.1103/PhysRevA.101.052316} {\bibfield  {journal} {\bibinfo  {journal} {Physical Review A}\ }\textbf {\bibinfo {volume} {101}},\ \bibinfo {pages}
  {052316} (\bibinfo {year} {2020})}\BibitemShut {NoStop}%
\bibitem [{\citenamefont {Belliardo}\ \emph {et~al.}(2026)\citenamefont {Belliardo}, \citenamefont {Gauger}, \citenamefont {Abobeih}, \citenamefont {Taminiau}, \citenamefont {Altmann},\ and\ \citenamefont {Bonato}}]{belliardo2025multi}%
  \BibitemOpen
  \bibfield  {author} {\bibinfo {author} {\bibfnamefont {F.}~\bibnamefont {Belliardo}}, \bibinfo {author} {\bibfnamefont {E.~M.}\ \bibnamefont {Gauger}}, \bibinfo {author} {\bibfnamefont {M.~H.}\ \bibnamefont {Abobeih}}, \bibinfo {author} {\bibfnamefont {T.~H.}\ \bibnamefont {Taminiau}}, \bibinfo {author} {\bibfnamefont {Y.}~\bibnamefont {Altmann}},\ and\ \bibinfo {author} {\bibfnamefont {C.}~\bibnamefont {Bonato}},\ }\bibfield  {title} {\bibinfo {title} {{Multidimensional quantum estimation and model learning framework based on variational Bayesian inference}},\ }\href {https://doi.org/10.1103/6yhz-9vr6} {\bibfield  {journal} {\bibinfo  {journal} {PRX Quantum}\ }\textbf {\bibinfo {volume} {7}},\ \bibinfo {pages} {020360} (\bibinfo {year} {2026})}\BibitemShut {NoStop}%
\bibitem [{\citenamefont {Moss}\ \emph {et~al.}(2026)\citenamefont {Moss}, \citenamefont {Muhle}, \citenamefont {Drews}, \citenamefont {Macke},\ and\ \citenamefont {Schr{\"o}der}}]{moss2026fnope}%
  \BibitemOpen
  \bibfield  {author} {\bibinfo {author} {\bibfnamefont {G.}~\bibnamefont {Moss}}, \bibinfo {author} {\bibfnamefont {L.}~\bibnamefont {Muhle}}, \bibinfo {author} {\bibfnamefont {R.}~\bibnamefont {Drews}}, \bibinfo {author} {\bibfnamefont {J.~H.}\ \bibnamefont {Macke}},\ and\ \bibinfo {author} {\bibfnamefont {C.}~\bibnamefont {Schr{\"o}der}},\ }\bibfield  {title} {\bibinfo {title} {{FNOPE: Simulation-based inference on function spaces with Fourier Neural Operators}},\ }in\ \href@noop {} {\emph {\bibinfo {booktitle} {Advances in Neural Information Processing Systems}}},\ Vol.~\bibinfo {volume} {38}\ (\bibinfo {year} {2026})\BibitemShut {NoStop}%
\bibitem [{\citenamefont {White}\ \emph {et~al.}(2022)\citenamefont {White}, \citenamefont {Pollock}, \citenamefont {Hollenberg}, \citenamefont {Modi},\ and\ \citenamefont {Hill}}]{WhiteNonMarkovian}%
  \BibitemOpen
  \bibfield  {author} {\bibinfo {author} {\bibfnamefont {G.}~\bibnamefont {White}}, \bibinfo {author} {\bibfnamefont {F.}~\bibnamefont {Pollock}}, \bibinfo {author} {\bibfnamefont {L.}~\bibnamefont {Hollenberg}}, \bibinfo {author} {\bibfnamefont {K.}~\bibnamefont {Modi}},\ and\ \bibinfo {author} {\bibfnamefont {C.}~\bibnamefont {Hill}},\ }\bibfield  {title} {\bibinfo {title} {{Non-Markovian quantum process tomography}},\ }\href {https://doi.org/10.1103/PRXQuantum.3.020344} {\bibfield  {journal} {\bibinfo  {journal} {PRX Quantum}\ }\textbf {\bibinfo {volume} {3}},\ \bibinfo {pages} {020344} (\bibinfo {year} {2022})}\BibitemShut {NoStop}%
\bibitem [{\citenamefont {Pollock}\ \emph {et~al.}(2018)\citenamefont {Pollock}, \citenamefont {Rodr\'{\i}guez-Rosario}, \citenamefont {Frauenheim}, \citenamefont {Paternostro},\ and\ \citenamefont {Modi}}]{PhysRevA.97.012127}%
  \BibitemOpen
  \bibfield  {author} {\bibinfo {author} {\bibfnamefont {F.~A.}\ \bibnamefont {Pollock}}, \bibinfo {author} {\bibfnamefont {C.}~\bibnamefont {Rodr\'{\i}guez-Rosario}}, \bibinfo {author} {\bibfnamefont {T.}~\bibnamefont {Frauenheim}}, \bibinfo {author} {\bibfnamefont {M.}~\bibnamefont {Paternostro}},\ and\ \bibinfo {author} {\bibfnamefont {K.}~\bibnamefont {Modi}},\ }\bibfield  {title} {\bibinfo {title} {{Non-Markovian quantum processes: Complete framework and efficient characterization}},\ }\href {https://doi.org/10.1103/PhysRevA.97.012127} {\bibfield  {journal} {\bibinfo  {journal} {Physical Review A}\ }\textbf {\bibinfo {volume} {97}},\ \bibinfo {pages} {012127} (\bibinfo {year} {2018})}\BibitemShut {NoStop}%
\bibitem [{\citenamefont {Keeling}\ \emph {et~al.}(2026)\citenamefont {Keeling}, \citenamefont {Stoudenmire}, \citenamefont {Ba\~nuls},\ and\ \citenamefont {Reichman}}]{1ncg-11hz}%
  \BibitemOpen
  \bibfield  {author} {\bibinfo {author} {\bibfnamefont {J.}~\bibnamefont {Keeling}}, \bibinfo {author} {\bibfnamefont {E.~M.}\ \bibnamefont {Stoudenmire}}, \bibinfo {author} {\bibfnamefont {M.-C.}\ \bibnamefont {Ba\~nuls}},\ and\ \bibinfo {author} {\bibfnamefont {D.~R.}\ \bibnamefont {Reichman}},\ }\bibfield  {title} {\bibinfo {title} {{Process tensor approaches to non-Markovian quantum dynamics}},\ }\href {https://doi.org/10.1103/1ncg-11hz} {\bibfield  {journal} {\bibinfo  {journal} {Physical Review X}\ }\textbf {\bibinfo {volume} {16}},\ \bibinfo {pages} {020502} (\bibinfo {year} {2026})}\BibitemShut {NoStop}%
\bibitem [{\citenamefont {Cimini}\ \emph {et~al.}(2019)\citenamefont {Cimini}, \citenamefont {Gianani}, \citenamefont {Spagnolo}, \citenamefont {Leccese}, \citenamefont {Sciarrino},\ and\ \citenamefont {Barbieri}}]{PhysRevLett.123.230502}%
  \BibitemOpen
  \bibfield  {author} {\bibinfo {author} {\bibfnamefont {V.}~\bibnamefont {Cimini}}, \bibinfo {author} {\bibfnamefont {I.}~\bibnamefont {Gianani}}, \bibinfo {author} {\bibfnamefont {N.}~\bibnamefont {Spagnolo}}, \bibinfo {author} {\bibfnamefont {F.}~\bibnamefont {Leccese}}, \bibinfo {author} {\bibfnamefont {F.}~\bibnamefont {Sciarrino}},\ and\ \bibinfo {author} {\bibfnamefont {M.}~\bibnamefont {Barbieri}},\ }\bibfield  {title} {\bibinfo {title} {Calibration of quantum sensors by neural networks},\ }\href {https://doi.org/10.1103/PhysRevLett.123.230502} {\bibfield  {journal} {\bibinfo  {journal} {Physical Review Letters}\ }\textbf {\bibinfo {volume} {123}},\ \bibinfo {pages} {230502} (\bibinfo {year} {2019})}\BibitemShut {NoStop}%
\bibitem [{\citenamefont {Cimini}\ \emph {et~al.}(2021)\citenamefont {Cimini}, \citenamefont {Polino}, \citenamefont {Valeri}, \citenamefont {Gianani}, \citenamefont {Spagnolo}, \citenamefont {Corrielli}, \citenamefont {Crespi}, \citenamefont {Osellame}, \citenamefont {Barbieri},\ and\ \citenamefont {Sciarrino}}]{PhysRevApplied.15.044003}%
  \BibitemOpen
  \bibfield  {author} {\bibinfo {author} {\bibfnamefont {V.}~\bibnamefont {Cimini}}, \bibinfo {author} {\bibfnamefont {E.}~\bibnamefont {Polino}}, \bibinfo {author} {\bibfnamefont {M.}~\bibnamefont {Valeri}}, \bibinfo {author} {\bibfnamefont {I.}~\bibnamefont {Gianani}}, \bibinfo {author} {\bibfnamefont {N.}~\bibnamefont {Spagnolo}}, \bibinfo {author} {\bibfnamefont {G.}~\bibnamefont {Corrielli}}, \bibinfo {author} {\bibfnamefont {A.}~\bibnamefont {Crespi}}, \bibinfo {author} {\bibfnamefont {R.}~\bibnamefont {Osellame}}, \bibinfo {author} {\bibfnamefont {M.}~\bibnamefont {Barbieri}},\ and\ \bibinfo {author} {\bibfnamefont {F.}~\bibnamefont {Sciarrino}},\ }\bibfield  {title} {\bibinfo {title} {Calibration of multiparameter sensors via machine learning at the single-photon level},\ }\href {https://doi.org/10.1103/PhysRevApplied.15.044003} {\bibfield  {journal} {\bibinfo  {journal} {Physical Review Applied}\ }\textbf {\bibinfo {volume} {15}},\ \bibinfo {pages} {044003} (\bibinfo {year} {2021})}\BibitemShut
  {NoStop}%
\bibitem [{\citenamefont {Veps{\"a}l{\"a}inen}\ \emph {et~al.}(2022)\citenamefont {Veps{\"a}l{\"a}inen}, \citenamefont {Winik}, \citenamefont {Karamlou}, \citenamefont {Braum{\"u}ller}, \citenamefont {Paolo}, \citenamefont {Sung}, \citenamefont {Kannan}, \citenamefont {Kjaergaard}, \citenamefont {Kim}, \citenamefont {Melville} \emph {et~al.}}]{vepsalainen2022improving}%
  \BibitemOpen
  \bibfield  {author} {\bibinfo {author} {\bibfnamefont {A.}~\bibnamefont {Veps{\"a}l{\"a}inen}}, \bibinfo {author} {\bibfnamefont {R.}~\bibnamefont {Winik}}, \bibinfo {author} {\bibfnamefont {A.~H.}\ \bibnamefont {Karamlou}}, \bibinfo {author} {\bibfnamefont {J.}~\bibnamefont {Braum{\"u}ller}}, \bibinfo {author} {\bibfnamefont {A.~D.}\ \bibnamefont {Paolo}}, \bibinfo {author} {\bibfnamefont {Y.}~\bibnamefont {Sung}}, \bibinfo {author} {\bibfnamefont {B.}~\bibnamefont {Kannan}}, \bibinfo {author} {\bibfnamefont {M.}~\bibnamefont {Kjaergaard}}, \bibinfo {author} {\bibfnamefont {D.~K.}\ \bibnamefont {Kim}}, \bibinfo {author} {\bibfnamefont {A.~J.}\ \bibnamefont {Melville}}, \emph {et~al.},\ }\bibfield  {title} {\bibinfo {title} {Improving qubit coherence using closed-loop feedback},\ }\href {https://doi.org/10.1038/s41467-022-29287-4} {\bibfield  {journal} {\bibinfo  {journal} {Nature Communications}\ }\textbf {\bibinfo {volume} {13}},\ \bibinfo {pages} {1932} (\bibinfo {year} {2022})}\BibitemShut {NoStop}%
\bibitem [{\citenamefont {Shulman}\ \emph {et~al.}(2014)\citenamefont {Shulman}, \citenamefont {Harvey}, \citenamefont {Nichol}, \citenamefont {Bartlett}, \citenamefont {Doherty}, \citenamefont {Umansky},\ and\ \citenamefont {Yacoby}}]{shulman2014suppressing}%
  \BibitemOpen
  \bibfield  {author} {\bibinfo {author} {\bibfnamefont {M.~D.}\ \bibnamefont {Shulman}}, \bibinfo {author} {\bibfnamefont {S.~P.}\ \bibnamefont {Harvey}}, \bibinfo {author} {\bibfnamefont {J.~M.}\ \bibnamefont {Nichol}}, \bibinfo {author} {\bibfnamefont {S.~D.}\ \bibnamefont {Bartlett}}, \bibinfo {author} {\bibfnamefont {A.~C.}\ \bibnamefont {Doherty}}, \bibinfo {author} {\bibfnamefont {V.}~\bibnamefont {Umansky}},\ and\ \bibinfo {author} {\bibfnamefont {A.}~\bibnamefont {Yacoby}},\ }\bibfield  {title} {\bibinfo {title} {{Suppressing qubit dephasing using real-time Hamiltonian estimation}},\ }\href {https://doi.org/10.1038/ncomms6156} {\bibfield  {journal} {\bibinfo  {journal} {Nature Communications}\ }\textbf {\bibinfo {volume} {5}},\ \bibinfo {pages} {5156} (\bibinfo {year} {2014})}\BibitemShut {NoStop}%
\bibitem [{\citenamefont {Dinh}\ \emph {et~al.}(2017)\citenamefont {Dinh}, \citenamefont {Sohl-Dickstein},\ and\ \citenamefont {Bengio}}]{dinh2017density}%
  \BibitemOpen
  \bibfield  {author} {\bibinfo {author} {\bibfnamefont {L.}~\bibnamefont {Dinh}}, \bibinfo {author} {\bibfnamefont {J.}~\bibnamefont {Sohl-Dickstein}},\ and\ \bibinfo {author} {\bibfnamefont {S.}~\bibnamefont {Bengio}},\ }\bibfield  {title} {\bibinfo {title} {{Density estimation using Real {NVP}}},\ }in\ \href@noop {} {\emph {\bibinfo {booktitle} {International Conference on Learning Representations}}}\ (\bibinfo {year} {2017})\BibitemShut {NoStop}%
\bibitem [{\citenamefont {Rezende}\ and\ \citenamefont {Mohamed}(2015)}]{10.5555/3045118.3045281}%
  \BibitemOpen
  \bibfield  {author} {\bibinfo {author} {\bibfnamefont {D.~J.}\ \bibnamefont {Rezende}}\ and\ \bibinfo {author} {\bibfnamefont {S.}~\bibnamefont {Mohamed}},\ }\bibfield  {title} {\bibinfo {title} {Variational inference with normalizing flows},\ }in\ \href@noop {} {\emph {\bibinfo {booktitle} {International Conference on Machine Learning}}},\ Vol.~\bibinfo {volume} {37}\ (\bibinfo {year} {2015})\ p.\ \bibinfo {pages} {1530–1538}\BibitemShut {NoStop}%
\bibitem [{\citenamefont {Papamakarios}\ \emph {et~al.}(2017)\citenamefont {Papamakarios}, \citenamefont {Pavlakou},\ and\ \citenamefont {Murray}}]{10.5555/3294771.3294994}%
  \BibitemOpen
  \bibfield  {author} {\bibinfo {author} {\bibfnamefont {G.}~\bibnamefont {Papamakarios}}, \bibinfo {author} {\bibfnamefont {T.}~\bibnamefont {Pavlakou}},\ and\ \bibinfo {author} {\bibfnamefont {I.}~\bibnamefont {Murray}},\ }\bibfield  {title} {\bibinfo {title} {Masked autoregressive flow for density estimation},\ }in\ \href@noop {} {\emph {\bibinfo {booktitle} {Advances in Neural Information Processing Systems}}},\ Vol.~\bibinfo {volume} {30}\ (\bibinfo {year} {2017})\BibitemShut {NoStop}%
\bibitem [{\citenamefont {Durkan}\ \emph {et~al.}(2019)\citenamefont {Durkan}, \citenamefont {Bekasov}, \citenamefont {Murray},\ and\ \citenamefont {Papamakarios}}]{10.5555/3454287.3454962}%
  \BibitemOpen
  \bibfield  {author} {\bibinfo {author} {\bibfnamefont {C.}~\bibnamefont {Durkan}}, \bibinfo {author} {\bibfnamefont {A.}~\bibnamefont {Bekasov}}, \bibinfo {author} {\bibfnamefont {I.}~\bibnamefont {Murray}},\ and\ \bibinfo {author} {\bibfnamefont {G.}~\bibnamefont {Papamakarios}},\ }\bibfield  {title} {\bibinfo {title} {Neural spline flows},\ }in\ \href@noop {} {\emph {\bibinfo {booktitle} {Advances in Neural Information Processing Systems}}},\ Vol.~\bibinfo {volume} {32}\ (\bibinfo {year} {2019})\BibitemShut {NoStop}%
\bibitem [{\citenamefont {No{\'e}}\ \emph {et~al.}(2019)\citenamefont {No{\'e}}, \citenamefont {Olsson}, \citenamefont {K{\"o}hler},\ and\ \citenamefont {Wu}}]{noe2019boltzmann}%
  \BibitemOpen
  \bibfield  {author} {\bibinfo {author} {\bibfnamefont {F.}~\bibnamefont {No{\'e}}}, \bibinfo {author} {\bibfnamefont {S.}~\bibnamefont {Olsson}}, \bibinfo {author} {\bibfnamefont {J.}~\bibnamefont {K{\"o}hler}},\ and\ \bibinfo {author} {\bibfnamefont {H.}~\bibnamefont {Wu}},\ }\bibfield  {title} {\bibinfo {title} {Boltzmann generators: Sampling equilibrium states of many-body systems with deep learning},\ }\href {https://doi.org/10.1126/science.aaw1147} {\bibfield  {journal} {\bibinfo  {journal} {Science}\ }\textbf {\bibinfo {volume} {365}},\ \bibinfo {pages} {eaaw1147} (\bibinfo {year} {2019})}\BibitemShut {NoStop}%
\bibitem [{\citenamefont {Asghar}\ \emph {et~al.}(2024)\citenamefont {Asghar}, \citenamefont {Pei}, \citenamefont {Volpe},\ and\ \citenamefont {Ni}}]{asghar2024efficient}%
  \BibitemOpen
  \bibfield  {author} {\bibinfo {author} {\bibfnamefont {S.}~\bibnamefont {Asghar}}, \bibinfo {author} {\bibfnamefont {Q.-X.}\ \bibnamefont {Pei}}, \bibinfo {author} {\bibfnamefont {G.}~\bibnamefont {Volpe}},\ and\ \bibinfo {author} {\bibfnamefont {R.}~\bibnamefont {Ni}},\ }\bibfield  {title} {\bibinfo {title} {Efficient rare event sampling with unsupervised normalizing flows},\ }\href {https://doi.org/https://doi.org/10.1038/s42256-024-00918-3} {\bibfield  {journal} {\bibinfo  {journal} {Nature Machine Intelligence}\ }\textbf {\bibinfo {volume} {6}},\ \bibinfo {pages} {1370} (\bibinfo {year} {2024})}\BibitemShut {NoStop}%
\bibitem [{\citenamefont {Albergo}\ \emph {et~al.}(2019)\citenamefont {Albergo}, \citenamefont {Kanwar},\ and\ \citenamefont {Shanahan}}]{PhysRevD.100.034515}%
  \BibitemOpen
  \bibfield  {author} {\bibinfo {author} {\bibfnamefont {M.~S.}\ \bibnamefont {Albergo}}, \bibinfo {author} {\bibfnamefont {G.}~\bibnamefont {Kanwar}},\ and\ \bibinfo {author} {\bibfnamefont {P.~E.}\ \bibnamefont {Shanahan}},\ }\bibfield  {title} {\bibinfo {title} {{Flow-based generative models for Markov chain Monte Carlo in lattice field theory}},\ }\href {https://doi.org/10.1103/PhysRevD.100.034515} {\bibfield  {journal} {\bibinfo  {journal} {Physical Review D}\ }\textbf {\bibinfo {volume} {100}},\ \bibinfo {pages} {034515} (\bibinfo {year} {2019})}\BibitemShut {NoStop}%
\bibitem [{\citenamefont {Zou}\ \emph {et~al.}(2026)\citenamefont {Zou}, \citenamefont {Rahm}, \citenamefont {Kockum},\ and\ \citenamefont {Olsson}}]{zou2025generative}%
  \BibitemOpen
  \bibfield  {author} {\bibinfo {author} {\bibfnamefont {H.}~\bibnamefont {Zou}}, \bibinfo {author} {\bibfnamefont {M.}~\bibnamefont {Rahm}}, \bibinfo {author} {\bibfnamefont {A.~F.}\ \bibnamefont {Kockum}},\ and\ \bibinfo {author} {\bibfnamefont {S.}~\bibnamefont {Olsson}},\ }\bibfield  {title} {\bibinfo {title} {Generative flow-based warm start of the variational quantum eigensolver},\ }\href {https://doi.org/https://doi.org/10.1038/s41534-025-01159-x} {\bibfield  {journal} {\bibinfo  {journal} {npj Quantum Information}\ }\textbf {\bibinfo {volume} {12}},\ \bibinfo {pages} {5} (\bibinfo {year} {2026})}\BibitemShut {NoStop}%
\bibitem [{\citenamefont {Wallman}\ and\ \citenamefont {Emerson}(2016)}]{PhysRevA.94.052325}%
  \BibitemOpen
  \bibfield  {author} {\bibinfo {author} {\bibfnamefont {J.~J.}\ \bibnamefont {Wallman}}\ and\ \bibinfo {author} {\bibfnamefont {J.}~\bibnamefont {Emerson}},\ }\bibfield  {title} {\bibinfo {title} {Noise tailoring for scalable quantum computation via randomized compiling},\ }\href {https://doi.org/10.1103/PhysRevA.94.052325} {\bibfield  {journal} {\bibinfo  {journal} {Physical Review A}\ }\textbf {\bibinfo {volume} {94}},\ \bibinfo {pages} {052325} (\bibinfo {year} {2016})}\BibitemShut {NoStop}%
\bibitem [{\citenamefont {Cai}\ \emph {et~al.}(2023)\citenamefont {Cai}, \citenamefont {Babbush}, \citenamefont {Benjamin}, \citenamefont {Endo}, \citenamefont {Huggins}, \citenamefont {Li}, \citenamefont {McClean},\ and\ \citenamefont {O'Brien}}]{RevModPhys.95.045005}%
  \BibitemOpen
  \bibfield  {author} {\bibinfo {author} {\bibfnamefont {Z.}~\bibnamefont {Cai}}, \bibinfo {author} {\bibfnamefont {R.}~\bibnamefont {Babbush}}, \bibinfo {author} {\bibfnamefont {S.~C.}\ \bibnamefont {Benjamin}}, \bibinfo {author} {\bibfnamefont {S.}~\bibnamefont {Endo}}, \bibinfo {author} {\bibfnamefont {W.~J.}\ \bibnamefont {Huggins}}, \bibinfo {author} {\bibfnamefont {Y.}~\bibnamefont {Li}}, \bibinfo {author} {\bibfnamefont {J.~R.}\ \bibnamefont {McClean}},\ and\ \bibinfo {author} {\bibfnamefont {T.~E.}\ \bibnamefont {O'Brien}},\ }\bibfield  {title} {\bibinfo {title} {Quantum error mitigation},\ }\href {https://doi.org/10.1103/RevModPhys.95.045005} {\bibfield  {journal} {\bibinfo  {journal} {Reviews of Modern Physics}\ }\textbf {\bibinfo {volume} {95}},\ \bibinfo {pages} {045005} (\bibinfo {year} {2023})}\BibitemShut {NoStop}%
\bibitem [{\citenamefont {Tsubouchi}\ \emph {et~al.}(2023)\citenamefont {Tsubouchi}, \citenamefont {Sagawa},\ and\ \citenamefont {Yoshioka}}]{PhysRevLett.131.210601}%
  \BibitemOpen
  \bibfield  {author} {\bibinfo {author} {\bibfnamefont {K.}~\bibnamefont {Tsubouchi}}, \bibinfo {author} {\bibfnamefont {T.}~\bibnamefont {Sagawa}},\ and\ \bibinfo {author} {\bibfnamefont {N.}~\bibnamefont {Yoshioka}},\ }\bibfield  {title} {\bibinfo {title} {Universal cost bound of quantum error mitigation based on quantum estimation theory},\ }\href {https://doi.org/10.1103/PhysRevLett.131.210601} {\bibfield  {journal} {\bibinfo  {journal} {Physical Review Letters}\ }\textbf {\bibinfo {volume} {131}},\ \bibinfo {pages} {210601} (\bibinfo {year} {2023})}\BibitemShut {NoStop}%
\bibitem [{\citenamefont {Takagi}\ \emph {et~al.}(2023)\citenamefont {Takagi}, \citenamefont {Tajima},\ and\ \citenamefont {Gu}}]{PhysRevLett.131.210602}%
  \BibitemOpen
  \bibfield  {author} {\bibinfo {author} {\bibfnamefont {R.}~\bibnamefont {Takagi}}, \bibinfo {author} {\bibfnamefont {H.}~\bibnamefont {Tajima}},\ and\ \bibinfo {author} {\bibfnamefont {M.}~\bibnamefont {Gu}},\ }\bibfield  {title} {\bibinfo {title} {Universal sampling lower bounds for quantum error mitigation},\ }\href {https://doi.org/10.1103/PhysRevLett.131.210602} {\bibfield  {journal} {\bibinfo  {journal} {Physical Review Letters}\ }\textbf {\bibinfo {volume} {131}},\ \bibinfo {pages} {210602} (\bibinfo {year} {2023})}\BibitemShut {NoStop}%
\bibitem [{\citenamefont {Lolur}\ \emph {et~al.}(2023)\citenamefont {Lolur}, \citenamefont {Skogh}, \citenamefont {Dobrautz}, \citenamefont {Warren}, \citenamefont {Bizn{\'a}rov{\'a}}, \citenamefont {Osman}, \citenamefont {Tancredi}, \citenamefont {Wendin}, \citenamefont {Bylander},\ and\ \citenamefont {Rahm}}]{lolur2023reference}%
  \BibitemOpen
  \bibfield  {author} {\bibinfo {author} {\bibfnamefont {P.}~\bibnamefont {Lolur}}, \bibinfo {author} {\bibfnamefont {M.}~\bibnamefont {Skogh}}, \bibinfo {author} {\bibfnamefont {W.}~\bibnamefont {Dobrautz}}, \bibinfo {author} {\bibfnamefont {C.}~\bibnamefont {Warren}}, \bibinfo {author} {\bibfnamefont {J.}~\bibnamefont {Bizn{\'a}rov{\'a}}}, \bibinfo {author} {\bibfnamefont {A.}~\bibnamefont {Osman}}, \bibinfo {author} {\bibfnamefont {G.}~\bibnamefont {Tancredi}}, \bibinfo {author} {\bibfnamefont {G.}~\bibnamefont {Wendin}}, \bibinfo {author} {\bibfnamefont {J.}~\bibnamefont {Bylander}},\ and\ \bibinfo {author} {\bibfnamefont {M.}~\bibnamefont {Rahm}},\ }\bibfield  {title} {\bibinfo {title} {{Reference-state error mitigation: A strategy for high accuracy quantum computation of chemistry}},\ }\href {https://doi.org/10.1021/acs.jctc.2c00807} {\bibfield  {journal} {\bibinfo  {journal} {Journal of Chemical Theory and Computation}\ }\textbf {\bibinfo {volume} {19}},\ \bibinfo {pages} {783} (\bibinfo {year}
  {2023})}\BibitemShut {NoStop}%
\bibitem [{\citenamefont {Zou}\ \emph {et~al.}(2025)\citenamefont {Zou}, \citenamefont {Magnusson}, \citenamefont {Brunander}, \citenamefont {Dobrautz},\ and\ \citenamefont {Rahm}}]{zou2025multireference}%
  \BibitemOpen
  \bibfield  {author} {\bibinfo {author} {\bibfnamefont {H.}~\bibnamefont {Zou}}, \bibinfo {author} {\bibfnamefont {E.}~\bibnamefont {Magnusson}}, \bibinfo {author} {\bibfnamefont {H.}~\bibnamefont {Brunander}}, \bibinfo {author} {\bibfnamefont {W.}~\bibnamefont {Dobrautz}},\ and\ \bibinfo {author} {\bibfnamefont {M.}~\bibnamefont {Rahm}},\ }\bibfield  {title} {\bibinfo {title} {Multireference error mitigation for quantum computation of chemistry},\ }\href {https://doi.org/10.1039/d5dd00202h} {\bibfield  {journal} {\bibinfo  {journal} {Digital Discovery}\ }\textbf {\bibinfo {volume} {4}},\ \bibinfo {pages} {2521} (\bibinfo {year} {2025})}\BibitemShut {NoStop}%
\bibitem [{\citenamefont {Angrisani}\ \emph {et~al.}(2025)\citenamefont {Angrisani}, \citenamefont {Schmidhuber}, \citenamefont {Rudolph}, \citenamefont {Cerezo}, \citenamefont {Holmes},\ and\ \citenamefont {Huang}}]{lh6x-7rc3}%
  \BibitemOpen
  \bibfield  {author} {\bibinfo {author} {\bibfnamefont {A.}~\bibnamefont {Angrisani}}, \bibinfo {author} {\bibfnamefont {A.}~\bibnamefont {Schmidhuber}}, \bibinfo {author} {\bibfnamefont {M.~S.}\ \bibnamefont {Rudolph}}, \bibinfo {author} {\bibfnamefont {M.}~\bibnamefont {Cerezo}}, \bibinfo {author} {\bibfnamefont {Z.}~\bibnamefont {Holmes}},\ and\ \bibinfo {author} {\bibfnamefont {H.-Y.}\ \bibnamefont {Huang}},\ }\bibfield  {title} {\bibinfo {title} {Classically estimating observables of noiseless quantum circuits},\ }\href {https://doi.org/10.1103/lh6x-7rc3} {\bibfield  {journal} {\bibinfo  {journal} {Physical Review Letters}\ }\textbf {\bibinfo {volume} {135}},\ \bibinfo {pages} {170602} (\bibinfo {year} {2025})}\BibitemShut {NoStop}%
\bibitem [{\citenamefont {Schollw\"ock}(2005)}]{RevModPhys.77.259}%
  \BibitemOpen
  \bibfield  {author} {\bibinfo {author} {\bibfnamefont {U.}~\bibnamefont {Schollw\"ock}},\ }\bibfield  {title} {\bibinfo {title} {The density-matrix renormalization group},\ }\href {https://doi.org/10.1103/RevModPhys.77.259} {\bibfield  {journal} {\bibinfo  {journal} {Reviews of Modern Physics}\ }\textbf {\bibinfo {volume} {77}},\ \bibinfo {pages} {259} (\bibinfo {year} {2005})}\BibitemShut {NoStop}%
\bibitem [{\citenamefont {Kurmapu}\ \emph {et~al.}(2023)\citenamefont {Kurmapu}, \citenamefont {Tiunova}, \citenamefont {Tiunov}, \citenamefont {Ringbauer}, \citenamefont {Maier}, \citenamefont {Blatt}, \citenamefont {Monz}, \citenamefont {Fedorov},\ and\ \citenamefont {Lvovsky}}]{PRXQuantum.4.040345}%
  \BibitemOpen
  \bibfield  {author} {\bibinfo {author} {\bibfnamefont {M.~K.}\ \bibnamefont {Kurmapu}}, \bibinfo {author} {\bibfnamefont {V.}~\bibnamefont {Tiunova}}, \bibinfo {author} {\bibfnamefont {E.}~\bibnamefont {Tiunov}}, \bibinfo {author} {\bibfnamefont {M.}~\bibnamefont {Ringbauer}}, \bibinfo {author} {\bibfnamefont {C.}~\bibnamefont {Maier}}, \bibinfo {author} {\bibfnamefont {R.}~\bibnamefont {Blatt}}, \bibinfo {author} {\bibfnamefont {T.}~\bibnamefont {Monz}}, \bibinfo {author} {\bibfnamefont {A.~K.}\ \bibnamefont {Fedorov}},\ and\ \bibinfo {author} {\bibfnamefont {A.}~\bibnamefont {Lvovsky}},\ }\bibfield  {title} {\bibinfo {title} {Reconstructing complex states of a $20$-qubit quantum simulator},\ }\href {https://doi.org/10.1103/PRXQuantum.4.040345} {\bibfield  {journal} {\bibinfo  {journal} {PRX Quantum}\ }\textbf {\bibinfo {volume} {4}},\ \bibinfo {pages} {040345} (\bibinfo {year} {2023})}\BibitemShut {NoStop}%
\bibitem [{\citenamefont {Carleo}\ and\ \citenamefont {Troyer}(2017)}]{carleo2017solving}%
  \BibitemOpen
  \bibfield  {author} {\bibinfo {author} {\bibfnamefont {G.}~\bibnamefont {Carleo}}\ and\ \bibinfo {author} {\bibfnamefont {M.}~\bibnamefont {Troyer}},\ }\bibfield  {title} {\bibinfo {title} {Solving the quantum many-body problem with artificial neural networks},\ }\href {https://doi.org/10.1126/science.aag2302} {\bibfield  {journal} {\bibinfo  {journal} {Science}\ }\textbf {\bibinfo {volume} {355}},\ \bibinfo {pages} {602} (\bibinfo {year} {2017})}\BibitemShut {NoStop}%
\bibitem [{\citenamefont {Lange}\ \emph {et~al.}(2024)\citenamefont {Lange}, \citenamefont {Van~de Walle}, \citenamefont {Abedinnia},\ and\ \citenamefont {Bohrdt}}]{lange2024architectures}%
  \BibitemOpen
  \bibfield  {author} {\bibinfo {author} {\bibfnamefont {H.}~\bibnamefont {Lange}}, \bibinfo {author} {\bibfnamefont {A.}~\bibnamefont {Van~de Walle}}, \bibinfo {author} {\bibfnamefont {A.}~\bibnamefont {Abedinnia}},\ and\ \bibinfo {author} {\bibfnamefont {A.}~\bibnamefont {Bohrdt}},\ }\bibfield  {title} {\bibinfo {title} {{From architectures to applications: A review of neural quantum states}},\ }\href {https://doi.org/10.1088/2058-9565/ad7168} {\bibfield  {journal} {\bibinfo  {journal} {Quantum Science and Technology}\ }\textbf {\bibinfo {volume} {9}},\ \bibinfo {pages} {040501} (\bibinfo {year} {2024})}\BibitemShut {NoStop}%
\bibitem [{\citenamefont {Schreiber}\ \emph {et~al.}(2025)\citenamefont {Schreiber}, \citenamefont {Eisert},\ and\ \citenamefont {Meyer}}]{PRXQuantum.6.020346}%
  \BibitemOpen
  \bibfield  {author} {\bibinfo {author} {\bibfnamefont {F.~J.}\ \bibnamefont {Schreiber}}, \bibinfo {author} {\bibfnamefont {J.}~\bibnamefont {Eisert}},\ and\ \bibinfo {author} {\bibfnamefont {J.~J.}\ \bibnamefont {Meyer}},\ }\bibfield  {title} {\bibinfo {title} {Tomography of parametrized quantum states},\ }\href {https://doi.org/10.1103/PRXQuantum.6.020346} {\bibfield  {journal} {\bibinfo  {journal} {PRX Quantum}\ }\textbf {\bibinfo {volume} {6}},\ \bibinfo {pages} {020346} (\bibinfo {year} {2025})}\BibitemShut {NoStop}%
\bibitem [{\citenamefont {Gaikwad}\ \emph {et~al.}(2025)\citenamefont {Gaikwad}, \citenamefont {Torres}, \citenamefont {Ahmed},\ and\ \citenamefont {Kockum}}]{gaikwad2025gradient}%
  \BibitemOpen
  \bibfield  {author} {\bibinfo {author} {\bibfnamefont {A.}~\bibnamefont {Gaikwad}}, \bibinfo {author} {\bibfnamefont {M.~S.}\ \bibnamefont {Torres}}, \bibinfo {author} {\bibfnamefont {S.}~\bibnamefont {Ahmed}},\ and\ \bibinfo {author} {\bibfnamefont {A.~F.}\ \bibnamefont {Kockum}},\ }\bibfield  {title} {\bibinfo {title} {Gradient-descent methods for fast quantum state tomography},\ }\href {https://doi.org/10.1088/2058-9565/ae0baa} {\bibfield  {journal} {\bibinfo  {journal} {Quantum Science and Technology}\ }\textbf {\bibinfo {volume} {10}},\ \bibinfo {pages} {045055} (\bibinfo {year} {2025})}\BibitemShut {NoStop}%
\bibitem [{\citenamefont {Bluvstein}\ \emph {et~al.}(2024)\citenamefont {Bluvstein}, \citenamefont {Evered}, \citenamefont {Geim}, \citenamefont {Li}, \citenamefont {Zhou}, \citenamefont {Manovitz}, \citenamefont {Ebadi}, \citenamefont {Cain}, \citenamefont {Kalinowski}, \citenamefont {Hangleiter} \emph {et~al.}}]{bluvstein2024logical}%
  \BibitemOpen
  \bibfield  {author} {\bibinfo {author} {\bibfnamefont {D.}~\bibnamefont {Bluvstein}}, \bibinfo {author} {\bibfnamefont {S.~J.}\ \bibnamefont {Evered}}, \bibinfo {author} {\bibfnamefont {A.~A.}\ \bibnamefont {Geim}}, \bibinfo {author} {\bibfnamefont {S.~H.}\ \bibnamefont {Li}}, \bibinfo {author} {\bibfnamefont {H.}~\bibnamefont {Zhou}}, \bibinfo {author} {\bibfnamefont {T.}~\bibnamefont {Manovitz}}, \bibinfo {author} {\bibfnamefont {S.}~\bibnamefont {Ebadi}}, \bibinfo {author} {\bibfnamefont {M.}~\bibnamefont {Cain}}, \bibinfo {author} {\bibfnamefont {M.}~\bibnamefont {Kalinowski}}, \bibinfo {author} {\bibfnamefont {D.}~\bibnamefont {Hangleiter}}, \emph {et~al.},\ }\bibfield  {title} {\bibinfo {title} {Logical quantum processor based on reconfigurable atom arrays},\ }\href {https://doi.org/10.1038/s41586-023-06927-3} {\bibfield  {journal} {\bibinfo  {journal} {Nature}\ }\textbf {\bibinfo {volume} {626}},\ \bibinfo {pages} {58} (\bibinfo {year} {2024})}\BibitemShut {NoStop}%
\bibitem [{\citenamefont {Bluvstein}\ \emph {et~al.}(2026)\citenamefont {Bluvstein}, \citenamefont {Geim}, \citenamefont {Li}, \citenamefont {Evered}, \citenamefont {Bonilla~Ataides}, \citenamefont {Baranes}, \citenamefont {Gu}, \citenamefont {Manovitz}, \citenamefont {Xu}, \citenamefont {Kalinowski} \emph {et~al.}}]{bluvstein2026fault}%
  \BibitemOpen
  \bibfield  {author} {\bibinfo {author} {\bibfnamefont {D.}~\bibnamefont {Bluvstein}}, \bibinfo {author} {\bibfnamefont {A.~A.}\ \bibnamefont {Geim}}, \bibinfo {author} {\bibfnamefont {S.~H.}\ \bibnamefont {Li}}, \bibinfo {author} {\bibfnamefont {S.~J.}\ \bibnamefont {Evered}}, \bibinfo {author} {\bibfnamefont {J.~P.}\ \bibnamefont {Bonilla~Ataides}}, \bibinfo {author} {\bibfnamefont {G.}~\bibnamefont {Baranes}}, \bibinfo {author} {\bibfnamefont {A.}~\bibnamefont {Gu}}, \bibinfo {author} {\bibfnamefont {T.}~\bibnamefont {Manovitz}}, \bibinfo {author} {\bibfnamefont {M.}~\bibnamefont {Xu}}, \bibinfo {author} {\bibfnamefont {M.}~\bibnamefont {Kalinowski}}, \emph {et~al.},\ }\bibfield  {title} {\bibinfo {title} {A fault-tolerant neutral-atom architecture for universal quantum computation},\ }\href {https://doi.org/10.1038/s41586-025-09848-5} {\bibfield  {journal} {\bibinfo  {journal} {Nature}\ }\textbf {\bibinfo {volume} {649}},\ \bibinfo {pages} {39} (\bibinfo {year} {2026})}\BibitemShut {NoStop}%
\bibitem [{\citenamefont {Lin}\ \emph {et~al.}(2025)\citenamefont {Lin}, \citenamefont {Zhong}, \citenamefont {Li}, \citenamefont {Zhao}, \citenamefont {Zheng}, \citenamefont {Hu}, \citenamefont {Wu}, \citenamefont {Wu}, \citenamefont {Ma}, \citenamefont {Gao}, \citenamefont {Zhu}, \citenamefont {Su}, \citenamefont {Ouyang}, \citenamefont {Zhang}, \citenamefont {Rui}, \citenamefont {Chen}, \citenamefont {Lu},\ and\ \citenamefont {Pan}}]{2ym8-vs82}%
  \BibitemOpen
  \bibfield  {author} {\bibinfo {author} {\bibfnamefont {R.}~\bibnamefont {Lin}}, \bibinfo {author} {\bibfnamefont {H.-S.}\ \bibnamefont {Zhong}}, \bibinfo {author} {\bibfnamefont {Y.}~\bibnamefont {Li}}, \bibinfo {author} {\bibfnamefont {Z.-R.}\ \bibnamefont {Zhao}}, \bibinfo {author} {\bibfnamefont {L.-T.}\ \bibnamefont {Zheng}}, \bibinfo {author} {\bibfnamefont {T.-R.}\ \bibnamefont {Hu}}, \bibinfo {author} {\bibfnamefont {H.-M.}\ \bibnamefont {Wu}}, \bibinfo {author} {\bibfnamefont {Z.}~\bibnamefont {Wu}}, \bibinfo {author} {\bibfnamefont {W.-J.}\ \bibnamefont {Ma}}, \bibinfo {author} {\bibfnamefont {Y.}~\bibnamefont {Gao}}, \bibinfo {author} {\bibfnamefont {Y.-K.}\ \bibnamefont {Zhu}}, \bibinfo {author} {\bibfnamefont {Z.-F.}\ \bibnamefont {Su}}, \bibinfo {author} {\bibfnamefont {W.-L.}\ \bibnamefont {Ouyang}}, \bibinfo {author} {\bibfnamefont {Y.-C.}\ \bibnamefont {Zhang}}, \bibinfo {author} {\bibfnamefont {J.}~\bibnamefont {Rui}}, \bibinfo {author} {\bibfnamefont {M.-C.}\ \bibnamefont {Chen}}, \bibinfo
  {author} {\bibfnamefont {C.-Y.}\ \bibnamefont {Lu}},\ and\ \bibinfo {author} {\bibfnamefont {J.-W.}\ \bibnamefont {Pan}},\ }\bibfield  {title} {\bibinfo {title} {{AI-enabled parallel assembly of thousands of defect-free neutral atom arrays}},\ }\href {https://doi.org/10.1103/2ym8-vs82} {\bibfield  {journal} {\bibinfo  {journal} {Physical Review Letters}\ }\textbf {\bibinfo {volume} {135}},\ \bibinfo {pages} {060602} (\bibinfo {year} {2025})}\BibitemShut {NoStop}%
\bibitem [{\citenamefont {Childs}\ \emph {et~al.}(2021)\citenamefont {Childs}, \citenamefont {Su}, \citenamefont {Tran}, \citenamefont {Wiebe},\ and\ \citenamefont {Zhu}}]{PhysRevX.11.011020}%
  \BibitemOpen
  \bibfield  {author} {\bibinfo {author} {\bibfnamefont {A.~M.}\ \bibnamefont {Childs}}, \bibinfo {author} {\bibfnamefont {Y.}~\bibnamefont {Su}}, \bibinfo {author} {\bibfnamefont {M.~C.}\ \bibnamefont {Tran}}, \bibinfo {author} {\bibfnamefont {N.}~\bibnamefont {Wiebe}},\ and\ \bibinfo {author} {\bibfnamefont {S.}~\bibnamefont {Zhu}},\ }\bibfield  {title} {\bibinfo {title} {{Theory of Trotter error with commutator scaling}},\ }\href {https://doi.org/10.1103/PhysRevX.11.011020} {\bibfield  {journal} {\bibinfo  {journal} {Physical Review X}\ }\textbf {\bibinfo {volume} {11}},\ \bibinfo {pages} {011020} (\bibinfo {year} {2021})}\BibitemShut {NoStop}%
\bibitem [{\citenamefont {Seubert}\ \emph {et~al.}(2025)\citenamefont {Seubert}, \citenamefont {Hartung}, \citenamefont {Welte}, \citenamefont {Rempe},\ and\ \citenamefont {Distante}}]{PRXQuantum.6.010322}%
  \BibitemOpen
  \bibfield  {author} {\bibinfo {author} {\bibfnamefont {M.}~\bibnamefont {Seubert}}, \bibinfo {author} {\bibfnamefont {L.}~\bibnamefont {Hartung}}, \bibinfo {author} {\bibfnamefont {S.}~\bibnamefont {Welte}}, \bibinfo {author} {\bibfnamefont {G.}~\bibnamefont {Rempe}},\ and\ \bibinfo {author} {\bibfnamefont {E.}~\bibnamefont {Distante}},\ }\bibfield  {title} {\bibinfo {title} {Tweezer-assisted subwavelength positioning of atomic arrays in an optical cavity},\ }\href {https://doi.org/10.1103/PRXQuantum.6.010322} {\bibfield  {journal} {\bibinfo  {journal} {PRX Quantum}\ }\textbf {\bibinfo {volume} {6}},\ \bibinfo {pages} {010322} (\bibinfo {year} {2025})}\BibitemShut {NoStop}%
\bibitem [{\citenamefont {Huan}\ \emph {et~al.}(2024)\citenamefont {Huan}, \citenamefont {Jagalur},\ and\ \citenamefont {Marzouk}}]{huan2024optimal}%
  \BibitemOpen
  \bibfield  {author} {\bibinfo {author} {\bibfnamefont {X.}~\bibnamefont {Huan}}, \bibinfo {author} {\bibfnamefont {J.}~\bibnamefont {Jagalur}},\ and\ \bibinfo {author} {\bibfnamefont {Y.}~\bibnamefont {Marzouk}},\ }\bibfield  {title} {\bibinfo {title} {{Optimal experimental design: Formulations and computations}},\ }\href {https://doi.org/10.1017/S0962492924000023} {\bibfield  {journal} {\bibinfo  {journal} {Acta Numerica}\ }\textbf {\bibinfo {volume} {33}},\ \bibinfo {pages} {715} (\bibinfo {year} {2024})}\BibitemShut {NoStop}%
\bibitem [{\citenamefont {Chen}\ \emph {et~al.}(2026{\natexlab{b}})\citenamefont {Chen}, \citenamefont {Zhang}, \citenamefont {Jiang},\ and\ \citenamefont {Flammia}}]{Chenselfconsistent}%
  \BibitemOpen
  \bibfield  {author} {\bibinfo {author} {\bibfnamefont {S.}~\bibnamefont {Chen}}, \bibinfo {author} {\bibfnamefont {Z.}~\bibnamefont {Zhang}}, \bibinfo {author} {\bibfnamefont {L.}~\bibnamefont {Jiang}},\ and\ \bibinfo {author} {\bibfnamefont {S.~T.}\ \bibnamefont {Flammia}},\ }\bibfield  {title} {\bibinfo {title} {{Efficient self-consistent learning of gate set Pauli noise}},\ }\href {https://doi.org/10.1103/1pnv-t9px} {\bibfield  {journal} {\bibinfo  {journal} {PRX Quantum}\ }\textbf {\bibinfo {volume} {7}},\ \bibinfo {pages} {010305} (\bibinfo {year} {2026}{\natexlab{b}})}\BibitemShut {NoStop}%
\bibitem [{\citenamefont {Govia}\ \emph {et~al.}(2025)\citenamefont {Govia}, \citenamefont {Majumder}, \citenamefont {Barron}, \citenamefont {Mitchell}, \citenamefont {Seif}, \citenamefont {Kim}, \citenamefont {Wood}, \citenamefont {Pritchett}, \citenamefont {Merkel},\ and\ \citenamefont {McKay}}]{PRXQuantum.6.010354}%
  \BibitemOpen
  \bibfield  {author} {\bibinfo {author} {\bibfnamefont {L.}~\bibnamefont {Govia}}, \bibinfo {author} {\bibfnamefont {S.}~\bibnamefont {Majumder}}, \bibinfo {author} {\bibfnamefont {S.}~\bibnamefont {Barron}}, \bibinfo {author} {\bibfnamefont {B.}~\bibnamefont {Mitchell}}, \bibinfo {author} {\bibfnamefont {A.}~\bibnamefont {Seif}}, \bibinfo {author} {\bibfnamefont {Y.}~\bibnamefont {Kim}}, \bibinfo {author} {\bibfnamefont {C.}~\bibnamefont {Wood}}, \bibinfo {author} {\bibfnamefont {E.}~\bibnamefont {Pritchett}}, \bibinfo {author} {\bibfnamefont {S.}~\bibnamefont {Merkel}},\ and\ \bibinfo {author} {\bibfnamefont {D.}~\bibnamefont {McKay}},\ }\bibfield  {title} {\bibinfo {title} {Bounding the systematic error in quantum error mitigation due to model violation},\ }\href {https://doi.org/10.1103/PRXQuantum.6.010354} {\bibfield  {journal} {\bibinfo  {journal} {PRX Quantum}\ }\textbf {\bibinfo {volume} {6}},\ \bibinfo {pages} {010354} (\bibinfo {year} {2025})}\BibitemShut {NoStop}%
\bibitem [{\citenamefont {Wehenkel}\ \emph {et~al.}(2025)\citenamefont {Wehenkel}, \citenamefont {Gamella}, \citenamefont {Sener}, \citenamefont {Behrmann}, \citenamefont {Sapiro}, \citenamefont {Jacobsen},\ and\ \citenamefont {Cuturi}}]{wehenkel2025addressing}%
  \BibitemOpen
  \bibfield  {author} {\bibinfo {author} {\bibfnamefont {A.}~\bibnamefont {Wehenkel}}, \bibinfo {author} {\bibfnamefont {J.~L.}\ \bibnamefont {Gamella}}, \bibinfo {author} {\bibfnamefont {O.}~\bibnamefont {Sener}}, \bibinfo {author} {\bibfnamefont {J.}~\bibnamefont {Behrmann}}, \bibinfo {author} {\bibfnamefont {G.}~\bibnamefont {Sapiro}}, \bibinfo {author} {\bibfnamefont {J.-H.}\ \bibnamefont {Jacobsen}},\ and\ \bibinfo {author} {\bibfnamefont {M.}~\bibnamefont {Cuturi}},\ }\bibfield  {title} {\bibinfo {title} {Addressing misspecification in simulation-based inference through data-driven calibration},\ }in\ \href@noop {} {\emph {\bibinfo {booktitle} {International Conference on Machine Learning}}}\ (\bibinfo {year} {2025})\BibitemShut {NoStop}%
\bibitem [{\citenamefont {Nott}\ \emph {et~al.}(2023)\citenamefont {Nott}, \citenamefont {Drovandi},\ and\ \citenamefont {Frazier}}]{nott2023bayesian}%
  \BibitemOpen
  \bibfield  {author} {\bibinfo {author} {\bibfnamefont {D.~J.}\ \bibnamefont {Nott}}, \bibinfo {author} {\bibfnamefont {C.}~\bibnamefont {Drovandi}},\ and\ \bibinfo {author} {\bibfnamefont {D.~T.}\ \bibnamefont {Frazier}},\ }\bibfield  {title} {\bibinfo {title} {Bayesian inference for misspecified generative models},\ }\href {https://doi.org/10.1146/annurev-statistics-040522-015915} {\bibfield  {journal} {\bibinfo  {journal} {Annual Review of Statistics and Its Application}\ }\textbf {\bibinfo {volume} {11}} (\bibinfo {year} {2023})}\BibitemShut {NoStop}%
\bibitem [{\citenamefont {Anau~Montel}\ \emph {et~al.}(2025)\citenamefont {Anau~Montel}, \citenamefont {Alvey},\ and\ \citenamefont {Weniger}}]{PhysRevD.111.083013}%
  \BibitemOpen
  \bibfield  {author} {\bibinfo {author} {\bibfnamefont {N.}~\bibnamefont {Anau~Montel}}, \bibinfo {author} {\bibfnamefont {J.}~\bibnamefont {Alvey}},\ and\ \bibinfo {author} {\bibfnamefont {C.}~\bibnamefont {Weniger}},\ }\bibfield  {title} {\bibinfo {title} {{Tests for model misspecification in simulation-based inference: From local distortions to global model checks}},\ }\href {https://doi.org/10.1103/PhysRevD.111.083013} {\bibfield  {journal} {\bibinfo  {journal} {Physical Review D}\ }\textbf {\bibinfo {volume} {111}},\ \bibinfo {pages} {083013} (\bibinfo {year} {2025})}\BibitemShut {NoStop}%
\bibitem [{\citenamefont {Patel}\ \emph {et~al.}(2026)\citenamefont {Patel}, \citenamefont {Gaikwad}, \citenamefont {Huang}, \citenamefont {Kockum},\ and\ \citenamefont {Abad}}]{hynl-kxl2}%
  \BibitemOpen
  \bibfield  {author} {\bibinfo {author} {\bibfnamefont {A.}~\bibnamefont {Patel}}, \bibinfo {author} {\bibfnamefont {A.}~\bibnamefont {Gaikwad}}, \bibinfo {author} {\bibfnamefont {T.}~\bibnamefont {Huang}}, \bibinfo {author} {\bibfnamefont {A.~F.}\ \bibnamefont {Kockum}},\ and\ \bibinfo {author} {\bibfnamefont {T.}~\bibnamefont {Abad}},\ }\bibfield  {title} {\bibinfo {title} {Selective and efficient quantum state tomography for multiqubit systems},\ }\href {https://doi.org/10.1103/hynl-kxl2} {\bibfield  {journal} {\bibinfo  {journal} {Physical Review Research}\ }\textbf {\bibinfo {volume} {8}},\ \bibinfo {pages} {013339} (\bibinfo {year} {2026})}\BibitemShut {NoStop}%
\bibitem [{\citenamefont {Cotler}\ and\ \citenamefont {Wilczek}(2020)}]{PhysRevLett.124.100401}%
  \BibitemOpen
  \bibfield  {author} {\bibinfo {author} {\bibfnamefont {J.}~\bibnamefont {Cotler}}\ and\ \bibinfo {author} {\bibfnamefont {F.}~\bibnamefont {Wilczek}},\ }\bibfield  {title} {\bibinfo {title} {Quantum overlapping tomography},\ }\href {https://doi.org/10.1103/PhysRevLett.124.100401} {\bibfield  {journal} {\bibinfo  {journal} {Physical Review Letters}\ }\textbf {\bibinfo {volume} {124}},\ \bibinfo {pages} {100401} (\bibinfo {year} {2020})}\BibitemShut {NoStop}%
\bibitem [{\citenamefont {Peng}\ \emph {et~al.}(2020)\citenamefont {Peng}, \citenamefont {Harrow}, \citenamefont {Ozols},\ and\ \citenamefont {Wu}}]{PhysRevLett.125.150504}%
  \BibitemOpen
  \bibfield  {author} {\bibinfo {author} {\bibfnamefont {T.}~\bibnamefont {Peng}}, \bibinfo {author} {\bibfnamefont {A.~W.}\ \bibnamefont {Harrow}}, \bibinfo {author} {\bibfnamefont {M.}~\bibnamefont {Ozols}},\ and\ \bibinfo {author} {\bibfnamefont {X.}~\bibnamefont {Wu}},\ }\bibfield  {title} {\bibinfo {title} {Simulating large quantum circuits on a small quantum computer},\ }\href {https://doi.org/10.1103/PhysRevLett.125.150504} {\bibfield  {journal} {\bibinfo  {journal} {Physical Review Letters}\ }\textbf {\bibinfo {volume} {125}},\ \bibinfo {pages} {150504} (\bibinfo {year} {2020})}\BibitemShut {NoStop}%
\bibitem [{\citenamefont {Liu}\ \emph {et~al.}(2022)\citenamefont {Liu}, \citenamefont {Gonzales},\ and\ \citenamefont {Saleem}}]{liu2022classicalsimu}%
  \BibitemOpen
  \bibfield  {author} {\bibinfo {author} {\bibfnamefont {J.}~\bibnamefont {Liu}}, \bibinfo {author} {\bibfnamefont {A.}~\bibnamefont {Gonzales}},\ and\ \bibinfo {author} {\bibfnamefont {Z.~H.}\ \bibnamefont {Saleem}},\ }\href@noop {} {\bibinfo {title} {Classical simulators as quantum error mitigators via circuit cutting}} (\bibinfo {year} {2022}),\ \Eprint {https://arxiv.org/abs/2212.07335} {arXiv:2212.07335} \BibitemShut {NoStop}%
\bibitem [{\citenamefont {Elben}\ \emph {et~al.}(2023)\citenamefont {Elben}, \citenamefont {Flammia}, \citenamefont {Huang}, \citenamefont {Kueng}, \citenamefont {Preskill}, \citenamefont {Vermersch},\ and\ \citenamefont {Zoller}}]{elben2023randomized}%
  \BibitemOpen
  \bibfield  {author} {\bibinfo {author} {\bibfnamefont {A.}~\bibnamefont {Elben}}, \bibinfo {author} {\bibfnamefont {S.~T.}\ \bibnamefont {Flammia}}, \bibinfo {author} {\bibfnamefont {H.-Y.}\ \bibnamefont {Huang}}, \bibinfo {author} {\bibfnamefont {R.}~\bibnamefont {Kueng}}, \bibinfo {author} {\bibfnamefont {J.}~\bibnamefont {Preskill}}, \bibinfo {author} {\bibfnamefont {B.}~\bibnamefont {Vermersch}},\ and\ \bibinfo {author} {\bibfnamefont {P.}~\bibnamefont {Zoller}},\ }\bibfield  {title} {\bibinfo {title} {The randomized measurement toolbox},\ }\href {https://doi.org/10.1038/s42254-022-00535-2} {\bibfield  {journal} {\bibinfo  {journal} {Nature Reviews Physics}\ }\textbf {\bibinfo {volume} {5}},\ \bibinfo {pages} {9} (\bibinfo {year} {2023})}\BibitemShut {NoStop}%
\bibitem [{\citenamefont {Sanchez-Gonzalez}\ \emph {et~al.}(2020)\citenamefont {Sanchez-Gonzalez}, \citenamefont {Godwin}, \citenamefont {Pfaff}, \citenamefont {Ying}, \citenamefont {Leskovec},\ and\ \citenamefont {Battaglia}}]{sanchez2020learning}%
  \BibitemOpen
  \bibfield  {author} {\bibinfo {author} {\bibfnamefont {A.}~\bibnamefont {Sanchez-Gonzalez}}, \bibinfo {author} {\bibfnamefont {J.}~\bibnamefont {Godwin}}, \bibinfo {author} {\bibfnamefont {T.}~\bibnamefont {Pfaff}}, \bibinfo {author} {\bibfnamefont {R.}~\bibnamefont {Ying}}, \bibinfo {author} {\bibfnamefont {J.}~\bibnamefont {Leskovec}},\ and\ \bibinfo {author} {\bibfnamefont {P.}~\bibnamefont {Battaglia}},\ }\bibfield  {title} {\bibinfo {title} {Learning to simulate complex physics with graph networks},\ }in\ \href@noop {} {\emph {\bibinfo {booktitle} {International Conference on Machine Learning}}}\ (\bibinfo {year} {2020})\ pp.\ \bibinfo {pages} {8459--8468}\BibitemShut {NoStop}%
\bibitem [{\citenamefont {Diez}\ \emph {et~al.}(2026)\citenamefont {Diez}, \citenamefont {Schreiner},\ and\ \citenamefont {Olsson}}]{diez2026transferable}%
  \BibitemOpen
  \bibfield  {author} {\bibinfo {author} {\bibfnamefont {J.~V.}\ \bibnamefont {Diez}}, \bibinfo {author} {\bibfnamefont {M.}~\bibnamefont {Schreiner}},\ and\ \bibinfo {author} {\bibfnamefont {S.}~\bibnamefont {Olsson}},\ }\bibfield  {title} {\bibinfo {title} {Transferable generative models bridge femtosecond to nanosecond time-step molecular dynamics},\ }\href {https://doi.org/10.1126/sciadv.aed2333} {\bibfield  {journal} {\bibinfo  {journal} {Science Advances}\ }\textbf {\bibinfo {volume} {12}},\ \bibinfo {pages} {eaed2333} (\bibinfo {year} {2026})}\BibitemShut {NoStop}%
\bibitem [{\citenamefont {Wang}\ \emph {et~al.}(2026)\citenamefont {Wang}, \citenamefont {Wu}, \citenamefont {Liu}, \citenamefont {He}, \citenamefont {Shang}, \citenamefont {Guo},\ and\ \citenamefont {Chen}}]{wang2026scalablequantumerrormitigation}%
  \BibitemOpen
  \bibfield  {author} {\bibinfo {author} {\bibfnamefont {H.}~\bibnamefont {Wang}}, \bibinfo {author} {\bibfnamefont {X.}~\bibnamefont {Wu}}, \bibinfo {author} {\bibfnamefont {J.}~\bibnamefont {Liu}}, \bibinfo {author} {\bibfnamefont {R.}~\bibnamefont {He}}, \bibinfo {author} {\bibfnamefont {J.}~\bibnamefont {Shang}}, \bibinfo {author} {\bibfnamefont {H.}~\bibnamefont {Guo}},\ and\ \bibinfo {author} {\bibfnamefont {Q.}~\bibnamefont {Chen}},\ }\href@noop {} {\bibinfo {title} {Scalable quantum error mitigation with physically informed graph neural networks}} (\bibinfo {year} {2026}),\ \Eprint {https://arxiv.org/abs/2604.16815} {arXiv:2604.16815} \BibitemShut {NoStop}%
\bibitem [{\citenamefont {Cohen}\ \emph {et~al.}(2019)\citenamefont {Cohen}, \citenamefont {Weiler}, \citenamefont {Kicanaoglu},\ and\ \citenamefont {Welling}}]{cohen2019gauge}%
  \BibitemOpen
  \bibfield  {author} {\bibinfo {author} {\bibfnamefont {T.}~\bibnamefont {Cohen}}, \bibinfo {author} {\bibfnamefont {M.}~\bibnamefont {Weiler}}, \bibinfo {author} {\bibfnamefont {B.}~\bibnamefont {Kicanaoglu}},\ and\ \bibinfo {author} {\bibfnamefont {M.}~\bibnamefont {Welling}},\ }\bibfield  {title} {\bibinfo {title} {{Gauge equivariant convolutional networks and the icosahedral CNN}},\ }in\ \href@noop {} {\emph {\bibinfo {booktitle} {International Conference on Machine Learning}}}\ (\bibinfo {year} {2019})\BibitemShut {NoStop}%
\bibitem [{\citenamefont {Robledo-Moreno}\ \emph {et~al.}(2025)\citenamefont {Robledo-Moreno}, \citenamefont {Motta}, \citenamefont {Haas}, \citenamefont {Javadi-Abhari}, \citenamefont {Jurcevic}, \citenamefont {Kirby}, \citenamefont {Martiel}, \citenamefont {Sharma}, \citenamefont {Sharma}, \citenamefont {Shirakawa} \emph {et~al.}}]{robledo2025chemistry}%
  \BibitemOpen
  \bibfield  {author} {\bibinfo {author} {\bibfnamefont {J.}~\bibnamefont {Robledo-Moreno}}, \bibinfo {author} {\bibfnamefont {M.}~\bibnamefont {Motta}}, \bibinfo {author} {\bibfnamefont {H.}~\bibnamefont {Haas}}, \bibinfo {author} {\bibfnamefont {A.}~\bibnamefont {Javadi-Abhari}}, \bibinfo {author} {\bibfnamefont {P.}~\bibnamefont {Jurcevic}}, \bibinfo {author} {\bibfnamefont {W.}~\bibnamefont {Kirby}}, \bibinfo {author} {\bibfnamefont {S.}~\bibnamefont {Martiel}}, \bibinfo {author} {\bibfnamefont {K.}~\bibnamefont {Sharma}}, \bibinfo {author} {\bibfnamefont {S.}~\bibnamefont {Sharma}}, \bibinfo {author} {\bibfnamefont {T.}~\bibnamefont {Shirakawa}}, \emph {et~al.},\ }\bibfield  {title} {\bibinfo {title} {Chemistry beyond the scale of exact diagonalization on a quantum-centric supercomputer},\ }\href {https://doi.org/10.1126/sciadv.adu9991} {\bibfield  {journal} {\bibinfo  {journal} {Science Advances}\ }\textbf {\bibinfo {volume} {11}},\ \bibinfo {pages} {eadu9991} (\bibinfo {year} {2025})}\BibitemShut
  {NoStop}%
\bibitem [{\citenamefont {Seelam}\ \emph {et~al.}(2026)\citenamefont {Seelam}, \citenamefont {Chow}, \citenamefont {Córcoles}, \citenamefont {Sheldon}, \citenamefont {Mittal}, \citenamefont {Kandala}, \citenamefont {Dague}, \citenamefont {Hincks}, \citenamefont {Horii}, \citenamefont {Johnson}, \citenamefont {Le}, \citenamefont {Jamjoom},\ and\ \citenamefont {Gambetta}}]{seelam2026referencearch}%
  \BibitemOpen
  \bibfield  {author} {\bibinfo {author} {\bibfnamefont {S.}~\bibnamefont {Seelam}}, \bibinfo {author} {\bibfnamefont {J.~M.}\ \bibnamefont {Chow}}, \bibinfo {author} {\bibfnamefont {A.}~\bibnamefont {Córcoles}}, \bibinfo {author} {\bibfnamefont {S.}~\bibnamefont {Sheldon}}, \bibinfo {author} {\bibfnamefont {T.}~\bibnamefont {Mittal}}, \bibinfo {author} {\bibfnamefont {A.}~\bibnamefont {Kandala}}, \bibinfo {author} {\bibfnamefont {S.}~\bibnamefont {Dague}}, \bibinfo {author} {\bibfnamefont {I.}~\bibnamefont {Hincks}}, \bibinfo {author} {\bibfnamefont {H.}~\bibnamefont {Horii}}, \bibinfo {author} {\bibfnamefont {B.}~\bibnamefont {Johnson}}, \bibinfo {author} {\bibfnamefont {M.}~\bibnamefont {Le}}, \bibinfo {author} {\bibfnamefont {H.}~\bibnamefont {Jamjoom}},\ and\ \bibinfo {author} {\bibfnamefont {J.~M.}\ \bibnamefont {Gambetta}},\ }\href@noop {} {\bibinfo {title} {Reference architecture of a quantum-centric supercomputer}} (\bibinfo {year} {2026}),\ \Eprint {https://arxiv.org/abs/2603.10970}
  {arXiv:2603.10970} \BibitemShut {NoStop}%
\bibitem [{\citenamefont {Katabarwa}\ \emph {et~al.}(2024)\citenamefont {Katabarwa}, \citenamefont {Gratsea}, \citenamefont {Caesura},\ and\ \citenamefont {Johnson}}]{PRXQuantum.5.020101}%
  \BibitemOpen
  \bibfield  {author} {\bibinfo {author} {\bibfnamefont {A.}~\bibnamefont {Katabarwa}}, \bibinfo {author} {\bibfnamefont {K.}~\bibnamefont {Gratsea}}, \bibinfo {author} {\bibfnamefont {A.}~\bibnamefont {Caesura}},\ and\ \bibinfo {author} {\bibfnamefont {P.~D.}\ \bibnamefont {Johnson}},\ }\bibfield  {title} {\bibinfo {title} {Early fault-tolerant quantum computing},\ }\href {https://doi.org/10.1103/PRXQuantum.5.020101} {\bibfield  {journal} {\bibinfo  {journal} {PRX Quantum}\ }\textbf {\bibinfo {volume} {5}},\ \bibinfo {pages} {020101} (\bibinfo {year} {2024})}\BibitemShut {NoStop}%
\bibitem [{\citenamefont {Wiebe}\ \emph {et~al.}(2014{\natexlab{b}})\citenamefont {Wiebe}, \citenamefont {Granade}, \citenamefont {Ferrie},\ and\ \citenamefont {Cory}}]{PhysRevA.89.042314}%
  \BibitemOpen
  \bibfield  {author} {\bibinfo {author} {\bibfnamefont {N.}~\bibnamefont {Wiebe}}, \bibinfo {author} {\bibfnamefont {C.}~\bibnamefont {Granade}}, \bibinfo {author} {\bibfnamefont {C.}~\bibnamefont {Ferrie}},\ and\ \bibinfo {author} {\bibfnamefont {D.}~\bibnamefont {Cory}},\ }\bibfield  {title} {\bibinfo {title} {{Quantum Hamiltonian learning using imperfect quantum resources}},\ }\href {https://doi.org/10.1103/PhysRevA.89.042314} {\bibfield  {journal} {\bibinfo  {journal} {Physical Review A}\ }\textbf {\bibinfo {volume} {89}},\ \bibinfo {pages} {042314} (\bibinfo {year} {2014}{\natexlab{b}})}\BibitemShut {NoStop}%
\bibitem [{\citenamefont {Benedetti}\ \emph {et~al.}(2021)\citenamefont {Benedetti}, \citenamefont {Coyle}, \citenamefont {Fiorentini}, \citenamefont {Lubasch},\ and\ \citenamefont {Rosenkranz}}]{PhysRevApplied.16.044057}%
  \BibitemOpen
  \bibfield  {author} {\bibinfo {author} {\bibfnamefont {M.}~\bibnamefont {Benedetti}}, \bibinfo {author} {\bibfnamefont {B.}~\bibnamefont {Coyle}}, \bibinfo {author} {\bibfnamefont {M.}~\bibnamefont {Fiorentini}}, \bibinfo {author} {\bibfnamefont {M.}~\bibnamefont {Lubasch}},\ and\ \bibinfo {author} {\bibfnamefont {M.}~\bibnamefont {Rosenkranz}},\ }\bibfield  {title} {\bibinfo {title} {Variational inference with a quantum computer},\ }\href {https://doi.org/10.1103/PhysRevApplied.16.044057} {\bibfield  {journal} {\bibinfo  {journal} {Physical Review Applied}\ }\textbf {\bibinfo {volume} {16}},\ \bibinfo {pages} {044057} (\bibinfo {year} {2021})}\BibitemShut {NoStop}%
\bibitem [{\citenamefont {Coyle}\ \emph {et~al.}(2020)\citenamefont {Coyle}, \citenamefont {Mills}, \citenamefont {Danos},\ and\ \citenamefont {Kashefi}}]{coyle2020born}%
  \BibitemOpen
  \bibfield  {author} {\bibinfo {author} {\bibfnamefont {B.}~\bibnamefont {Coyle}}, \bibinfo {author} {\bibfnamefont {D.}~\bibnamefont {Mills}}, \bibinfo {author} {\bibfnamefont {V.}~\bibnamefont {Danos}},\ and\ \bibinfo {author} {\bibfnamefont {E.}~\bibnamefont {Kashefi}},\ }\bibfield  {title} {\bibinfo {title} {{The Born supremacy: quantum advantage and training of an Ising Born machine}},\ }\href {https://doi.org/10.1038/s41534-020-00288-9} {\bibfield  {journal} {\bibinfo  {journal} {npj Quantum Information}\ }\textbf {\bibinfo {volume} {6}},\ \bibinfo {pages} {60} (\bibinfo {year} {2020})}\BibitemShut {NoStop}%
\bibitem [{\citenamefont {Rozet}\ \emph {et~al.}(2022)\citenamefont {Rozet} \emph {et~al.}}]{rozet2022zuko}%
  \BibitemOpen
  \bibfield  {author} {\bibinfo {author} {\bibfnamefont {F.}~\bibnamefont {Rozet}} \emph {et~al.},\ }\href {https://doi.org/10.5281/zenodo.7625672} {\bibinfo {title} {{{Zuko}: Normalizing flows in PyTorch}}} (\bibinfo {year} {2022})\BibitemShut {NoStop}%
\bibitem [{\citenamefont {Boelts}\ \emph {et~al.}(2025)\citenamefont {Boelts}, \citenamefont {Deistler}, \citenamefont {Gloeckler}, \citenamefont {Álvaro Tejero-Cantero}, \citenamefont {Lueckmann}, \citenamefont {Moss}, \citenamefont {Steinbach}, \citenamefont {Moreau}, \citenamefont {Muratore}, \citenamefont {Linhart}, \citenamefont {Durkan}, \citenamefont {Vetter}, \citenamefont {Miller}, \citenamefont {Herold}, \citenamefont {Ziaeemehr}, \citenamefont {Pals}, \citenamefont {Gruner}, \citenamefont {Bischoff}, \citenamefont {Krouglova}, \citenamefont {Gao}, \citenamefont {Lappalainen}, \citenamefont {Mucsányi}, \citenamefont {Pei}, \citenamefont {Schulz}, \citenamefont {Stefanidi}, \citenamefont {Rodrigues}, \citenamefont {Schröder}, \citenamefont {Zaid}, \citenamefont {Beck}, \citenamefont {Kapoor}, \citenamefont {Greenberg}, \citenamefont {Gonçalves},\ and\ \citenamefont {Macke}}]{BoeltsDeistler_sbi_2025}%
  \BibitemOpen
  \bibfield  {author} {\bibinfo {author} {\bibfnamefont {J.}~\bibnamefont {Boelts}}, \bibinfo {author} {\bibfnamefont {M.}~\bibnamefont {Deistler}}, \bibinfo {author} {\bibfnamefont {M.}~\bibnamefont {Gloeckler}}, \bibinfo {author} {\bibnamefont {Álvaro Tejero-Cantero}}, \bibinfo {author} {\bibfnamefont {J.-M.}\ \bibnamefont {Lueckmann}}, \bibinfo {author} {\bibfnamefont {G.}~\bibnamefont {Moss}}, \bibinfo {author} {\bibfnamefont {P.}~\bibnamefont {Steinbach}}, \bibinfo {author} {\bibfnamefont {T.}~\bibnamefont {Moreau}}, \bibinfo {author} {\bibfnamefont {F.}~\bibnamefont {Muratore}}, \bibinfo {author} {\bibfnamefont {J.}~\bibnamefont {Linhart}}, \bibinfo {author} {\bibfnamefont {C.}~\bibnamefont {Durkan}}, \bibinfo {author} {\bibfnamefont {J.}~\bibnamefont {Vetter}}, \bibinfo {author} {\bibfnamefont {B.~K.}\ \bibnamefont {Miller}}, \bibinfo {author} {\bibfnamefont {M.}~\bibnamefont {Herold}}, \bibinfo {author} {\bibfnamefont {A.}~\bibnamefont {Ziaeemehr}}, \bibinfo {author} {\bibfnamefont {M.}~\bibnamefont
  {Pals}}, \bibinfo {author} {\bibfnamefont {T.}~\bibnamefont {Gruner}}, \bibinfo {author} {\bibfnamefont {S.}~\bibnamefont {Bischoff}}, \bibinfo {author} {\bibfnamefont {N.}~\bibnamefont {Krouglova}}, \bibinfo {author} {\bibfnamefont {R.}~\bibnamefont {Gao}}, \bibinfo {author} {\bibfnamefont {J.~K.}\ \bibnamefont {Lappalainen}}, \bibinfo {author} {\bibfnamefont {B.}~\bibnamefont {Mucsányi}}, \bibinfo {author} {\bibfnamefont {F.}~\bibnamefont {Pei}}, \bibinfo {author} {\bibfnamefont {A.}~\bibnamefont {Schulz}}, \bibinfo {author} {\bibfnamefont {Z.}~\bibnamefont {Stefanidi}}, \bibinfo {author} {\bibfnamefont {P.}~\bibnamefont {Rodrigues}}, \bibinfo {author} {\bibfnamefont {C.}~\bibnamefont {Schröder}}, \bibinfo {author} {\bibfnamefont {F.~A.}\ \bibnamefont {Zaid}}, \bibinfo {author} {\bibfnamefont {J.}~\bibnamefont {Beck}}, \bibinfo {author} {\bibfnamefont {J.}~\bibnamefont {Kapoor}}, \bibinfo {author} {\bibfnamefont {D.~S.}\ \bibnamefont {Greenberg}}, \bibinfo {author} {\bibfnamefont {P.~J.}\ \bibnamefont
  {Gonçalves}},\ and\ \bibinfo {author} {\bibfnamefont {J.~H.}\ \bibnamefont {Macke}},\ }\bibfield  {title} {\bibinfo {title} {sbi reloaded: a toolkit for simulation-based inference workflows},\ }\href {https://doi.org/10.21105/joss.07754} {\bibfield  {journal} {\bibinfo  {journal} {Journal of Open Source Software}\ }\textbf {\bibinfo {volume} {10}},\ \bibinfo {pages} {7754} (\bibinfo {year} {2025})}\BibitemShut {NoStop}%
\bibitem [{\citenamefont {Gidney}(2021)}]{gidney2021stim}%
  \BibitemOpen
  \bibfield  {author} {\bibinfo {author} {\bibfnamefont {C.}~\bibnamefont {Gidney}},\ }\bibfield  {title} {\bibinfo {title} {Stim: a fast stabilizer circuit simulator},\ }\href {https://doi.org/10.22331/q-2021-07-06-497} {\bibfield  {journal} {\bibinfo  {journal} {{Quantum}}\ }\textbf {\bibinfo {volume} {5}},\ \bibinfo {pages} {497} (\bibinfo {year} {2021})}\BibitemShut {NoStop}%
\bibitem [{\citenamefont {Javadi-Abhari}\ \emph {et~al.}(2024)\citenamefont {Javadi-Abhari}, \citenamefont {Treinish}, \citenamefont {Krsulich}, \citenamefont {Wood}, \citenamefont {Lishman}, \citenamefont {Gacon}, \citenamefont {Martiel}, \citenamefont {Nation}, \citenamefont {Bishop}, \citenamefont {Cross}, \citenamefont {Johnson},\ and\ \citenamefont {Gambetta}}]{qiskit2024}%
  \BibitemOpen
  \bibfield  {author} {\bibinfo {author} {\bibfnamefont {A.}~\bibnamefont {Javadi-Abhari}}, \bibinfo {author} {\bibfnamefont {M.}~\bibnamefont {Treinish}}, \bibinfo {author} {\bibfnamefont {K.}~\bibnamefont {Krsulich}}, \bibinfo {author} {\bibfnamefont {C.~J.}\ \bibnamefont {Wood}}, \bibinfo {author} {\bibfnamefont {J.}~\bibnamefont {Lishman}}, \bibinfo {author} {\bibfnamefont {J.}~\bibnamefont {Gacon}}, \bibinfo {author} {\bibfnamefont {S.}~\bibnamefont {Martiel}}, \bibinfo {author} {\bibfnamefont {P.~D.}\ \bibnamefont {Nation}}, \bibinfo {author} {\bibfnamefont {L.~S.}\ \bibnamefont {Bishop}}, \bibinfo {author} {\bibfnamefont {A.~W.}\ \bibnamefont {Cross}}, \bibinfo {author} {\bibfnamefont {B.~R.}\ \bibnamefont {Johnson}},\ and\ \bibinfo {author} {\bibfnamefont {J.~M.}\ \bibnamefont {Gambetta}},\ }\href@noop {} {\bibinfo {title} {Quantum computing with {Q}iskit}} (\bibinfo {year} {2024}),\ \Eprint {https://arxiv.org/abs/2405.08810} {arXiv:2405.08810} \BibitemShut {NoStop}%
\bibitem [{\citenamefont {Paszke}\ \emph {et~al.}(2019)\citenamefont {Paszke}, \citenamefont {Gross}, \citenamefont {Massa}, \citenamefont {Lerer}, \citenamefont {Bradbury}, \citenamefont {Chanan}, \citenamefont {Killeen}, \citenamefont {Lin}, \citenamefont {Gimelshein}, \citenamefont {Antiga}, \citenamefont {Desmaison}, \citenamefont {Kopf}, \citenamefont {Yang}, \citenamefont {DeVito}, \citenamefont {Raison}, \citenamefont {Tejani}, \citenamefont {Chilamkurthy}, \citenamefont {Steiner}, \citenamefont {Fang}, \citenamefont {Bai},\ and\ \citenamefont {Chintala}}]{NEURIPS2019_bdbca288}%
  \BibitemOpen
  \bibfield  {author} {\bibinfo {author} {\bibfnamefont {A.}~\bibnamefont {Paszke}}, \bibinfo {author} {\bibfnamefont {S.}~\bibnamefont {Gross}}, \bibinfo {author} {\bibfnamefont {F.}~\bibnamefont {Massa}}, \bibinfo {author} {\bibfnamefont {A.}~\bibnamefont {Lerer}}, \bibinfo {author} {\bibfnamefont {J.}~\bibnamefont {Bradbury}}, \bibinfo {author} {\bibfnamefont {G.}~\bibnamefont {Chanan}}, \bibinfo {author} {\bibfnamefont {T.}~\bibnamefont {Killeen}}, \bibinfo {author} {\bibfnamefont {Z.}~\bibnamefont {Lin}}, \bibinfo {author} {\bibfnamefont {N.}~\bibnamefont {Gimelshein}}, \bibinfo {author} {\bibfnamefont {L.}~\bibnamefont {Antiga}}, \bibinfo {author} {\bibfnamefont {A.}~\bibnamefont {Desmaison}}, \bibinfo {author} {\bibfnamefont {A.}~\bibnamefont {Kopf}}, \bibinfo {author} {\bibfnamefont {E.}~\bibnamefont {Yang}}, \bibinfo {author} {\bibfnamefont {Z.}~\bibnamefont {DeVito}}, \bibinfo {author} {\bibfnamefont {M.}~\bibnamefont {Raison}}, \bibinfo {author} {\bibfnamefont {A.}~\bibnamefont {Tejani}}, \bibinfo
  {author} {\bibfnamefont {S.}~\bibnamefont {Chilamkurthy}}, \bibinfo {author} {\bibfnamefont {B.}~\bibnamefont {Steiner}}, \bibinfo {author} {\bibfnamefont {L.}~\bibnamefont {Fang}}, \bibinfo {author} {\bibfnamefont {J.}~\bibnamefont {Bai}},\ and\ \bibinfo {author} {\bibfnamefont {S.}~\bibnamefont {Chintala}},\ }\bibfield  {title} {\bibinfo {title} {{PyTorch: An imperative style, high-performance deep learning library}},\ }in\ \href@noop {} {\emph {\bibinfo {booktitle} {Advances in Neural Information Processing Systems}}},\ Vol.~\bibinfo {volume} {32}\ (\bibinfo {year} {2019})\BibitemShut {NoStop}%
\bibitem [{\citenamefont {Germain}\ \emph {et~al.}(2015)\citenamefont {Germain}, \citenamefont {Gregor}, \citenamefont {Murray},\ and\ \citenamefont {Larochelle}}]{10.5555/3045118.3045213}%
  \BibitemOpen
  \bibfield  {author} {\bibinfo {author} {\bibfnamefont {M.}~\bibnamefont {Germain}}, \bibinfo {author} {\bibfnamefont {K.}~\bibnamefont {Gregor}}, \bibinfo {author} {\bibfnamefont {I.}~\bibnamefont {Murray}},\ and\ \bibinfo {author} {\bibfnamefont {H.}~\bibnamefont {Larochelle}},\ }\bibfield  {title} {\bibinfo {title} {{MADE: masked autoencoder for distribution estimation}},\ }in\ \href@noop {} {\emph {\bibinfo {booktitle} {International Conference on Machine Learning}}},\ Vol.~\bibinfo {volume} {37}\ (\bibinfo {year} {2015})\ p.\ \bibinfo {pages} {881–889}\BibitemShut {NoStop}%
\bibitem [{\citenamefont {Rall}\ \emph {et~al.}(2019)\citenamefont {Rall}, \citenamefont {Liang}, \citenamefont {Cook},\ and\ \citenamefont {Kretschmer}}]{PhysRevA.99.062337}%
  \BibitemOpen
  \bibfield  {author} {\bibinfo {author} {\bibfnamefont {P.}~\bibnamefont {Rall}}, \bibinfo {author} {\bibfnamefont {D.}~\bibnamefont {Liang}}, \bibinfo {author} {\bibfnamefont {J.}~\bibnamefont {Cook}},\ and\ \bibinfo {author} {\bibfnamefont {W.}~\bibnamefont {Kretschmer}},\ }\bibfield  {title} {\bibinfo {title} {{Simulation of qubit quantum circuits via Pauli propagation}},\ }\href {https://doi.org/10.1103/PhysRevA.99.062337} {\bibfield  {journal} {\bibinfo  {journal} {Physical Review A}\ }\textbf {\bibinfo {volume} {99}},\ \bibinfo {pages} {062337} (\bibinfo {year} {2019})}\BibitemShut {NoStop}%
\bibitem [{\citenamefont {Giurgica-Tiron}\ \emph {et~al.}(2020)\citenamefont {Giurgica-Tiron}, \citenamefont {Hindy}, \citenamefont {LaRose}, \citenamefont {Mari},\ and\ \citenamefont {Zeng}}]{giurgica2020digital}%
  \BibitemOpen
  \bibfield  {author} {\bibinfo {author} {\bibfnamefont {T.}~\bibnamefont {Giurgica-Tiron}}, \bibinfo {author} {\bibfnamefont {Y.}~\bibnamefont {Hindy}}, \bibinfo {author} {\bibfnamefont {R.}~\bibnamefont {LaRose}}, \bibinfo {author} {\bibfnamefont {A.}~\bibnamefont {Mari}},\ and\ \bibinfo {author} {\bibfnamefont {W.~J.}\ \bibnamefont {Zeng}},\ }\bibfield  {title} {\bibinfo {title} {Digital zero noise extrapolation for quantum error mitigation},\ }in\ \href@noop {} {\emph {\bibinfo {booktitle} {IEEE International Conference on Quantum Computing and Engineering}}}\ (\bibinfo {year} {2020})\ pp.\ \bibinfo {pages} {306--316}\BibitemShut {NoStop}%
\bibitem [{\citenamefont {James}\ \emph {et~al.}(2001)\citenamefont {James}, \citenamefont {Kwiat}, \citenamefont {Munro},\ and\ \citenamefont {White}}]{PhysRevA.64.052312}%
  \BibitemOpen
  \bibfield  {author} {\bibinfo {author} {\bibfnamefont {D.~F.~V.}\ \bibnamefont {James}}, \bibinfo {author} {\bibfnamefont {P.~G.}\ \bibnamefont {Kwiat}}, \bibinfo {author} {\bibfnamefont {W.~J.}\ \bibnamefont {Munro}},\ and\ \bibinfo {author} {\bibfnamefont {A.~G.}\ \bibnamefont {White}},\ }\bibfield  {title} {\bibinfo {title} {Measurement of qubits},\ }\href {https://doi.org/10.1103/PhysRevA.64.052312} {\bibfield  {journal} {\bibinfo  {journal} {Physical Review A}\ }\textbf {\bibinfo {volume} {64}},\ \bibinfo {pages} {052312} (\bibinfo {year} {2001})}\BibitemShut {NoStop}%
\end{thebibliography}%

\end{document}